\documentclass[review,3p]{elsarticle}
\journal{Neural Networks}

\usepackage{times}
\usepackage{xcolor}
\usepackage{graphicx}
\usepackage{amssymb}
\usepackage{multicol,multirow}
\usepackage{booktabs}
\usepackage{bigstrut}
\usepackage{floatrow}
\usepackage{rotating}
\usepackage{adjustbox}
\usepackage{amsmath}
\usepackage{hyperref}
\usepackage{color,soul}

\begin{document}

\begin{frontmatter}

\title{Retinal Vessel Segmentation via a Multi-resolution Contextual Network and Adversarial Learning}

\author[inst1]{Tariq M. Khan\corref{cor1}}
\cortext[cor1]{Corresponding author}
\ead{tariq045@gmail.com}
\author[inst2]{Syed S. Naqvi}
\author[inst3]{Antonio Robles-Kelly}
\author[inst1]{Imran Razzak}
\address[inst1]{School of Computer Science and Engineering, University of New South Wales, Sydney, NSW, Australia}
\address[inst2]{Department of Electrical and Computer Engineering, COMSATS University Islamabad, Pakistan}
\address[inst3]{School of Information Technology, Faculty of Science Engineering \& Built Environment, Deakin University, Locked Bag 20000, Geelong}
\begin{abstract}
Timely and affordable computer-aided diagnosis of retinal diseases is pivotal in precluding blindness. Accurate retinal vessel segmentation plays an important role in disease progression and diagnosis of such vision-threatening diseases. 
To this end, we propose a Multi-resolution Contextual Network (MRC-Net) that addresses these issues by extracting multi-scale features to learn contextual dependencies between semantically different features and using bi-directional recurrent learning to model former-latter and latter-former dependencies. Another key idea is training in adversarial settings for foreground segmentation improvement through optimization of the region-based scores. This novel strategy boosts the performance of the segmentation network in terms of the Dice score (and correspondingly Jaccard index) while keeping the number of trainable parameters comparatively low. We have evaluated our method on three benchmark datasets, including DRIVE, STARE, and CHASE, demonstrating its superior performance as compared with competitive approaches elsewhere in the literature.
\end{abstract}


\end{frontmatter}

\section{Introduction}
Diabetic retinopathy (DR) \cite{M.Khan2020} primarily affects the working-age population around the world \cite{Khawaja2019, khan2021residual} and is the principal cause of blindness. Recent investigations reveal that a substantial number of patients face vision deterioration due to delayed follow-ups, referrals, and treatment \cite{Foot2017loss,iqbal2022recent}. 
Computer-aided screening and diagnosis of such diseases is promising, can augment the deficient screening resources and help clinicians effectively use their time \cite{Fraz2012}.
The role of retinal vascular morphology for the detection and progression of DR has been established by various studies \cite{crosby2012retinal,habib2014association}. The studies suggest that the measurements of vessel structure can be predictive of disease progression and can be utilized to determine disease severity \cite{crosby2012retinal}.

The segmentation of the retinal vessels is an important and challenging step in the construction of an automated diagnosis system. The key signs of DR including hemorrhages and microaneurysms normally occur in the vessel surroundings, thus urging the need for robust vessel segmentation. Moreover, the topology of the vascular tree can be estimated by using convincing retinal vessel segmentation \cite{Zhao2018}. Physiological characteristics of the retinal vasculature (including shape, length, branching and diameter) have other important applications as well, such as identification, fundus image registration and categorization of retinal arteries and veins \cite{Zhao2018, Zheng2016}.

Important challenges of vessel segmentation are intensity inhomogeneity, contrast issues and vessel characteristics variations. The presence of exudates, lesions, and hemorrhages can further complicate the task at hand. Many studies for automatic segmentation of vessels by means of computer vision with either supervised or unsupervised algorithms \cite{iqbal2022g,Khawaja2019a,khan2018robust,arsalan2022prompt,khan2023simple,Guo2019,Zhao2019Synthesis,naqvi2023glan} have been reported. Studies using deep learning architectures have, in particular, been found to be more effective than alternatives \cite{Soomro2019,Soomro19,Xiuqin19}.

In the literature, numerous DL-based approaches with promising results have been proposed for retinal vessel segmentation \cite{M.Khan2020,iqbal2022g,khan2022t,khan2021rc,arsalan2022prompt}. For instance, U-Net \cite{Ronneberger2015} and its variants have proven to be useful for various medical image segmentation tasks. Gu \textit{et~al.} \cite{Gu2019CENetCE} presented a vessel segmentation network that preserved spatial information by capturing high-level contextual information. In another approach, Yan \textit{et~al.} \cite{Yan2018} introduced a joint loss including both a pixel-wise and a segmentation-level cost in U-Net for improved segmentation. DEU-Net \cite{Wang2019a} incorporated a function fusion module to capture semantic information. In \cite{Wang2019a}, a context path with a multi-scale convolution block is used. DeepVessel \cite{Fu2016} employed a multi-scale convolution neural network (CNN) at multiple layers to learn a rich hierarchical representation. Moreover, a conditional random field was employed to capture pixel interactions.

Despite, the efficacy of encoder-decoder network variants, these methods generally struggle in retention or reassurance of feature information between the encoder and the decoder. This can be generally attributed to their feature extraction strategy at the encoder and feature fusion policy at the decoder. There have been attempts to preserve the spatial information \cite{Gu2019CENetCE,Wang2019a}, through effective feature extraction and contextual information. Moreover, standalone adversarial learning has been explored before in the literature \cite{son2019towards} with a vanilla U-shaped generator. However, a systematic association of these design stages and their contribution to overall performance improvement still needs more exploration.

\begin{table}[!b]
  \centering
  \caption{Computational requirements of the proposed MRC-Net with current state methods.}
    \begin{tabular}{lcc}
    \toprule
    \multicolumn{1}{c}{Network} & \# of Parameters (M) & Model Size (MB) \\
    \midrule
    MobileNet-V3-small \cite{Howard2019MobileNet} & 2.5  & 11.0 \\
    ERFNet \cite{Romera2018ERFNet} & 2.06 & 8.0 \\
    MultiRes UNet \cite{IBTEHAZ202074} & 7.2  & -- \\
    VessNet \cite{Arsalan2019} & 9.3  & 36.6  \\
    MRC-Net & 0.9  & 3.85 \\
    \bottomrule
    \end{tabular}%
  \label{tab:timingComparison}%
\end{table}%

In this work, a Multi-resolution Contextual Network (MRC-Net) that extracts multi-scale features for spatial information preservation and learning contextual dependencies between semantically different features is introduced. The computational requirements are kept under check through adversarial learning-based performance enhancement.
To this end, to tackle the challenge of the adversarial nature of smartphone-based retinal image data, we incorporate a bi-directional recurrent block that captures former-to-latter and latter-to-former dependencies. By adding this block, the method proposed here is able to effectively capture both contextual dependencies in time between semantically varying features. Multi-scale features are extracted to effectively capture vessel information at all scales, ensuring special attention is paid to tiny vessels with degraded quality. To further boost the performance of the segmentation network against the noisy nature of the data, while simultaneously preventing it from exploding in terms of trainable parameters, the training of the network is performed in an adversarial end-to-end setting. The general architecture of the proposed MRC-Net is designed to be shallow by carefully setting the number of encoder-decoder blocks according to the representational requirements of the features.

The contributions of this study are as follows:
\begin{itemize}
    \item Multi-resolution feature extraction preserves rich information on vessels occurring at varying scales. Loss of information during the merging of information from multiple scales is prevented through residual learning, resulting in improved performance.
    \item Robust context-based feature fusion is introduced by capturing space-time relationships between encoder and decoder features that aid the network to recuperate information that may be lost during the convolution process.
    \item Adversarial learning is employed to improve the segmentation performance by explicit optimization of the Dice score of the output foreground maps. 
    \item The proposed approach demonstrated competitive performance with only 0.9M parameters. 
    \item A systematic assessment of the various design choices at network stages including feature extraction, feature fusion, loss function, and training strategy is presented for improved retinal vessel segmentation.
\end{itemize}

\section{Multi-resolution Contextual Network (MRC-Net)}\label{method}

In this work, we propose an adversarial learning-based network with multi-scale contextual features for robust vessel segmentation. The architectural details, the loss function, and the training strategy are presented in detail in the following sections.

\subsection{Overall Architecture}\label{overallArch}
\begin{figure*}
\centering
  \begin{tabular}{cc}
  \includegraphics[width=0.7\textwidth]{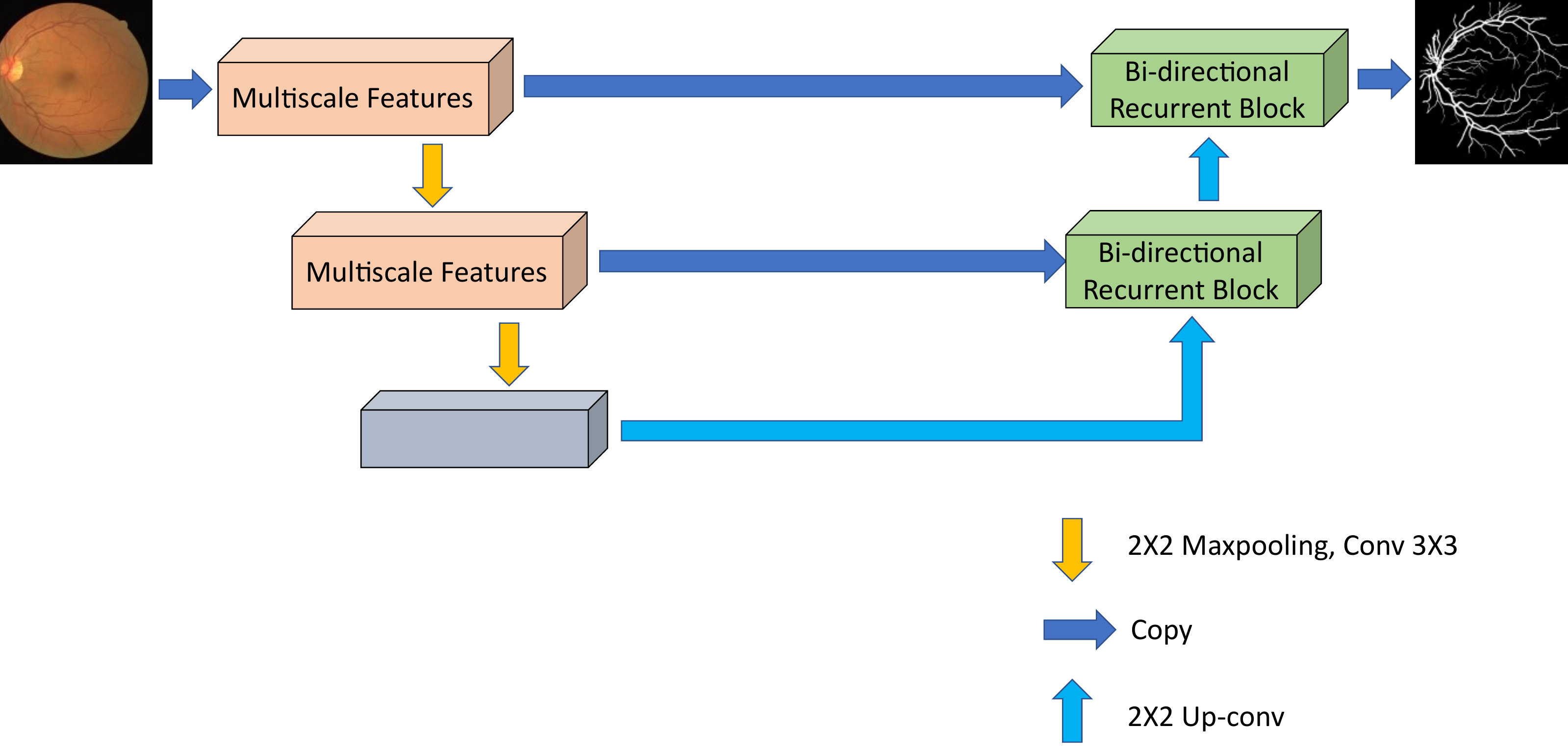} \\ &  (a) \\
  \includegraphics[width=0.7\textwidth]{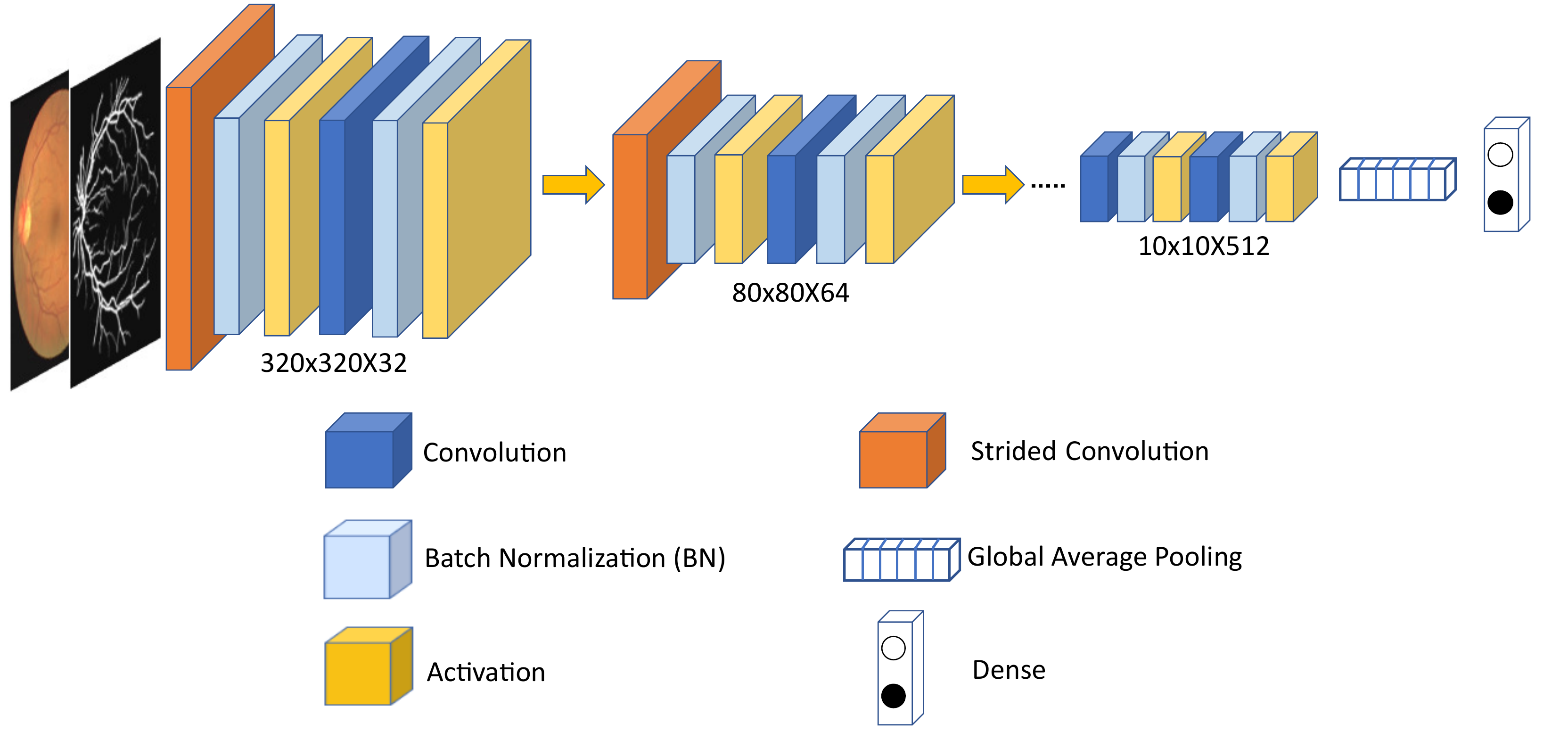} \\
   \\ & (b)  \\
  \end{tabular}
  \caption{a) Block level diagram of the generator network; b) Illustration of the discriminator network. }\label{blockDiagram}
\end{figure*}

The proposed architecture is a generative adversarial network (GAN) with generator and discriminator networks as shown in Figure~\ref{blockDiagram} (a) and (b), respectively. During training, the generator network generates a map with probabilities assigned to retinal vessels from an original fundus image. These maps range from 0 to 1 indicating the likelihood of a particular pixel belonging to a vessel. Given a fundus image and its corresponding vessel image, the job of the discriminator is to determine whether the vessel image corresponds to a hand-labelled vessel map or the generator output. The two networks compete in an adversarial manner during training to get better at their respective tasks. During inference, the trained generator network takes fundus images as input and generates retinal probability maps. The key idea behind the application of the GAN in this work is to make the generator more robust to the adversarial nature of the vessel segmentation problem without making it deeper and, hence, more computationally expensive. The generator network obtained through this setting is more robust as compared with those networks yielded via stand-alone training.

The proposed generator is an encoder-decoder fully convolutional network as shown in Figure~\ref{blockDiagram} (a). The key ideas of the generator network are the following:
\begin{enumerate}
    \item To achieve multi-scale feature extraction through the multi-resolution feature extraction module at the encoder as described in Section~\ref{MRFEM}. 
    \item To achieve robust context-based feature fusion at the decoder stage as described in Section~\ref{RFF}. Contextual learning has been employed in the past for improving performance \cite{Gu2019CENetCE,Wang2019a}, however, in contrast, we devise contextual learning for robust feature fusion by exploring the time dependencies between features at different network stages. In turn, the network latently recovers the information lost due to the convolution process. Moreover, the proposed context-aware fusion is systematically evaluated with other key design stages including multi-resolution feature extraction and adversarial training.
\end{enumerate}
Note that the two multi-resolution modules encompass all important scales thus capturing the required variations including the thickest and the thinnest vessel structures. As opposed to residual learning for bridging the semantic gap between encoder and decoder features, the multi-scale encoder features are fused with the decoder ones through respective bi-directional recurrent blocks.

The architecture of the discriminator network is presented in Figure~\ref{blockDiagram} (b). In the discriminator network, the features are extracted from the tensor by nine successive convolutions followed by batch normalization and activation. The feature maps are downsampled by max-pooling and strided convolution operations with a $2 \times 2$ filter size. The dimensionality-reduced feature maps are passed through global average pooling followed by dense prediction. The output of the network is either of the two classes, i.e., a human-annotated vessel map or a machine-generated segmentation.

During adversarial training, we specifically optimize the Dice loss. Moreover, during multiple rounds of adversarial learning our objective is to maximize the Dice score and  indirectly the Jaccard index. Specifically, we select the training round that maximizes the Dice or F1 score as
\begin{equation}
\mathrm{arg} \; \max_{F} \mathrm{GAN} (r),
\end{equation}
where $r$ is the number of training rounds. Our choice of optimizing the region-based scores (Dice score, Jaccard index) originates from prior works that demonstrate the superiority of the region-based type measures as compared to AUC and AP measures  for evaluation of foreground maps \cite{Margolin2014}.

\subsection{Multi-resolution Feature Extraction Module}\label{MRFEM}
The architectural detail of the multi-resolution feature extraction module is depicted in Figure~\ref{multResBlock}. It can be observed that we employed 3$\times$3, 5$\times$5, and 7$\times$7 receptive fields to extract multi-scale contextual information. During our experiments, we found that these three receptive field sizes capture the wide variation of vessels in terms of scale, and going narrower or deeper does not affect the performance. The shortcut connection is introduced to ensure that residual information lost during the filtering process is preserved and fed to the later stages and to also ease the training process. The proposed multi-resolution feature extraction approach derives the idea of factorizing convolutions from the well-known work of Szegedy and colleagues \cite{Szegedy2017}. This promotes parameter reduction as a result of weight sharing.

\subsection{Robust Feature Fusion}\label{RFF}
Recall that an inherent problem in the design of encoder-decoder architectures is the semantic gap between the encoder and decoder features. There are a number of options to consider in this regard, including residual learning \cite{Szegedy2017}, dense learning \cite{Zongwei2018}, and recurrent learning \cite{JWang2016}. Here we have opted to employ bi-directional recurrent learning to learn the dependencies between semantically lagging encoder and leading decoder features. The motivation for this resides in our architecture, which lends itself naturally to fusing encoder and decoder features. Thus, here we employ a two-dimensional bi-directional long-short-term memory convolutional layer at each encoder-decoder stage of the proposed generator network.

\subsection{Loss Function}
Let G be the generator that performs a mapping of an image $f$ to a vessel map $v$, i.e., $G:f\rightarrow v$. The discriminator $D:{f,v}\rightarrow{0,1}$ maps the image and vessel map to a binary value as a classification task, where 0 is for human-generated versus 1 for machine-generated $v$. Then the objective of the GAN can be treated as a minimax formulation with respect to the generator and the discriminator, given as:
\begin{equation}
\begin{split}
L_{\mathrm{GAN}}(G,D) = &\mathbb{E}_{f,v \sim p_d(f,v)}[\mathrm{log}D(f,v)]+
\\&
\mathbb{E}_{f \sim p_d(f)}\left [ \mathrm{log}(1-D(f,G(f))) \right ]
\end{split}
\end{equation}

For the segmentation task, the simplest objective is to penalize the distance between the ground truth $y$ and the predicted output $v$, i.e. the binary cross entropy
\begin{equation}
\begin{split}
L_{\mathrm{SEG}_1}(G) = \mathbb{E}_{y,v\sim p_d(l,v)}-y&\mathrm{log}(G(v))-
\\&
(1-y)\mathrm{log}(1-G(v))
\end{split}
\end{equation}
In order to evaluate the similarity between the predicted and ground truth maps, we formulate two more loss functions based on the employed similarity metric. The second loss is a sum of the binary cross entropy and the Dice loss, given as
\begin{equation}
L_{\mathrm{SEG}_2}(G) = L_{\mathrm{SEG}_1}(G)+ L_{\mathrm{Dice}},
\end{equation}
where $L_{\mathrm{Dice}} = 1 - \frac{2\left | f \bigcap v \right |}{\left | f \right |+\left |  v\right |}$.
By incorporating Intersection over Union (IoU) or the Jaccard loss $L_{\mathrm{Jaccard}}=1-\left ( \frac{\left | f \bigcap v \right |}{\left | f \right |+\left |  v\right |-\left | f \bigcap v \right |} \right )$,   
our third loss is formulated.
\begin{equation}
L_{\mathrm{SEG}}(G) = L_{\mathrm{SEG}_2}(G)+ L_{\mathrm{Jaccard}}
\end{equation}

\begin{figure*}
\centering
   \includegraphics[width=0.65\textwidth]{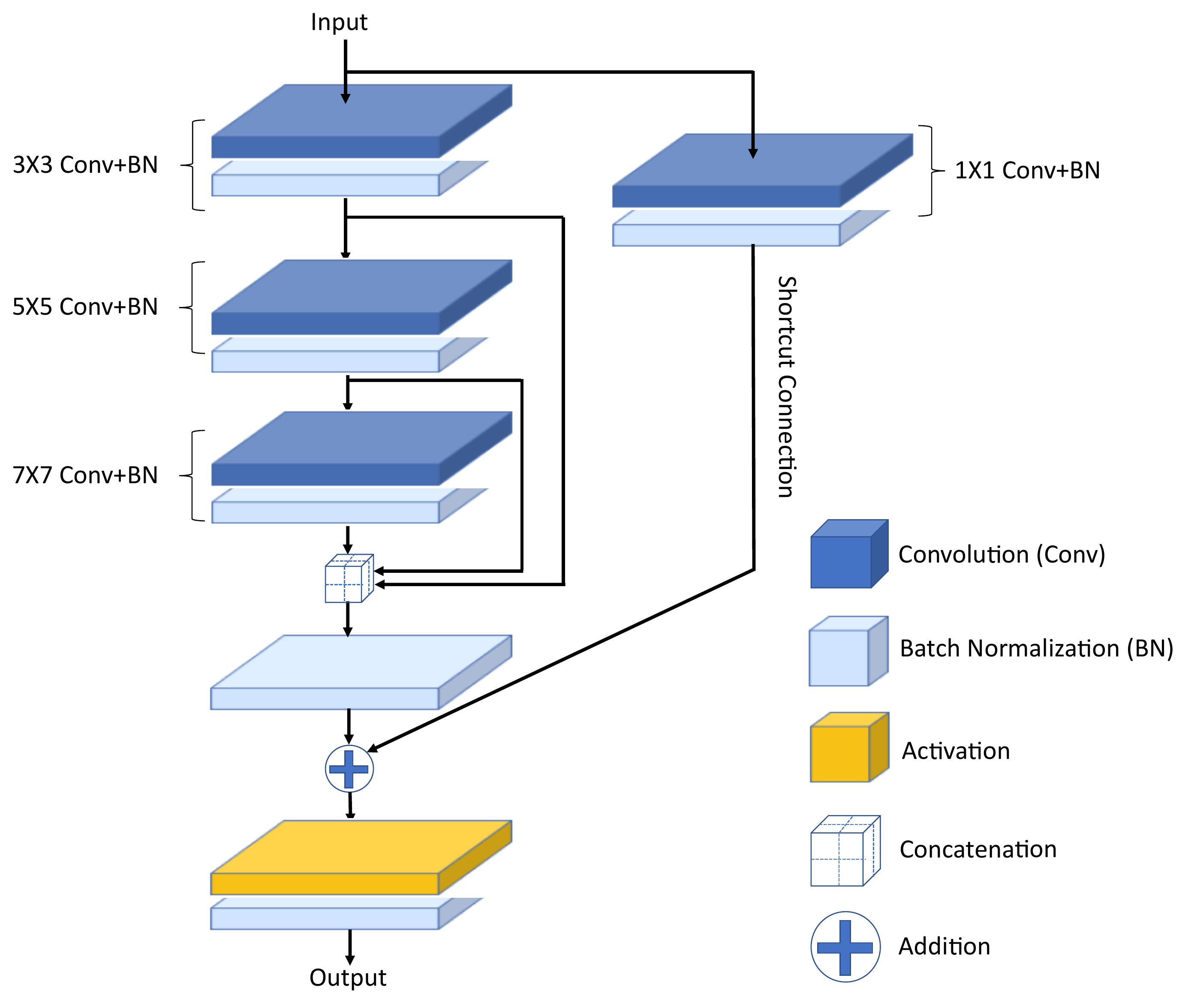}
    \caption{Architectural overview of the multi-resolution feature extraction module.}
    \label{multResBlock}
\end{figure*}

Finally, the overall loss function of the GAN is a composite loss expressed as
\begin{equation}
L_{\mathrm{COMP}} = \mathrm{arg} \; \min_{G} \left [ \mathrm{arg} \; \max_{D} L_{\mathrm{GAN}}(G,D)\right ] + \beta  L_{\mathrm{SEG}}(G),
\end{equation}
where $\beta$ is a parameter that balances the two diverse objectives.

\subsection{Datasets}

The proposed network is evaluated using CHASE \cite{Fraz2012c} 
\footnote{The dataset can found at \url{https://blogs.kingston.ac.uk/retinal/chasedb1/}}, 
DRIVE \cite{Staal2004} \footnote{The dataset is widely available at \url{https://drive.grand-challenge.org}} and 
STARE \cite{Hoover2000} \footnote{More information regarding the STARE project can be found at \url{https://cecas.clemson.edu/~ahoover/stare/}} databases. 
The DRIVE dataset came from a screening program for diabetic retinopathy in the Netherlands. It includes a total of 40 colour images, 20 for training and 20 for testing with an image size of 584$\times$565 pixels. Only seven images show signs of mild early diabetic retinopathy. Manual annotations, generated by experts, are used as the ground truth. The STARE dataset consists of 20 colour retinal images with a FOV of $35^{\circ}$ and a resolution of 700$\times$605 pixels. There are pathologies in ten of the twenty images. Two manual annotations are available as the gold standard. The annotation from the first expert is employed in this work.

The CHASE dataset includes 28 colour images, where each image is captured with a $30^{\circ}$ FOV centred at the optic disc and a resolution of 999$\times$960 pixels. As a ground truth, two different expert annotations are available. For our experiments, we use the first expert's segmentation ground truth. There are no separate training or testing sets in the CHASE dataset. The first 20 images were used for training, and the last eight images were used for testing.
%
\begin{table}[htbp]
  \centering
  \caption{Contribution of multi-resolution contextual feature extraction and bi-directional recurrent feature fusion.}
  \resizebox{0.7\columnwidth}{!}{%
    \begin{tabular}{lcccccc}
    \hline
    \multicolumn{1}{c}{\textbf{Method}} & \textbf{Acc} & \textbf{F1} & \textbf{Bacc} & \textbf{J} & \textbf{E} & \textbf{AUC} \bigstrut\\
    \hline
    MRC-Net without Multi-resolution+BConvLSTM & 0.9678 & 0.8160 & 0.8960 & 0.6895 & 0.3105 & 0.9747 \bigstrut[t]\\
    MRC-Net without Multi-resolution & 0.9680 & 0.8162 & 0.9015 & 0.6900 & 0.3100 & 0.9787 \\
    MRC-Net without BConvLSTM & 0.9696 & 0.8204 & 0.9019 & 0.6958 & 0.3042 & \textbf{0.9847} \\
    MRC-Net & \textbf{0.9698} & \textbf{0.8270} & \textbf{0.9044} & \textbf{0.7055} & \textbf{0.2945} & 0.9812 \bigstrut[b]\\
    \hline
    \end{tabular}%
    }
  \label{tab:ablationMB}%
\end{table}%
\section{Results and Discussion}\label{experimentalResults}

In this section, we compare the performance of our method on widely available datasets with current state benchmark methods.
\begin{table}[htbp]
  \centering  
  \caption{Ablation study of the adversarial training strategy.}
  \resizebox{0.7\columnwidth}{!}{%
    \begin{tabular}{rlcccccc}
    \hline
    \multicolumn{1}{c}{\textbf{Dataset}} & \multicolumn{1}{c}{\textbf{Method}} & \textbf{Acc} & \textbf{F1} & \textbf{Bacc} & \textbf{J} & \textbf{E} & \textbf{AUC} \bigstrut\\
    \hline
    \multicolumn{1}{r}{\multirow{2}[2]{*}{DRIVE}} & MRC-Net without Adversarial Training & 0.9701 & 0.8076 & 0.8966 & 0.7026 & 0.2974 & 0.9857 \bigstrut[t]\\
          & MRC-Net & 0.9698 & 0.8270 & 0.9044 & 0.7055 & 0.2945 & 0.9812 \bigstrut[b]\\
    \hline
    \multicolumn{1}{r}{\multirow{2}[2]{*}{STARE}} & MRC-Net without Adversarial Training & 0.9738 & 0.8208 & 0.9000 & 0.7000 & 0.3000 & 0.9780 \bigstrut[t]\\
          & MRC-Net & 0.9747 & 0.8286 & 0.9032 & 0.7105 & 0.2895 & 0.9684 \bigstrut[b]\\
    \hline
    \multicolumn{1}{r}{\multirow{2}[2]{*}{CHASE}} & MRC-Net without Adversarial Training & 0.9781 & 0.8269 & 0.9147 & 0.7067 & 0.2933 & 0.9852 \bigstrut[t]\\
          & MRC-Net & 0.9779 & 0.8548 & 0.9186 & 0.7466 & 0.2534 & 0.9850 \bigstrut[b]\\
    \hline
    \end{tabular}%
    }
  \label{ablation1}%
\end{table}%

\begin{table}[htbp]
  \centering
  \caption{Effect of loss functions on segmentation performance.}
    \begin{tabular}{lcccccc}
    \hline
    \multicolumn{1}{c}{\textbf{Method}} & \textbf{Acc} & \textbf{F1} & \textbf{Bacc} & \textbf{J} & \textbf{E} & \textbf{AUC} \bigstrut\\
    \hline
    MRC-Net+Dice & 0.9693 & 0.8257 & 0.9071 & 0.7036 & 0.2964 & 0.9742 \bigstrut[t]\\
    MRC-Net+IoU & 0.9700 & 0.8285 & 0.9062 & 0.7077 & 0.2923 & 0.9761 \bigstrut[b]\\
    \hline
    \end{tabular}%
  \label{losses}%
\end{table}%

\subsection{Experimental Setup}
The tests were performed on a computer system running Ubuntu 20.04.1 LTS and equipped with an Intel(R) Xeon(R) Gold 5120 CPU @ 2.20GHz processor, 260GB of RAM, and a GeForce GTX1080TI GPU \footnote{The code can be downloaded at: \textcolor{magenta}{We will make our code publicly available upon acceptance of this paper}}. For our implementation, the Keras 2.2.4 library was employed. The ADAM was employed and the learning rate was initialized to $2 \times 10^-5$ with an exponential decay rate of 0.9. The images were resized to 640 pixels as a pre-processing step for all datasets. The z-score statistic normalization and augmentation techniques including contrast enhancement, random flipping, and random rotation between 1-360 degrees were applied. The total number of training rounds $r$ for the GAN was set to 10\footnote{Our adversarial training implementation is adapted from the Keras implementation \url{https://github.com/jaeminSon/V-GAN}}. The balancing parameter $\beta$ in $L_{\mathrm{COMP}}$ was fixed to 10 in our experiments so as to place more emphasis on the segmentation task. In $L_{\mathrm{SEG}_4}(G)$, $\gamma$ was set to 0.5 to balance the cross entropy and the Dice loss. The total trainable parameters of the proposed generator network for input size of 640$\times$640$\times$3 is $\sim$ 0.9M.

\begin{figure*}[h!]
	\centering
	\resizebox{0.85\textwidth}{!}{%
		\begin{tabular}{ccccccc}
            Image & Ground Truth & BCDUNet & MultiResUNet & SegNet & U-Net++ & MRC-Net \\
			\includegraphics[width=0.14\textwidth]{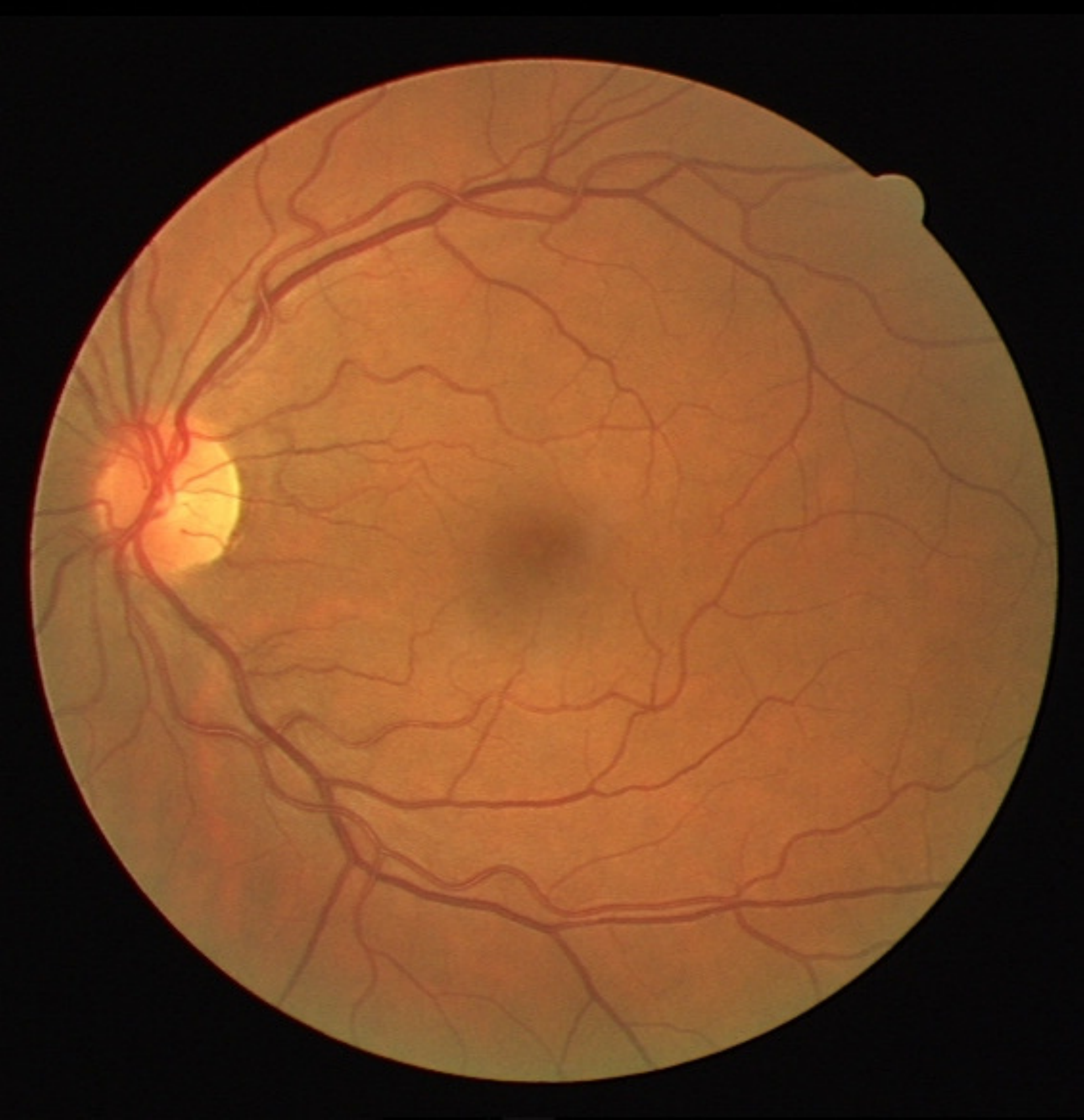}   & \includegraphics[width=0.14\textwidth]{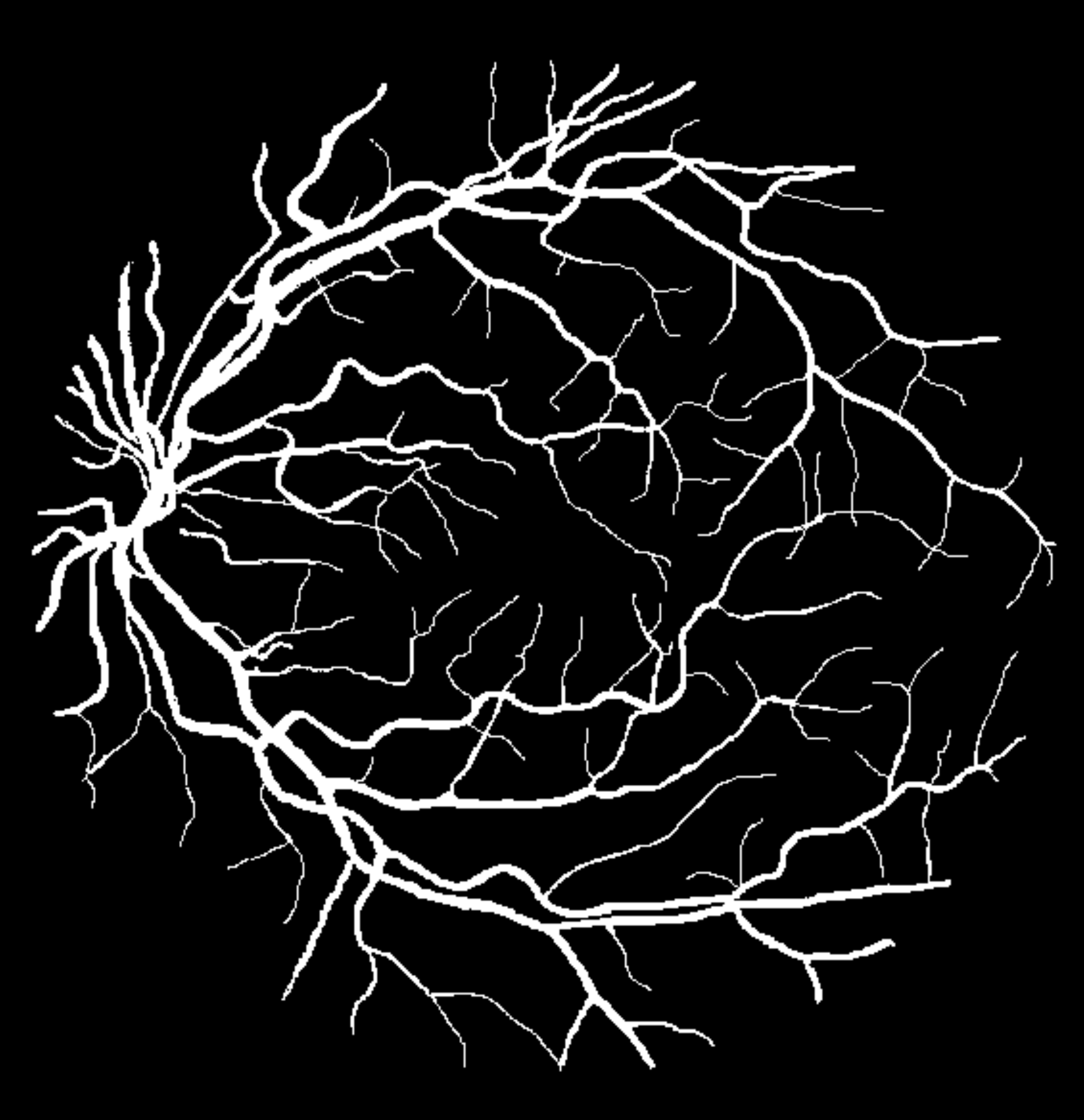}    & \includegraphics[width=0.14\textwidth]{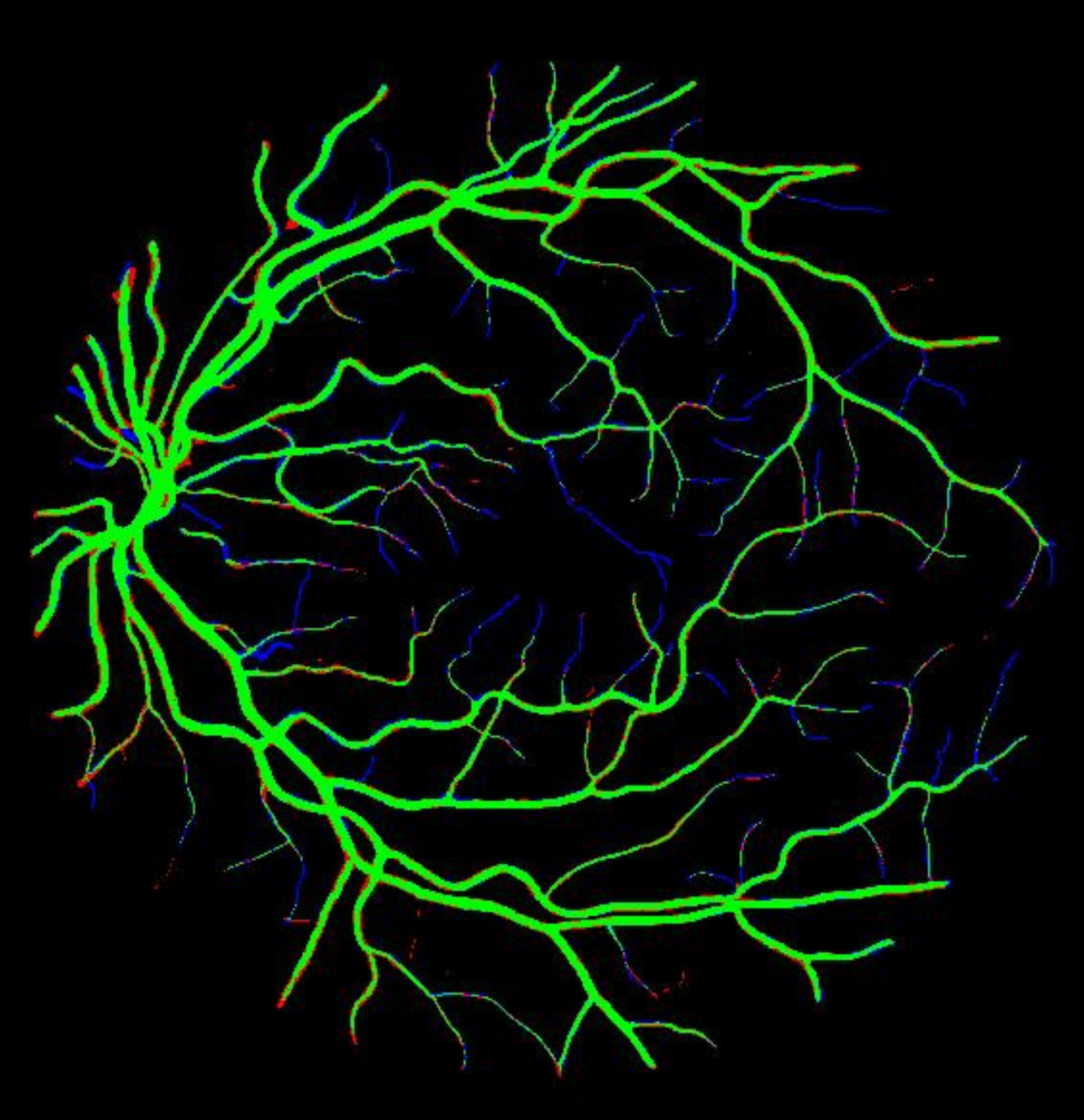}       & \includegraphics[width=0.14\textwidth]{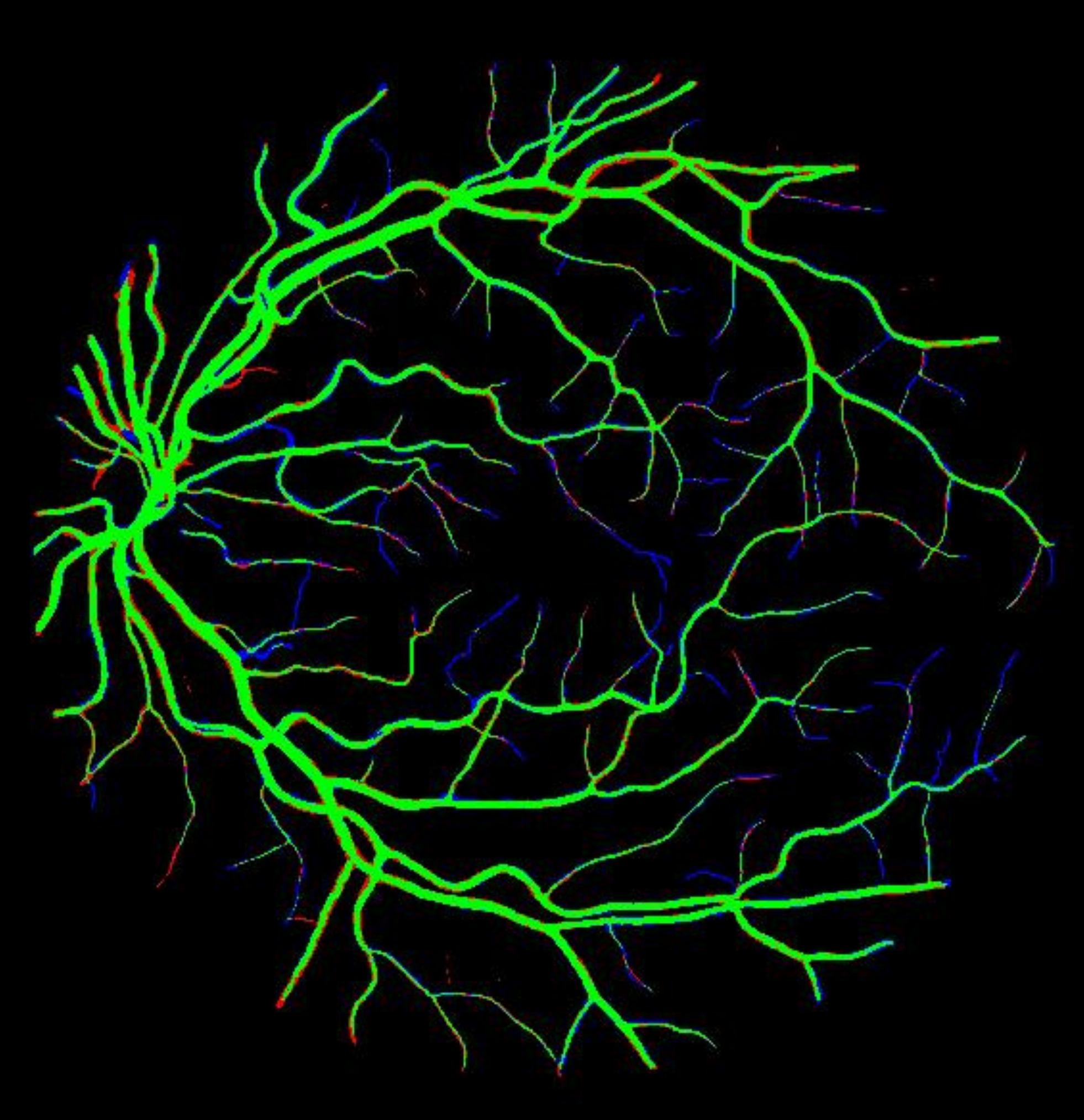}       & \includegraphics[width=0.14\textwidth]{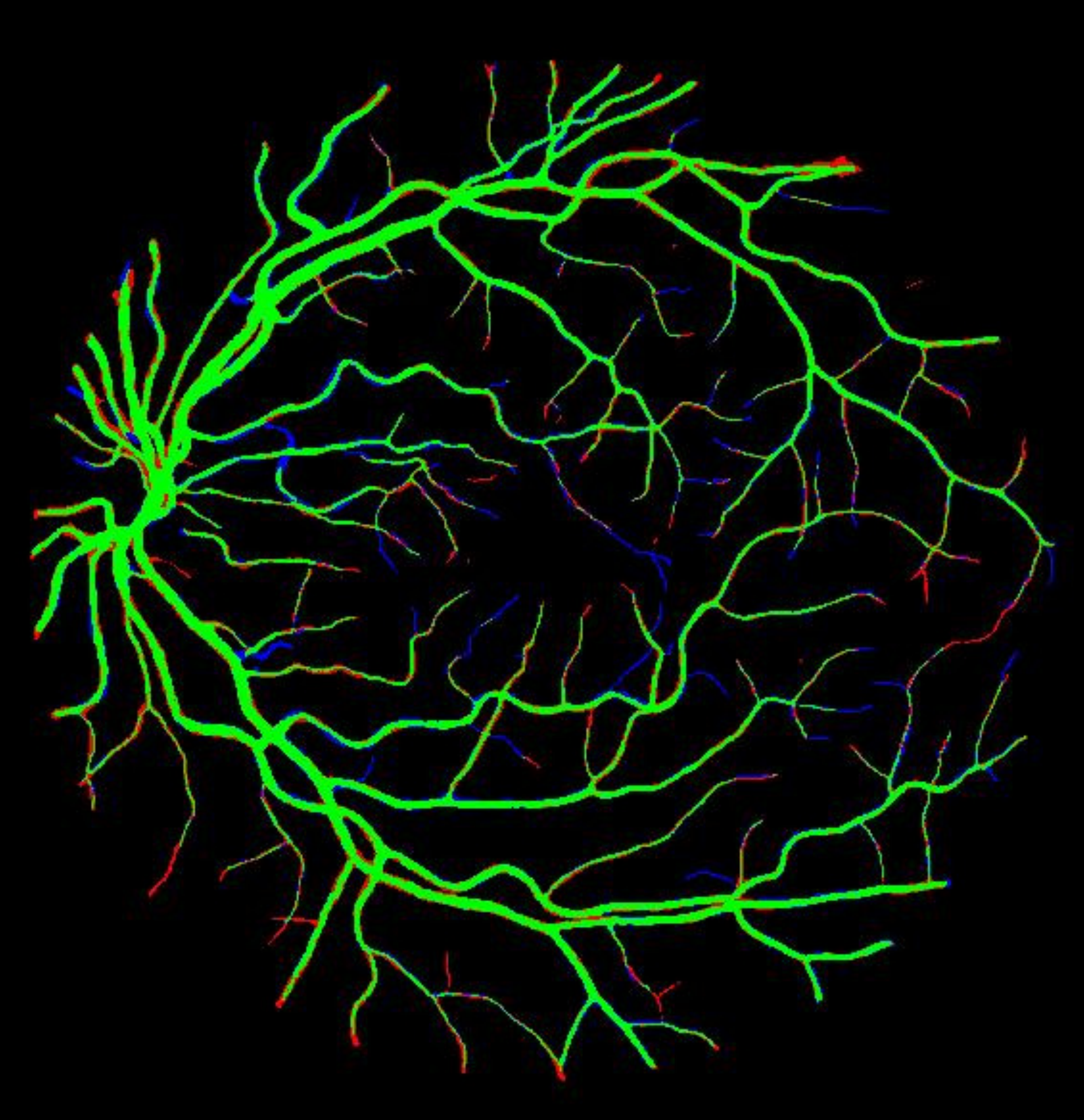}      & \includegraphics[width=0.14\textwidth]{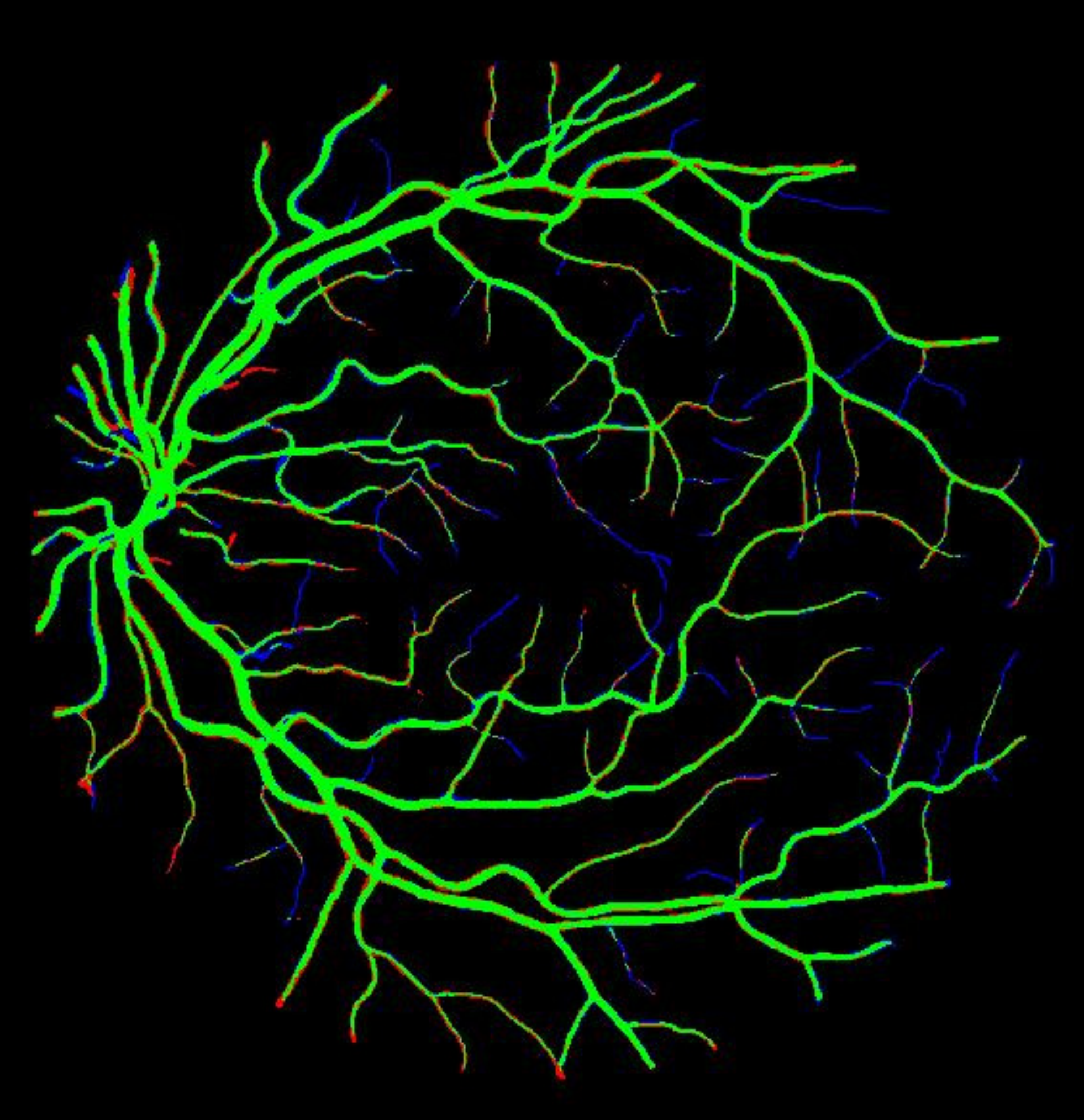}      & \includegraphics[width=0.14\textwidth]{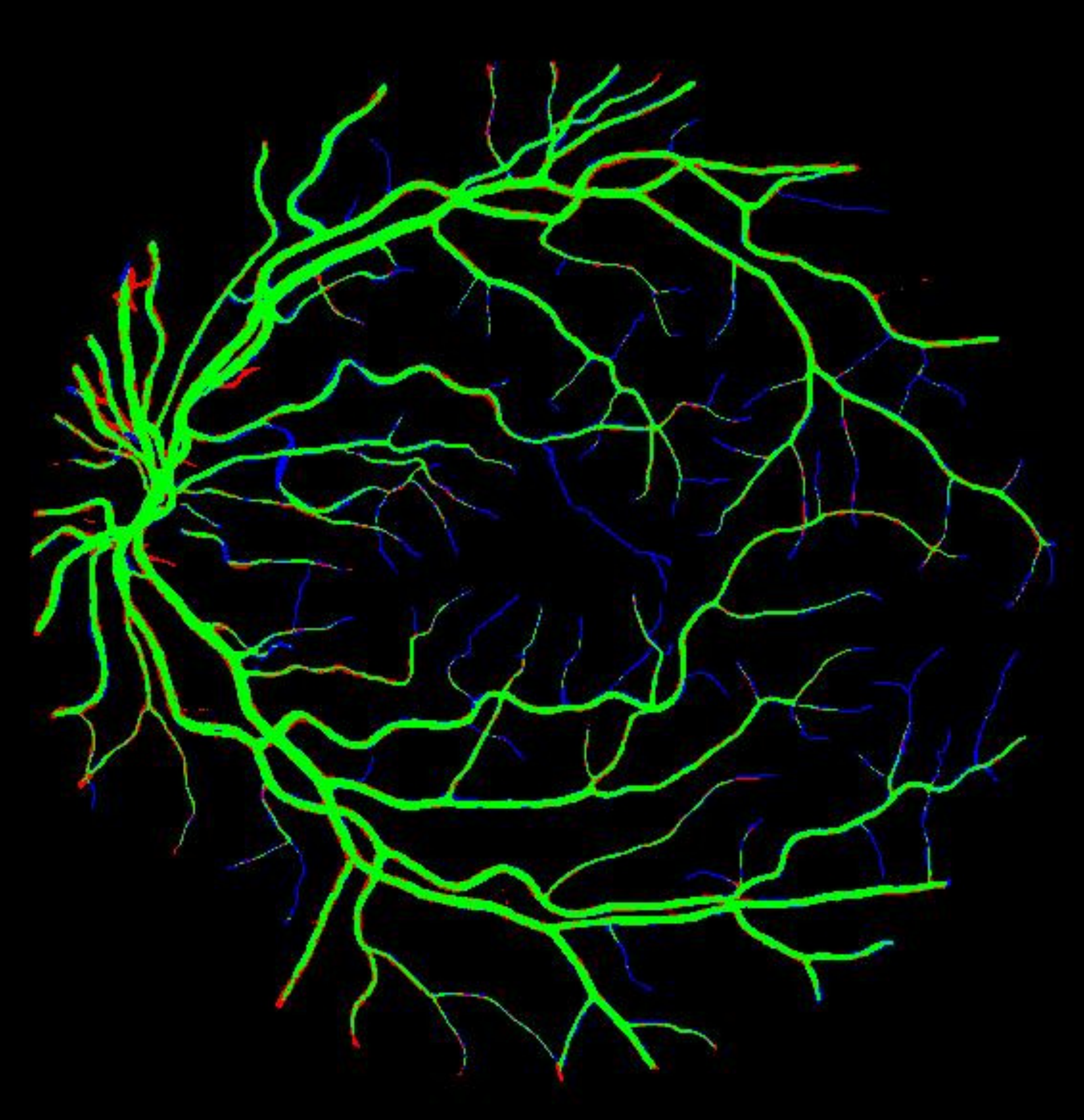} \\
			\includegraphics[width=0.14\textwidth]{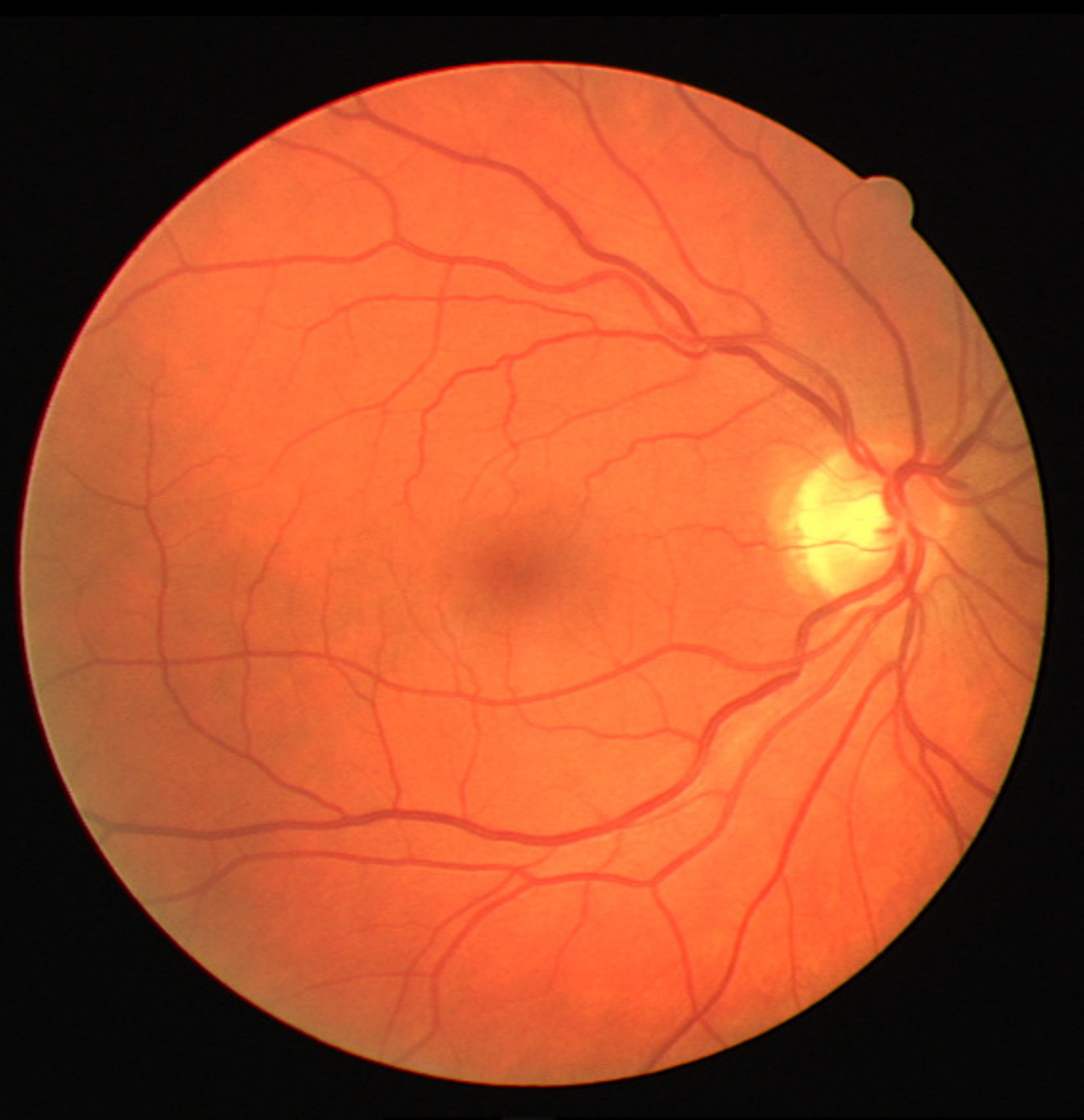}   & \includegraphics[width=0.14\textwidth]{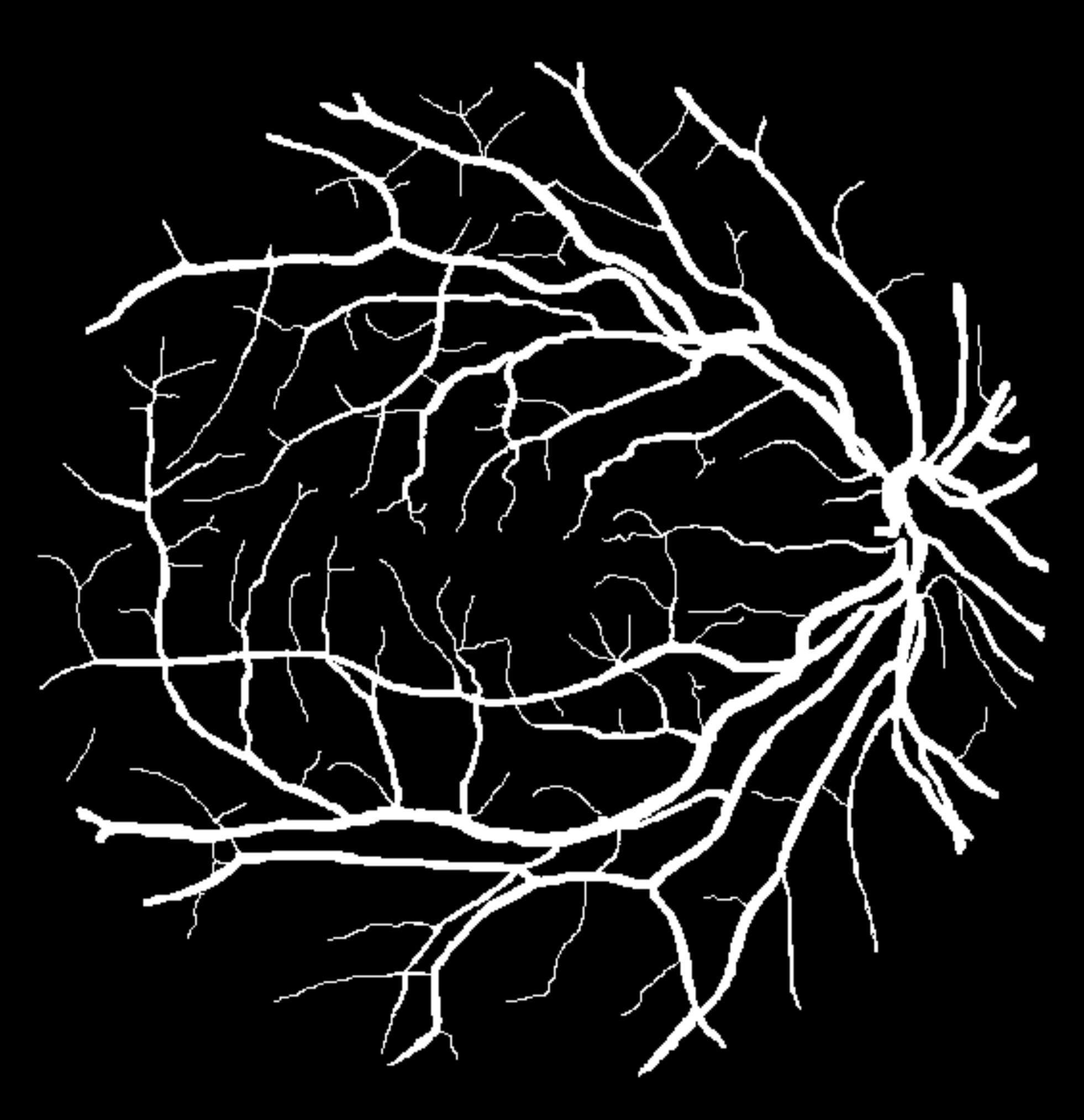}      &  \includegraphics[width=0.14\textwidth]{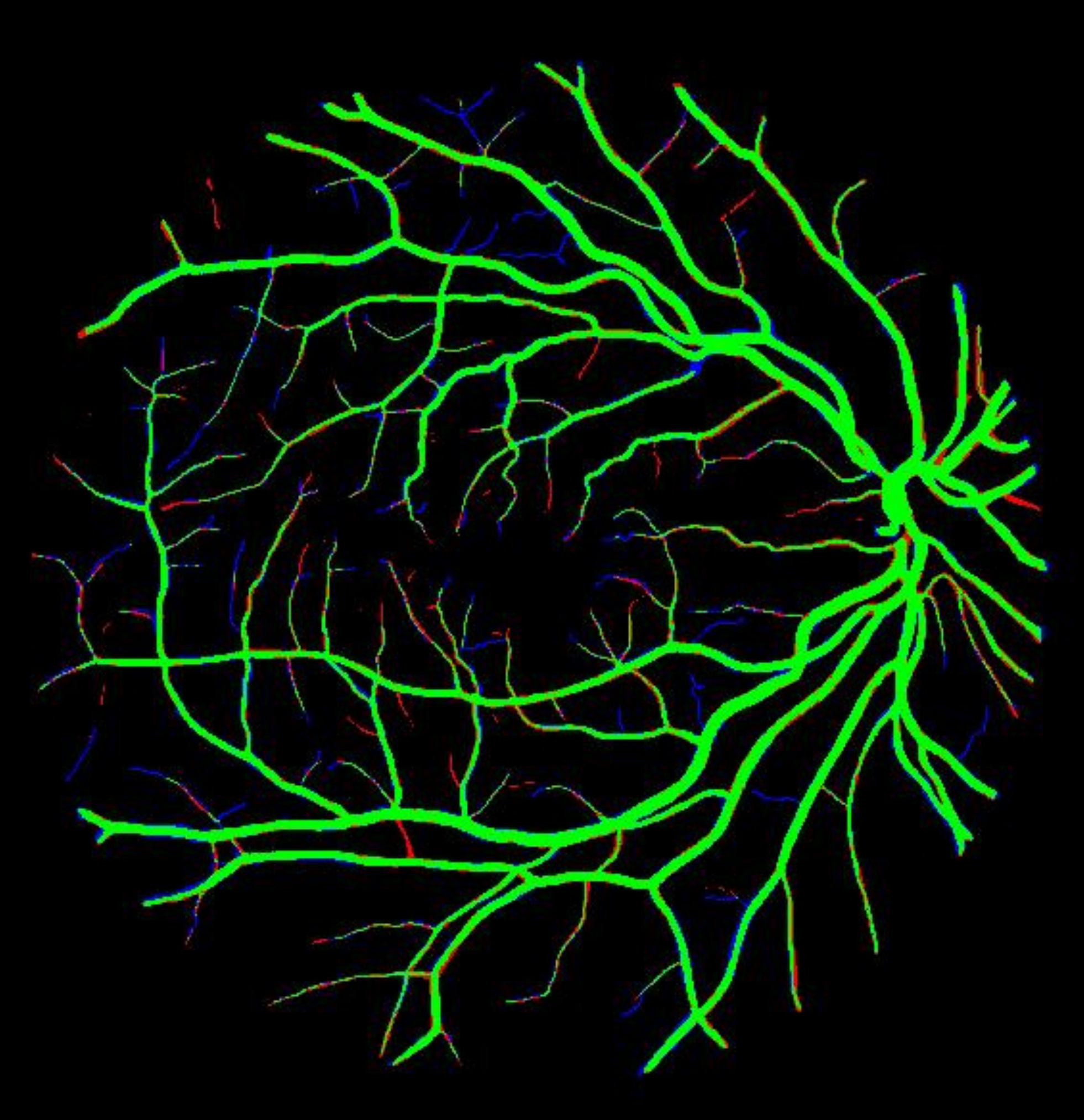}       & \includegraphics[width=0.14\textwidth]{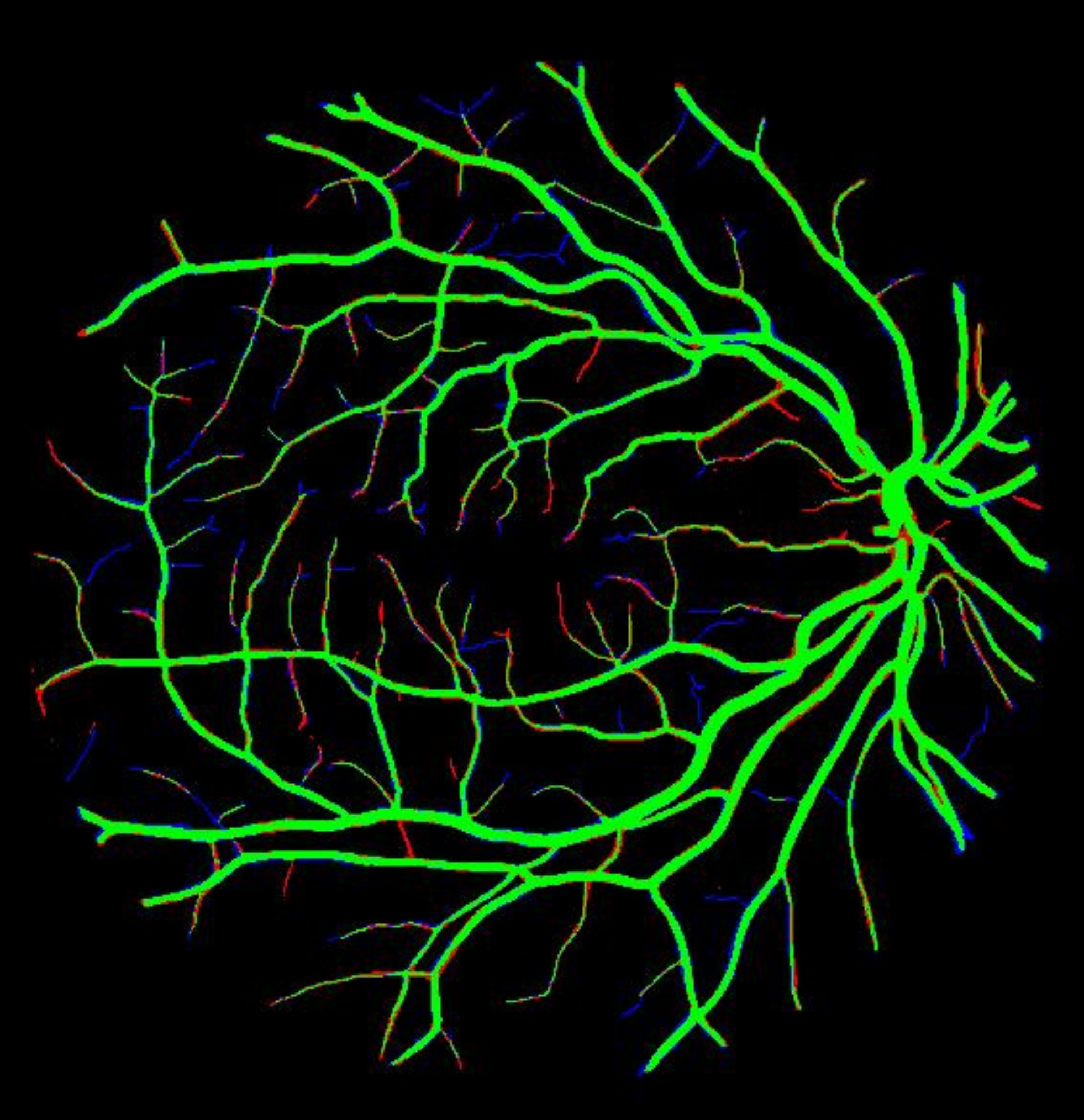}       & \includegraphics[width=0.14\textwidth]{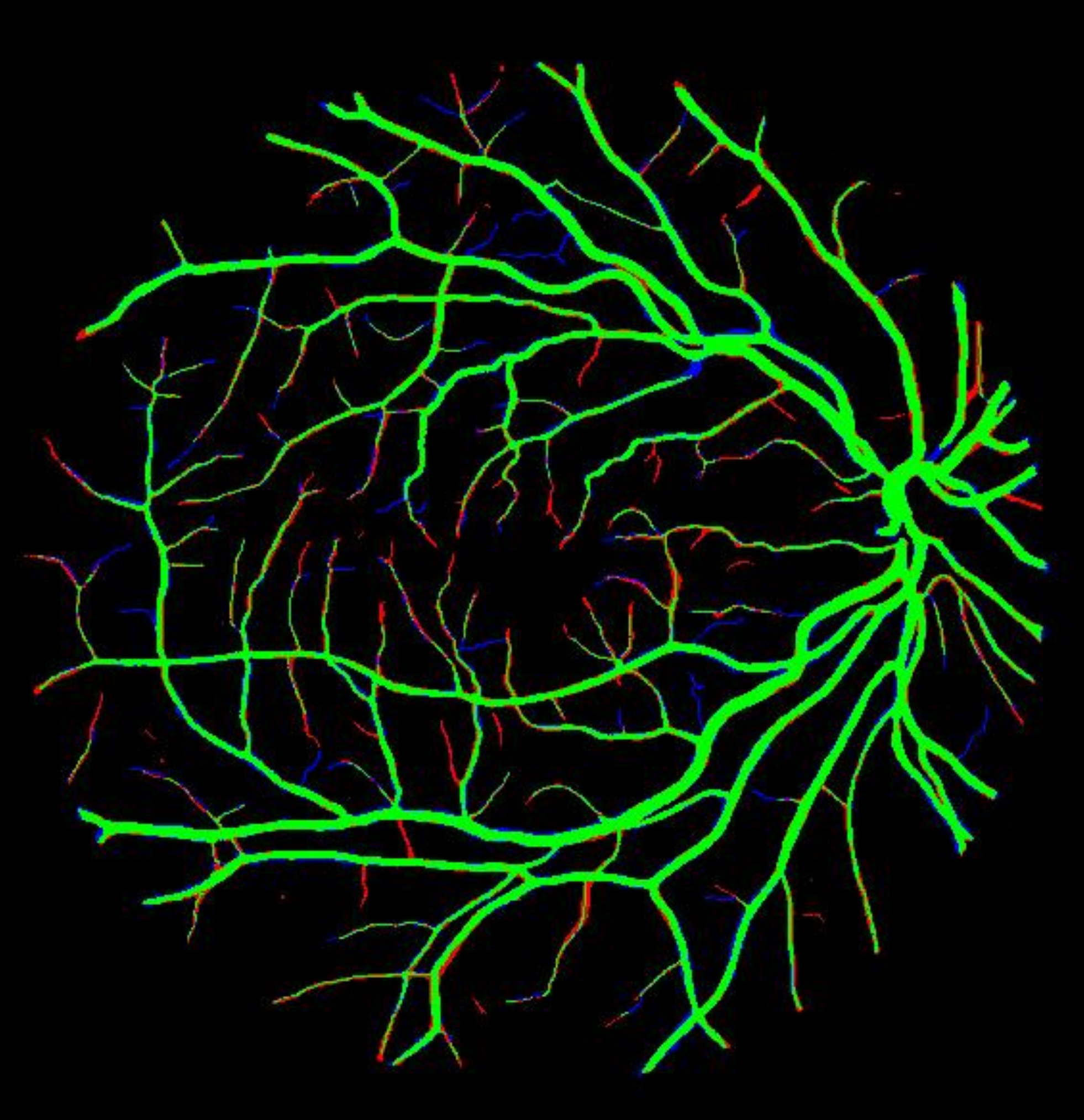}      & \includegraphics[width=0.14\textwidth]{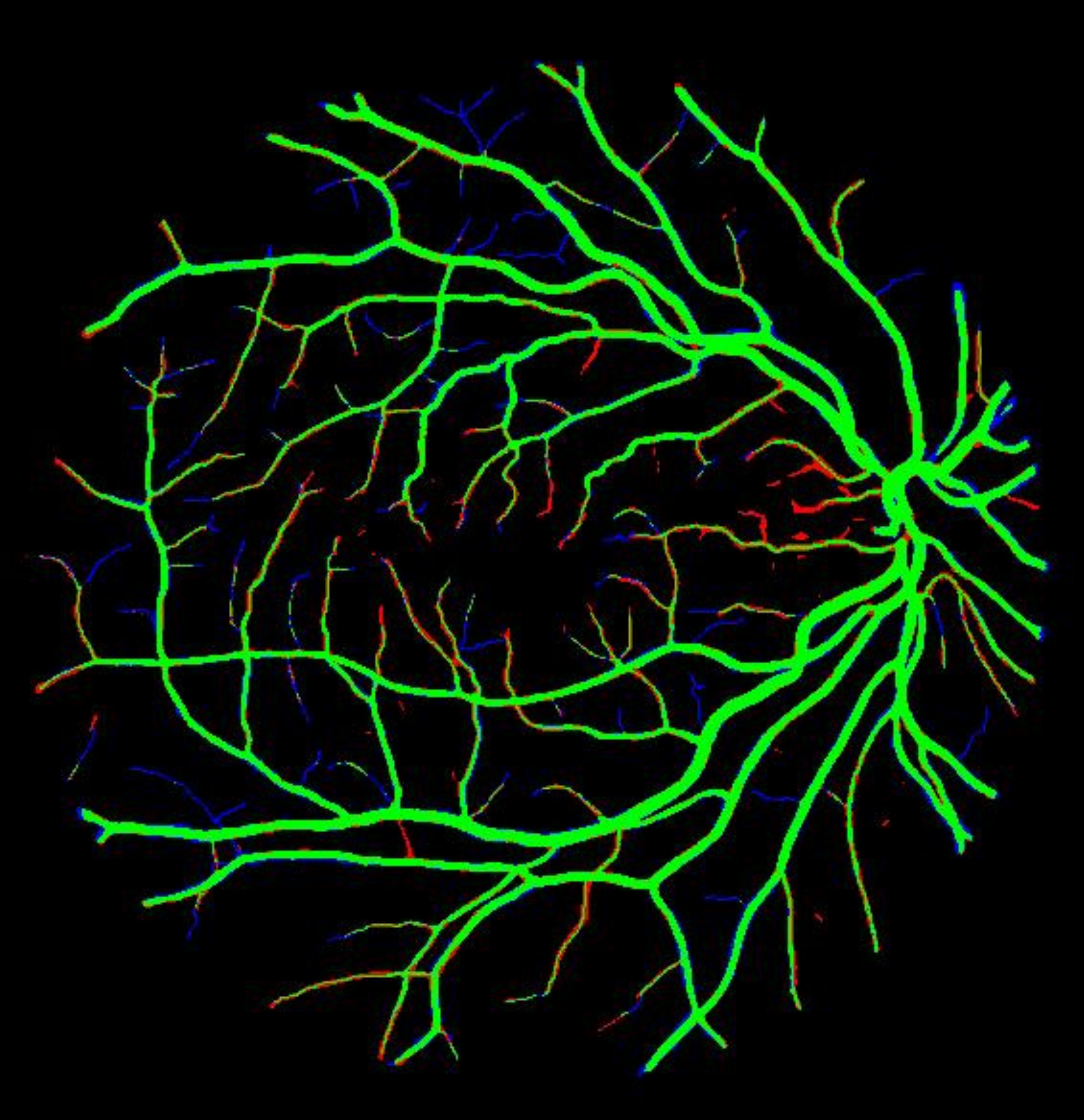}      & \includegraphics[width=0.14\textwidth]{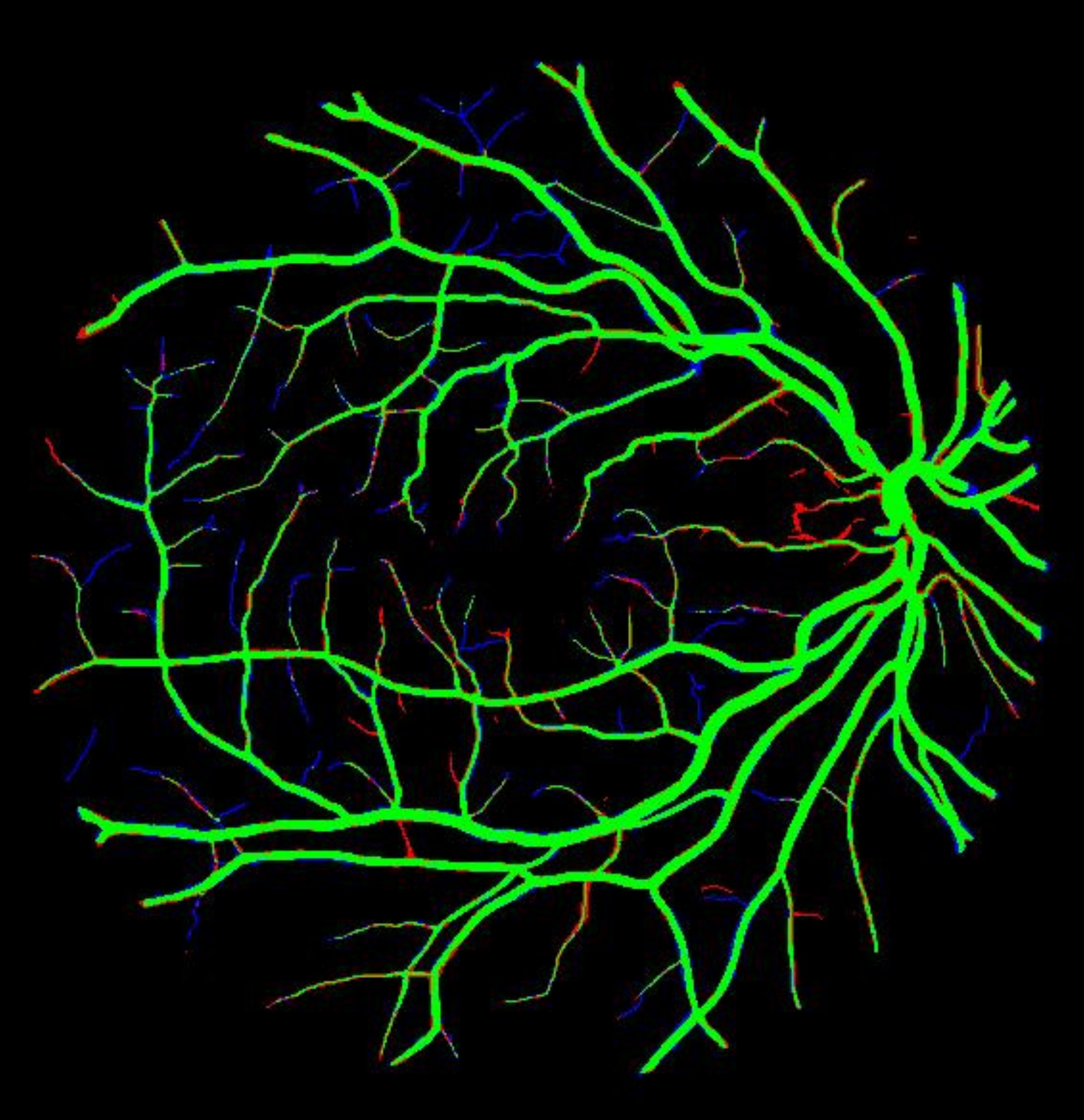} \\
			\includegraphics[width=0.14\textwidth]{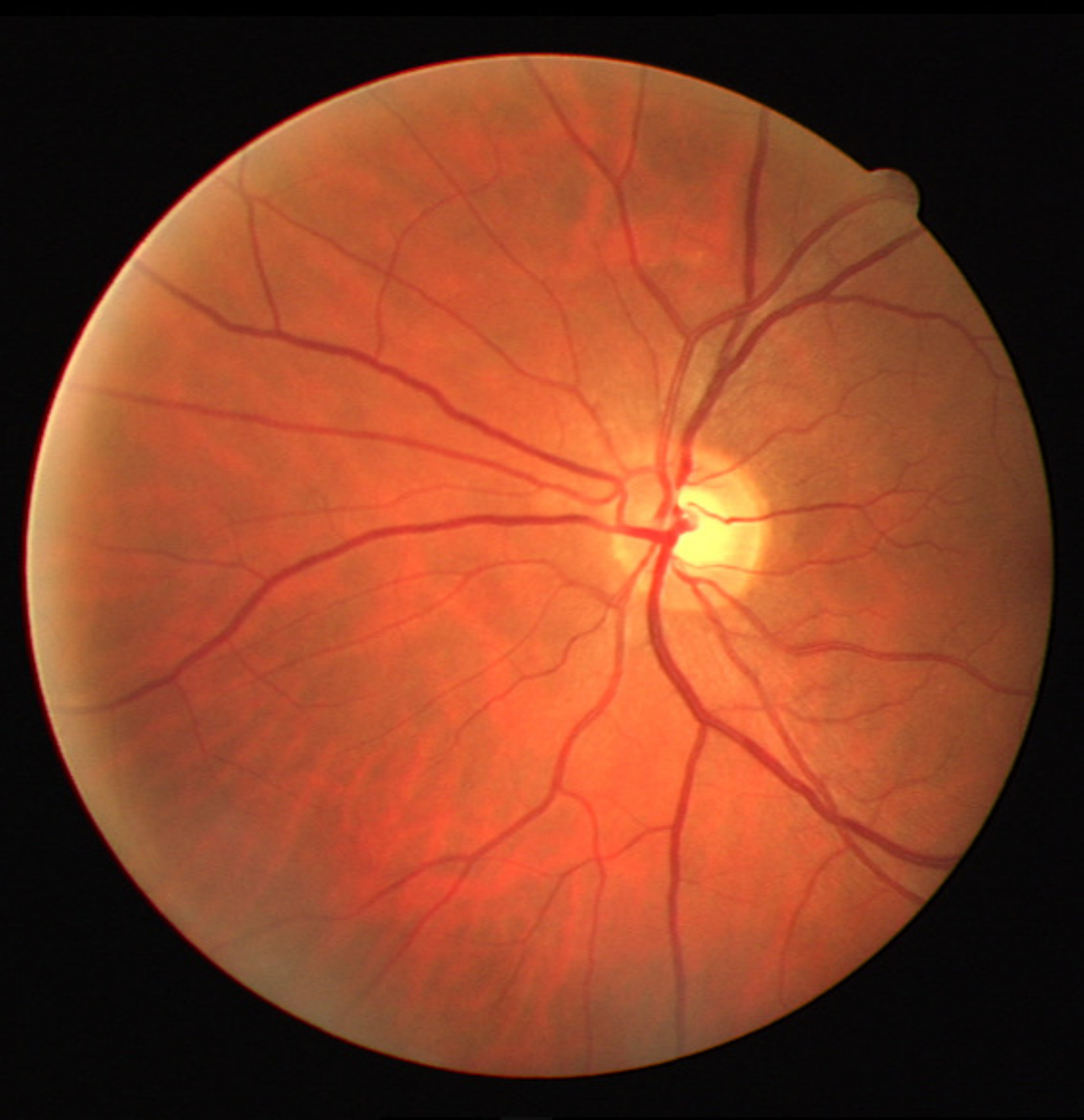}   & \includegraphics[width=0.14\textwidth]{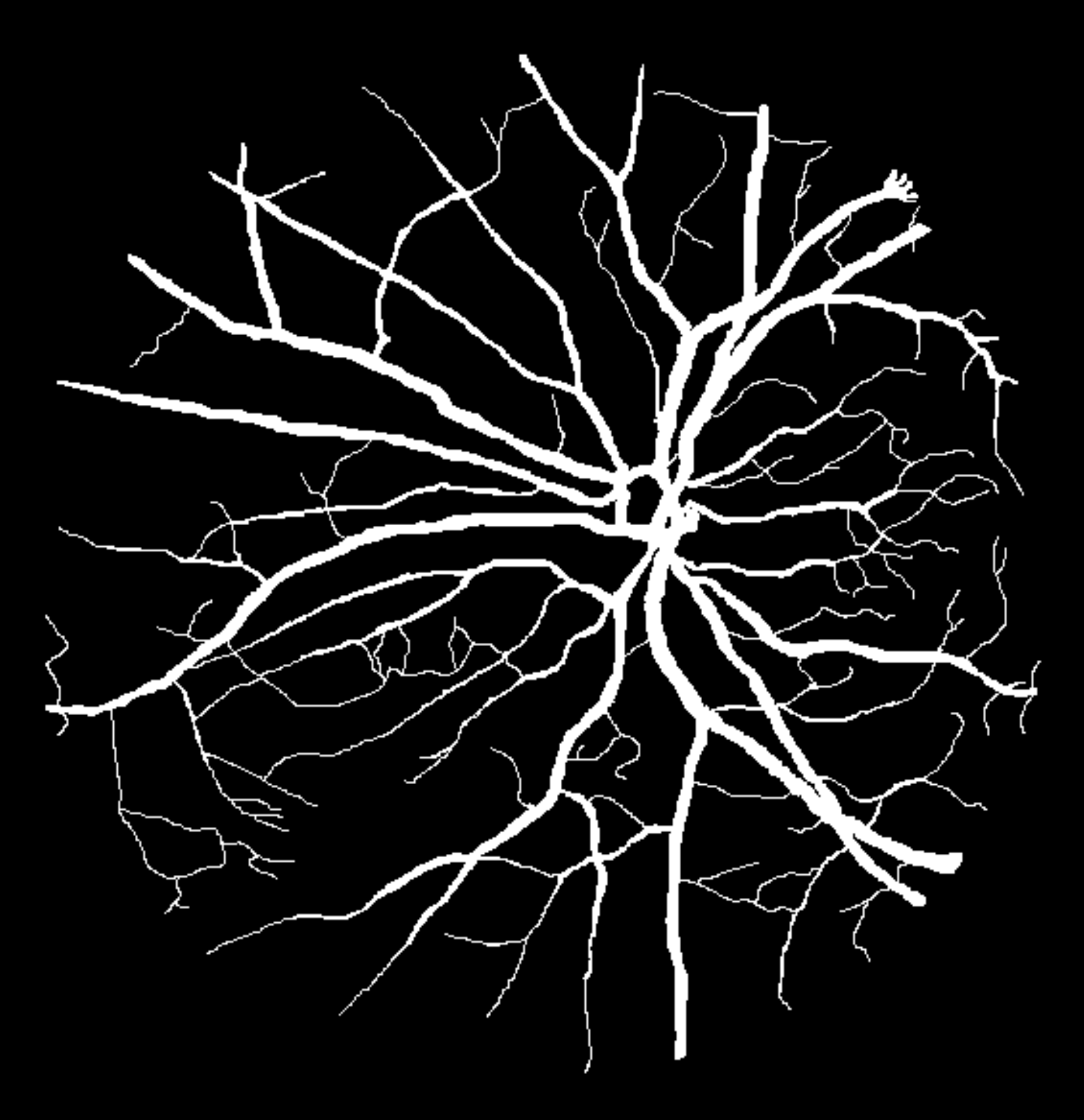}      &     \includegraphics[width=0.14\textwidth]{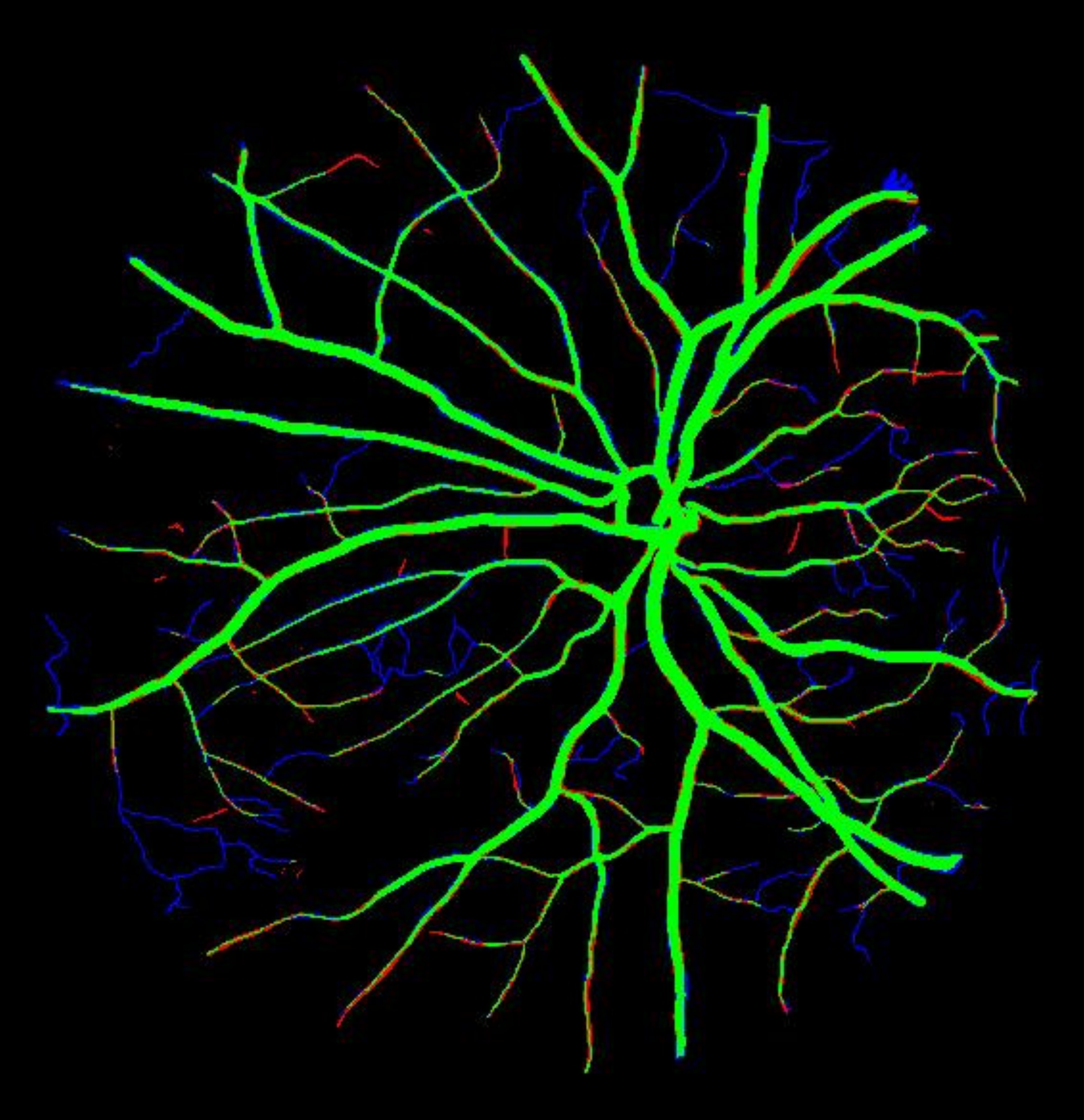}       & \includegraphics[width=0.14\textwidth]{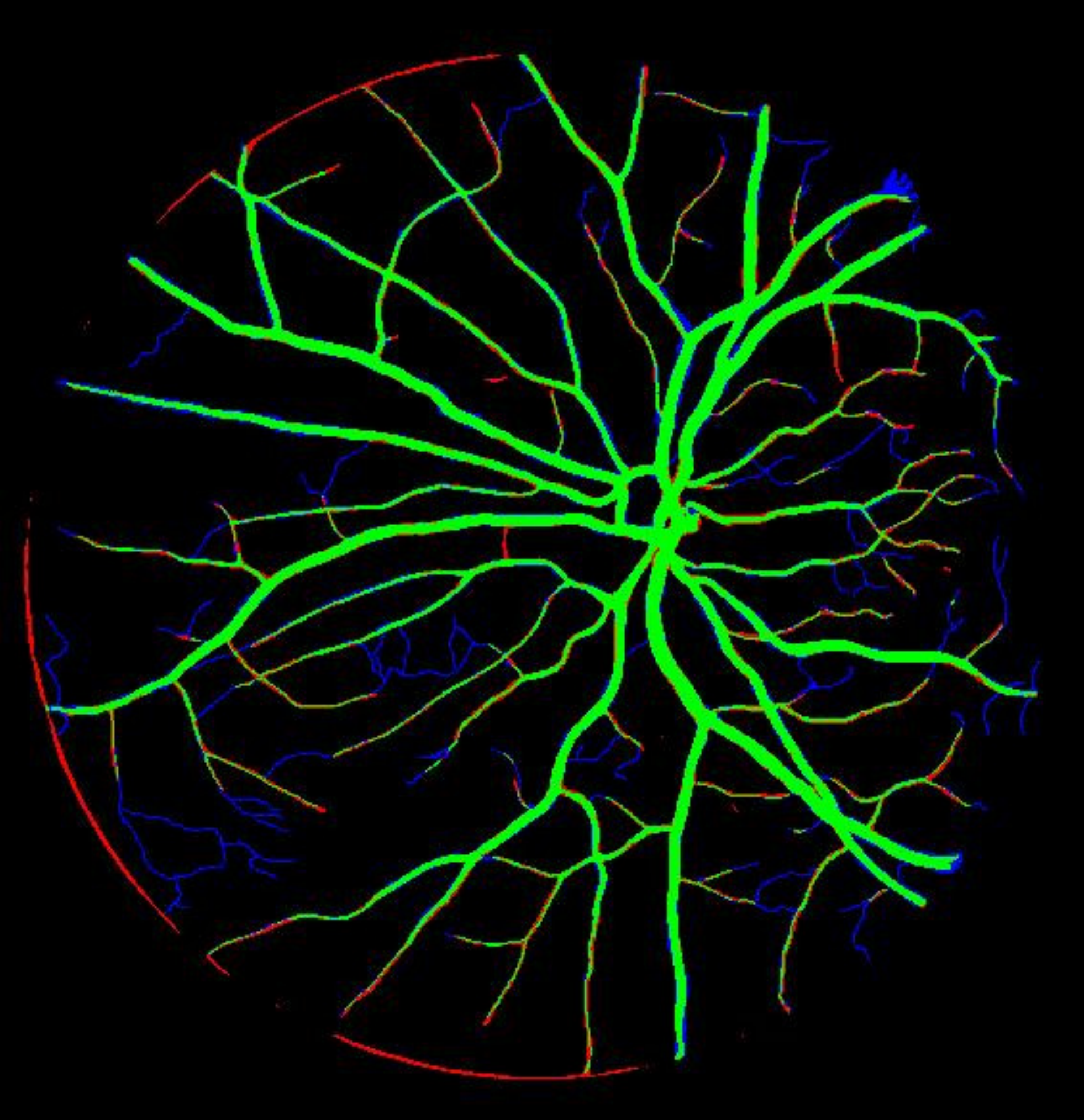}       & \includegraphics[width=0.14\textwidth]{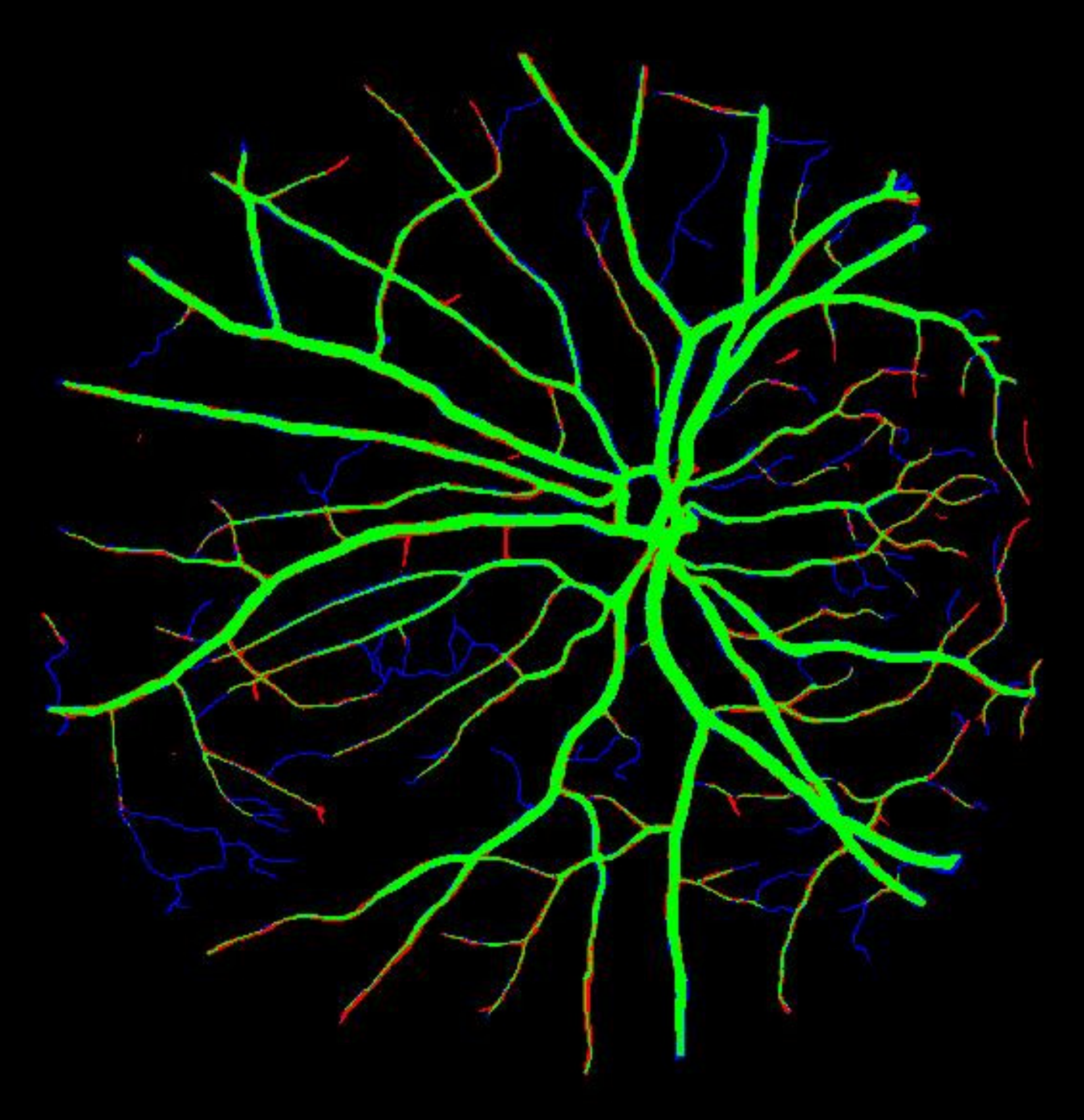}      & \includegraphics[width=0.14\textwidth]{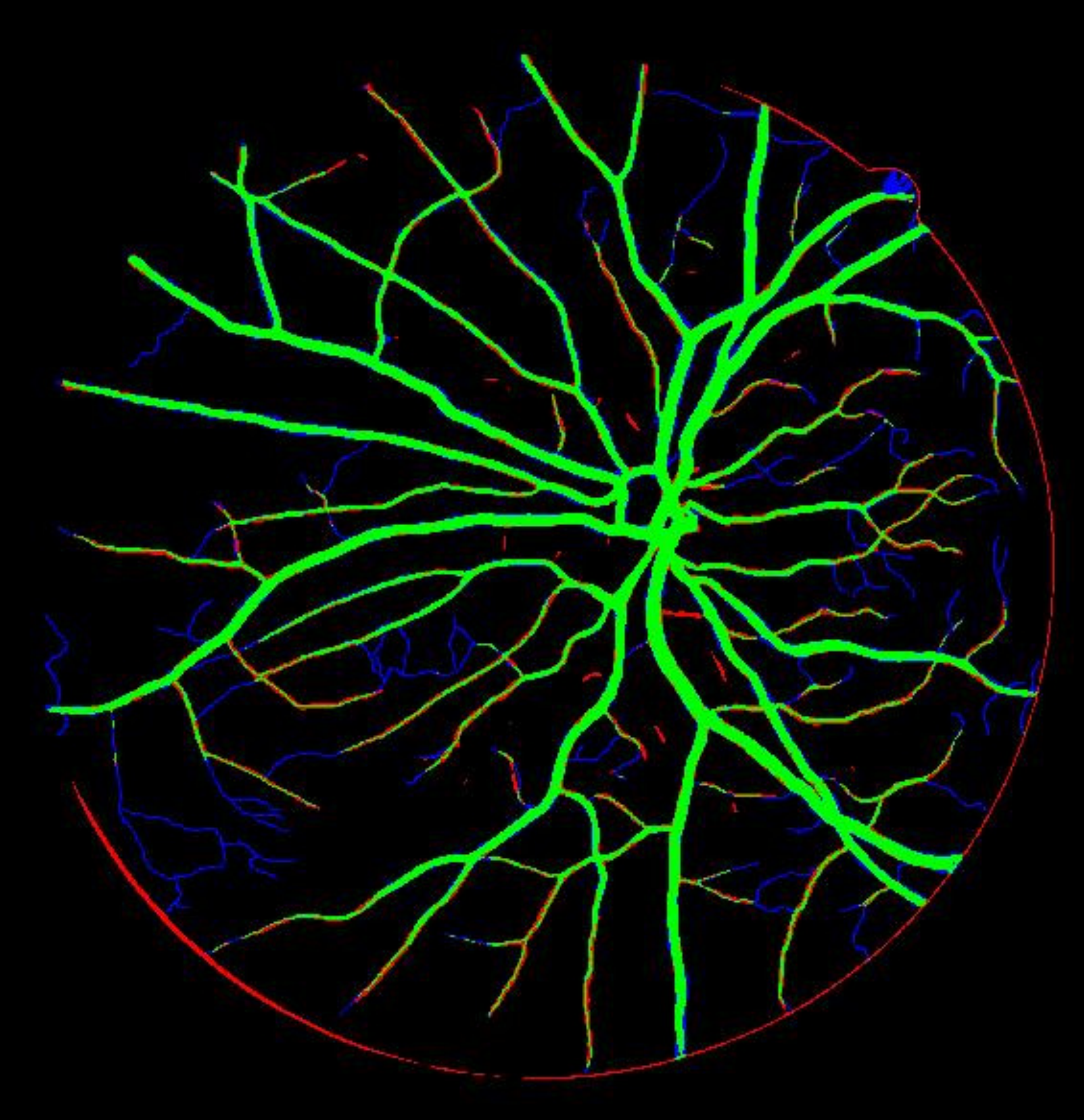}      & \includegraphics[width=0.14\textwidth]{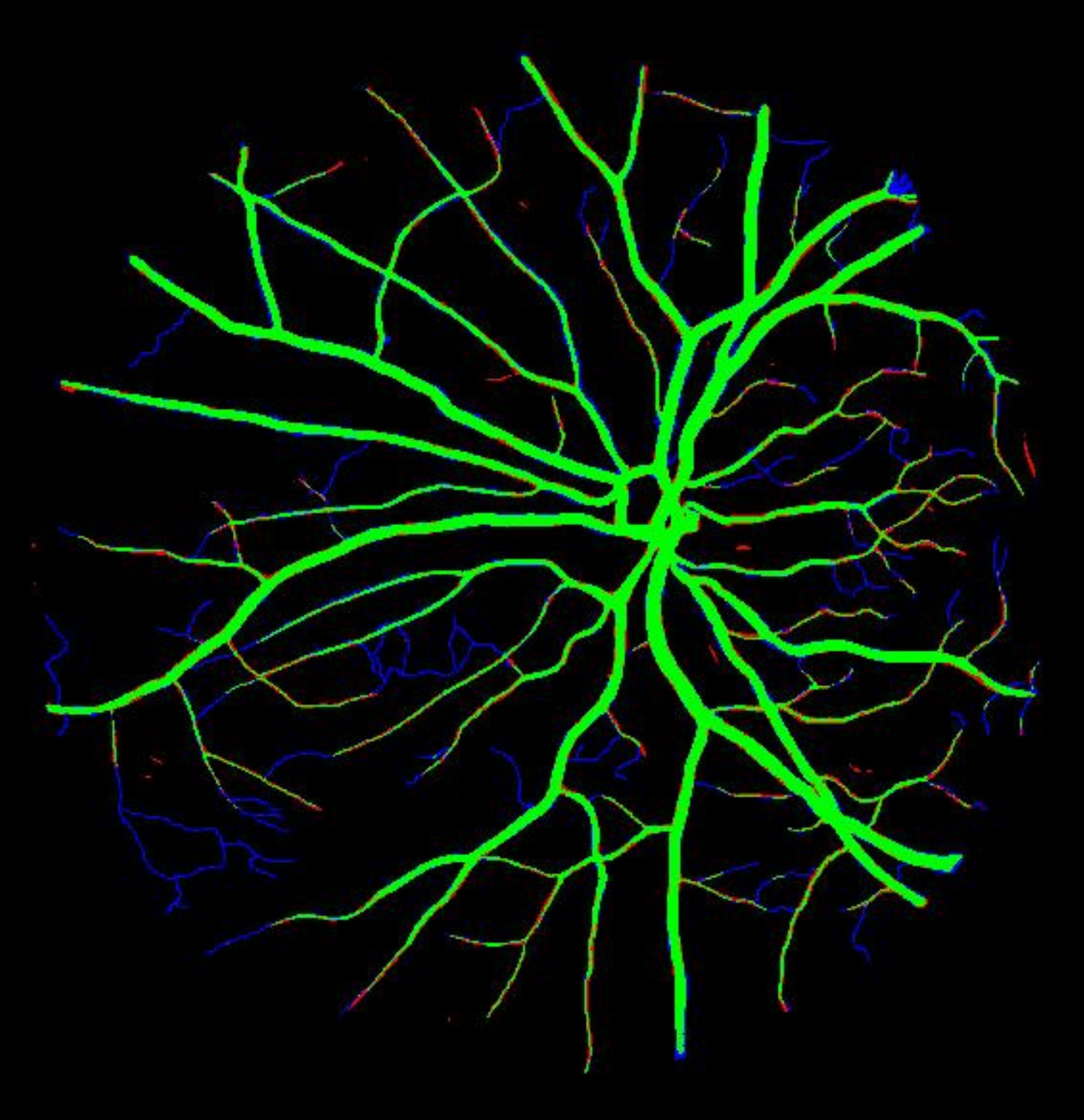} \\
			\includegraphics[width=0.14\textwidth]{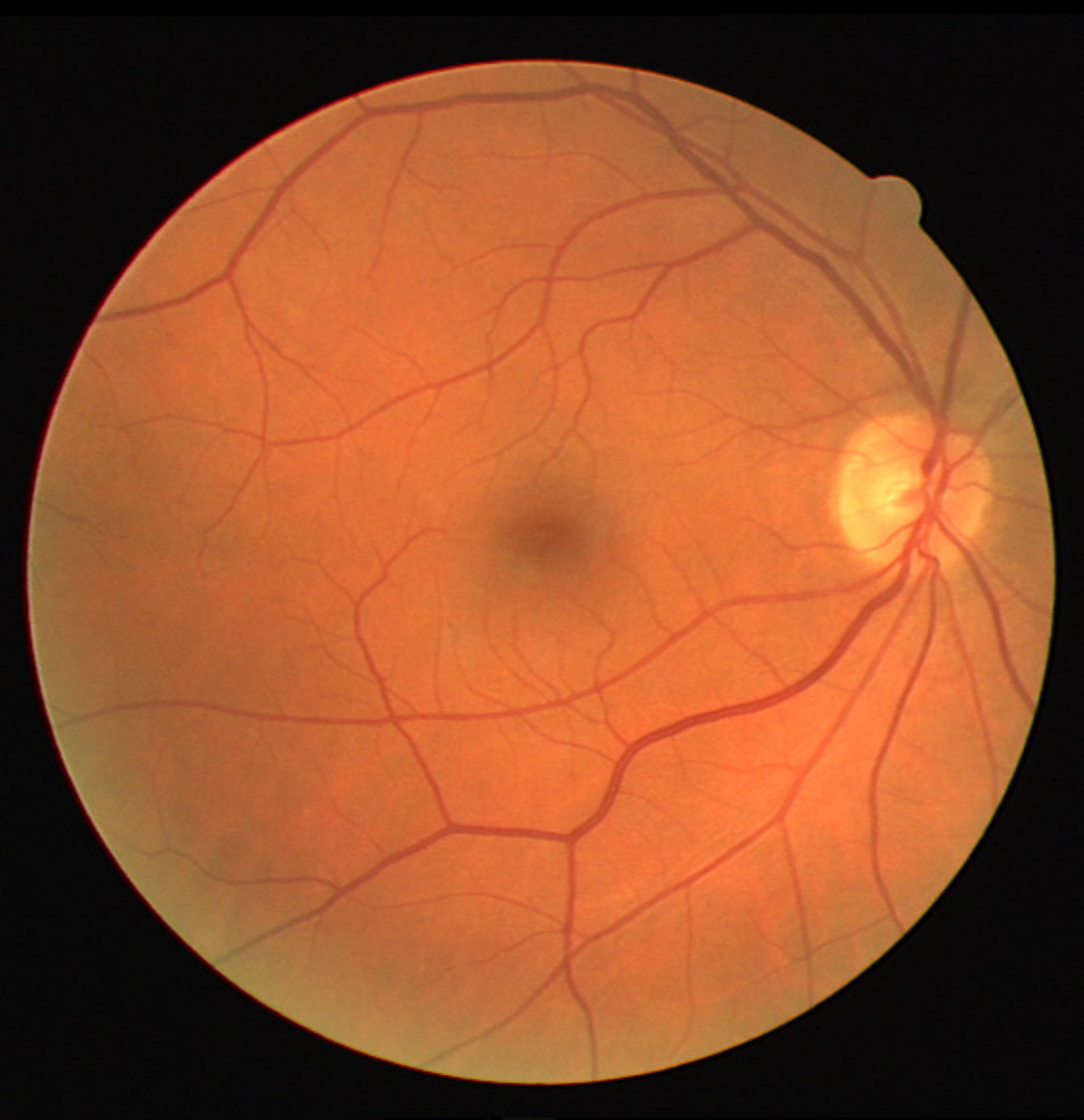}   & \includegraphics[width=0.14\textwidth]{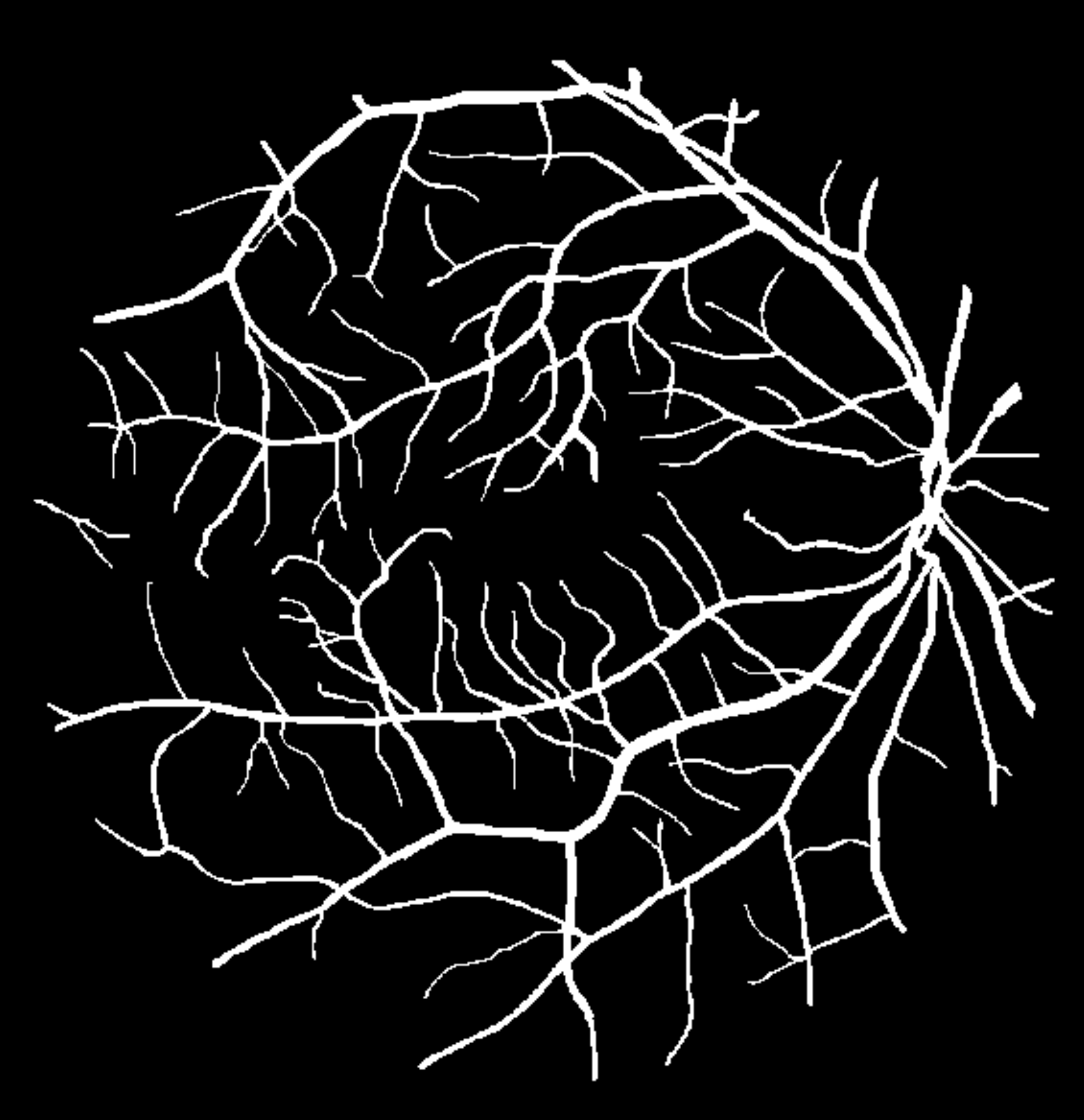}      &     \includegraphics[width=0.14\textwidth]{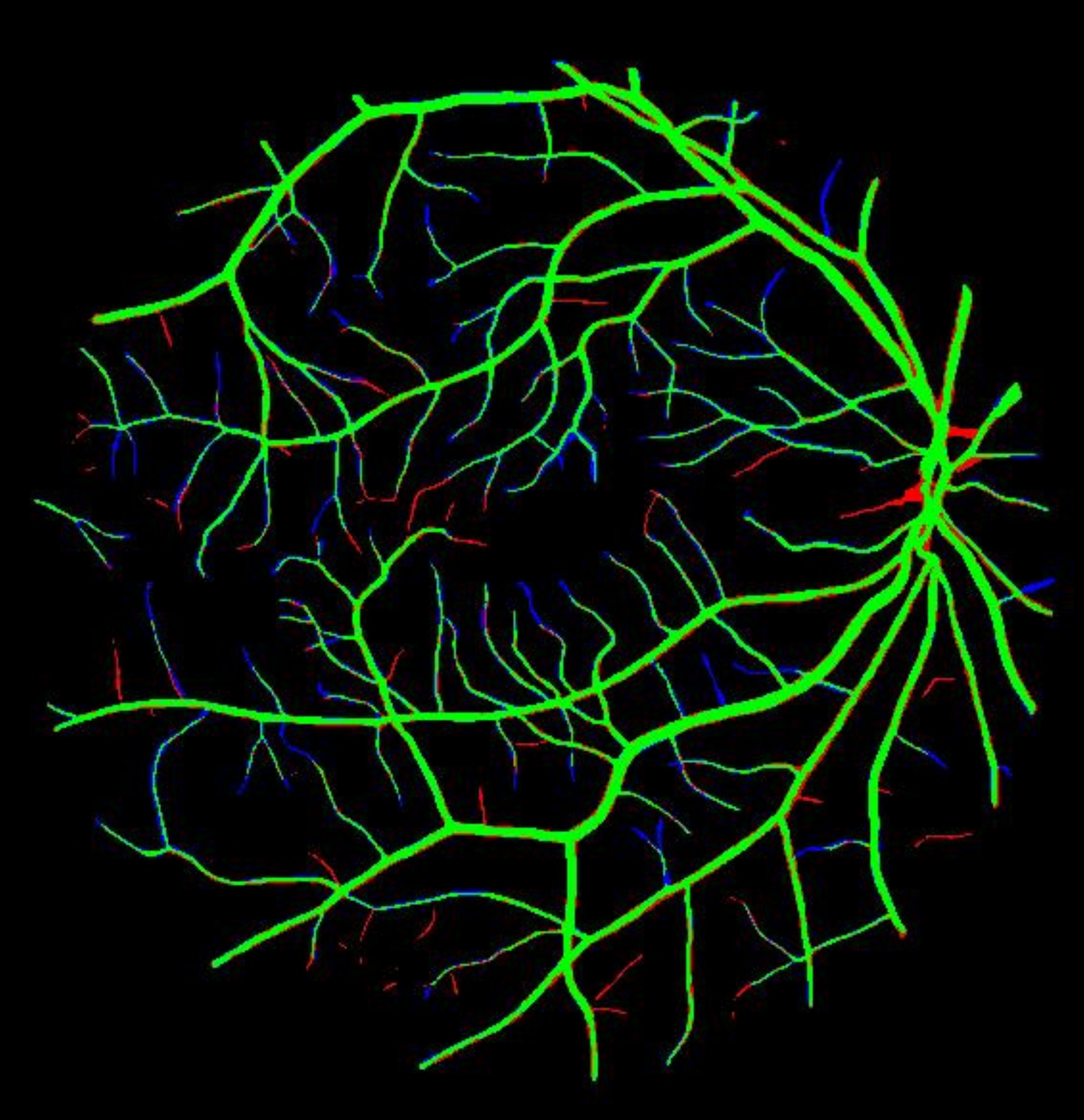}       & \includegraphics[width=0.14\textwidth]{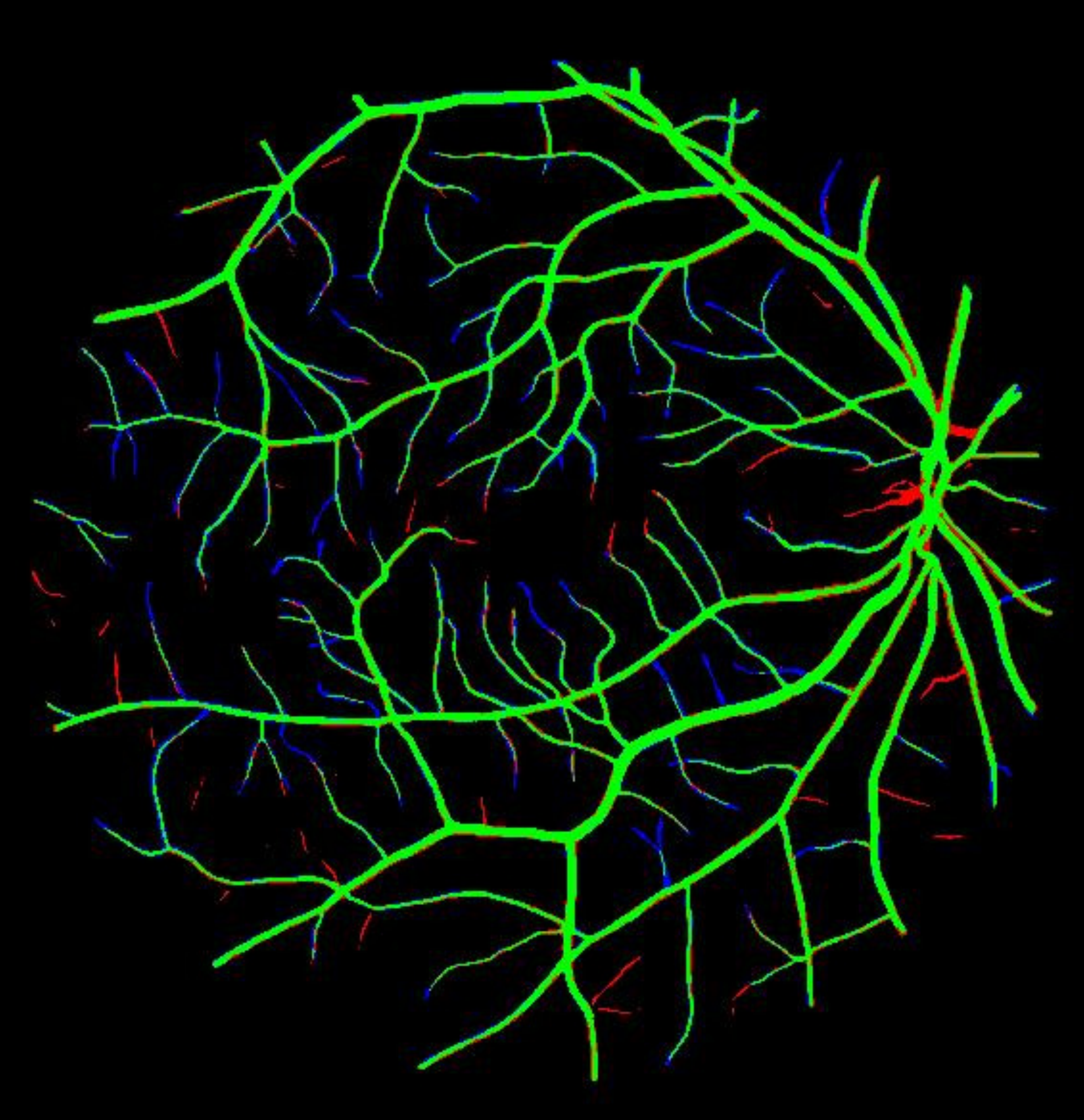}       & \includegraphics[width=0.14\textwidth]{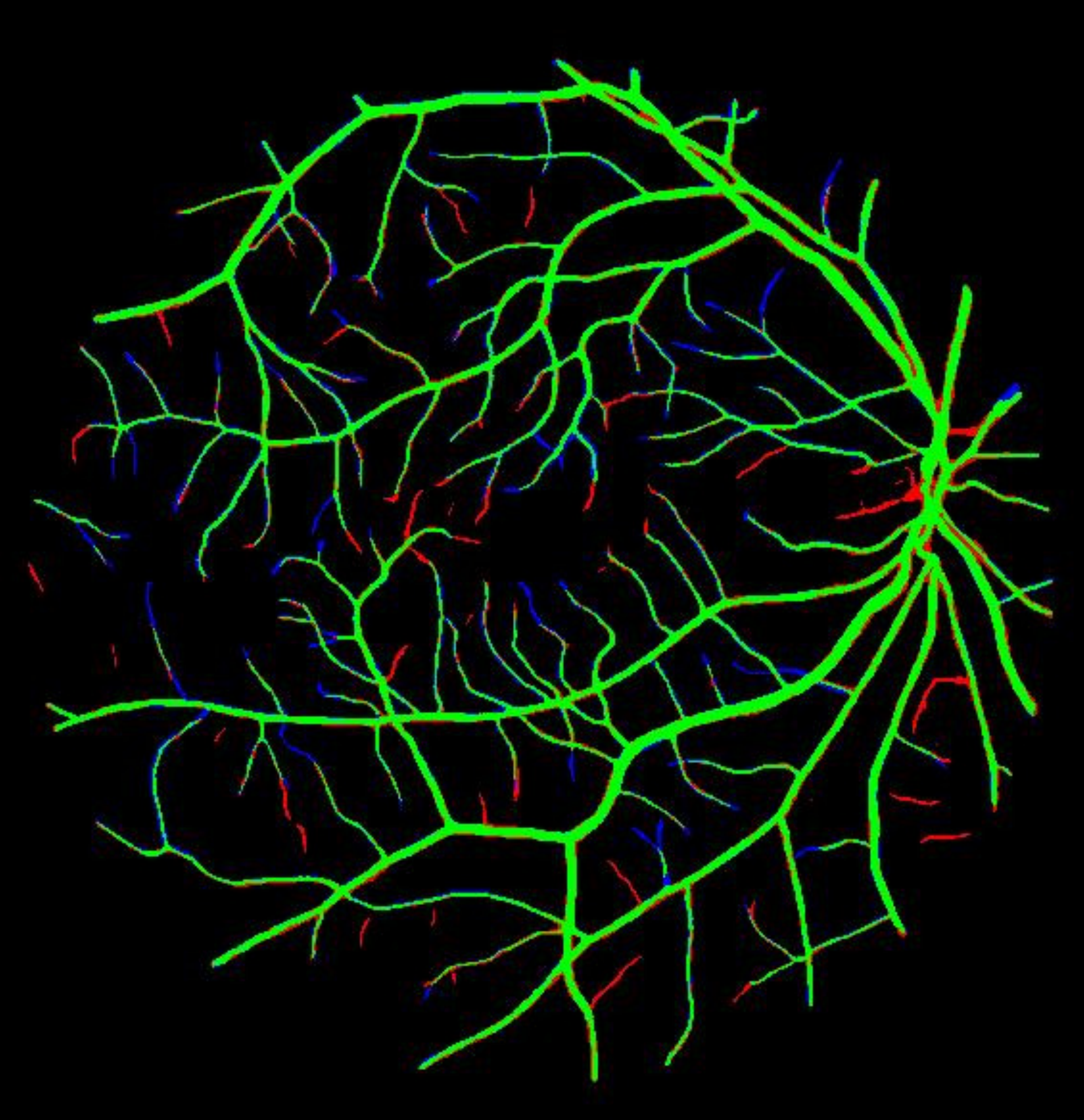}      & \includegraphics[width=0.14\textwidth]{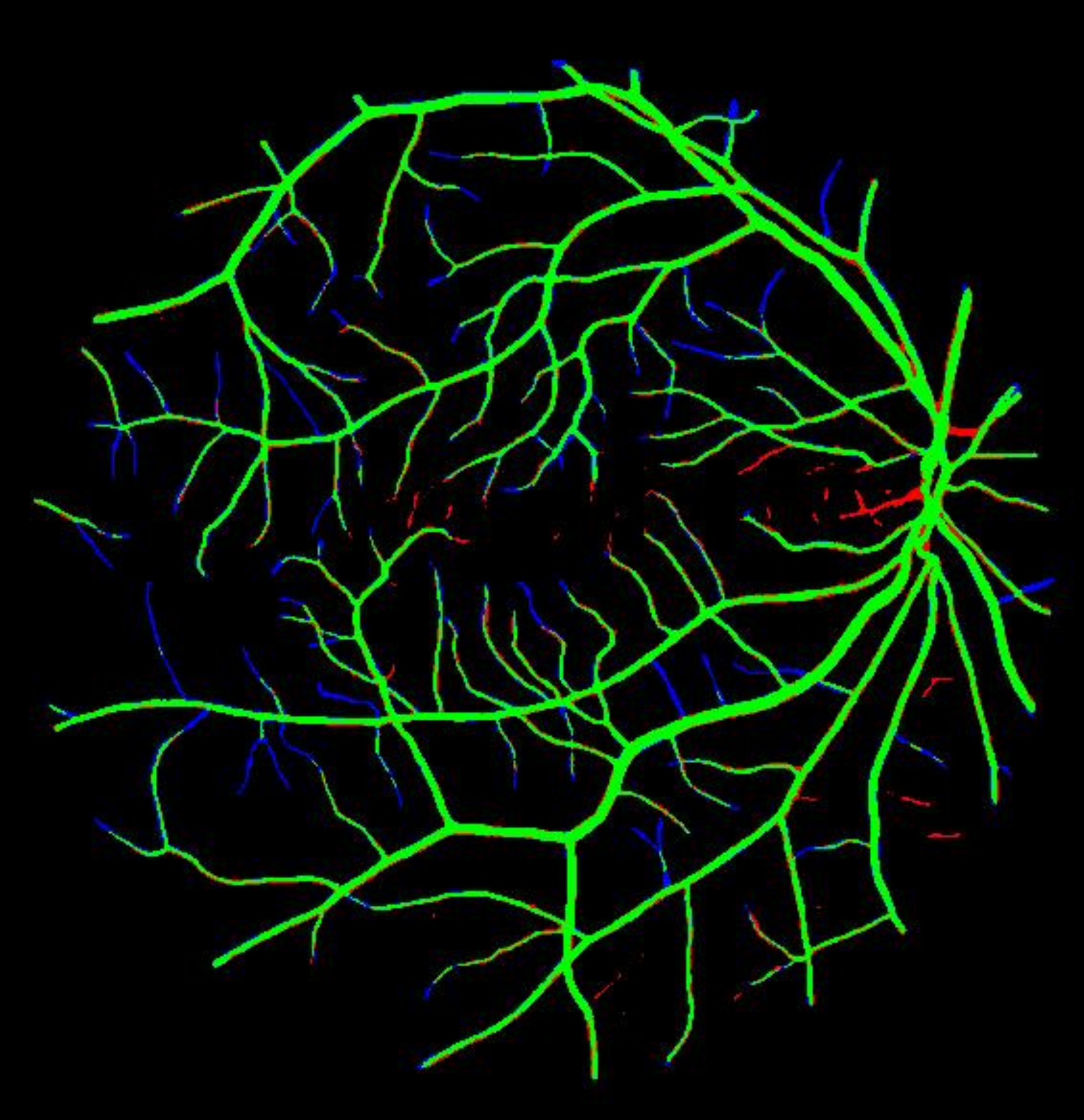}      & \includegraphics[width=0.14\textwidth]{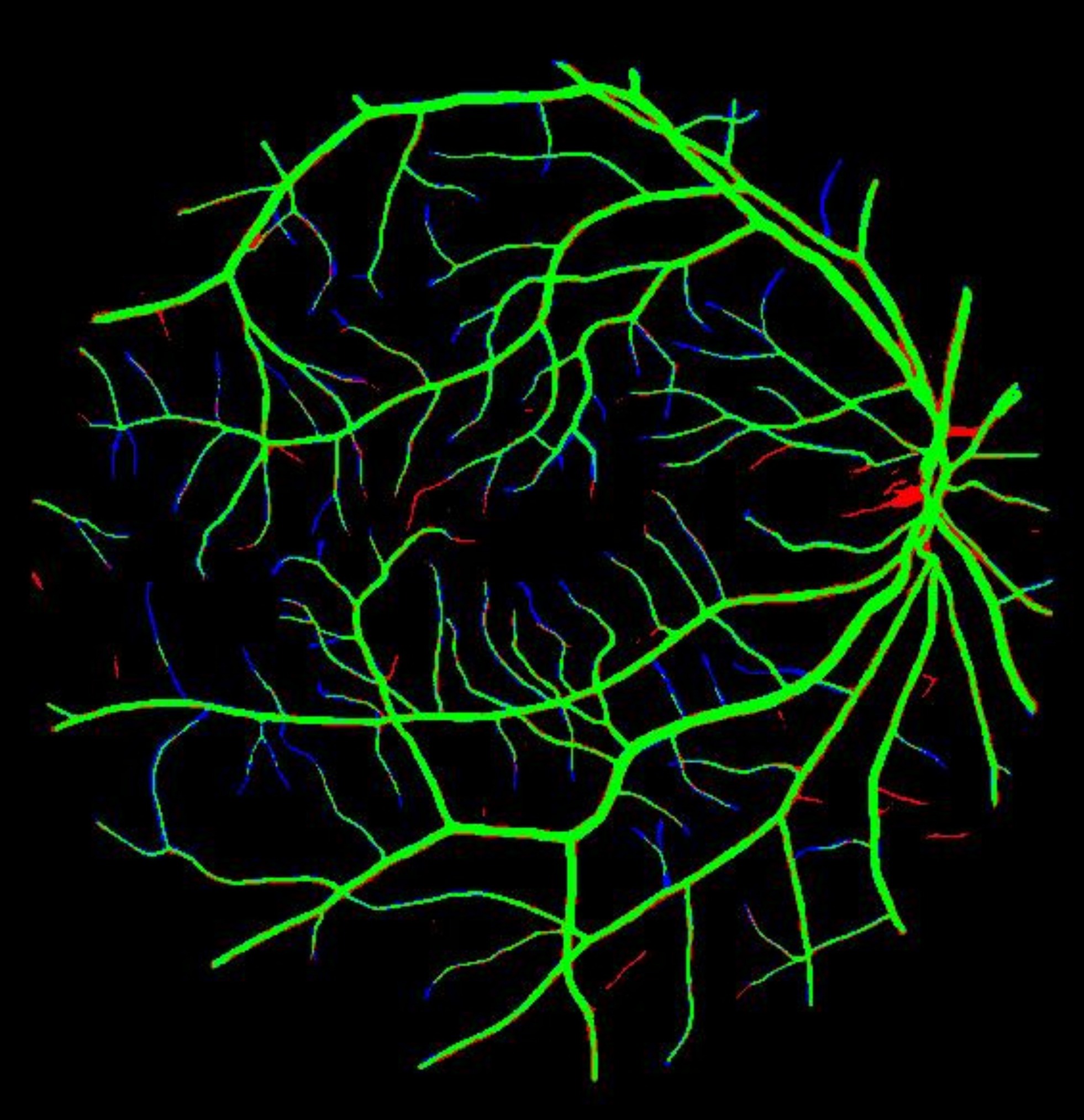} \\
			\includegraphics[width=0.14\textwidth]{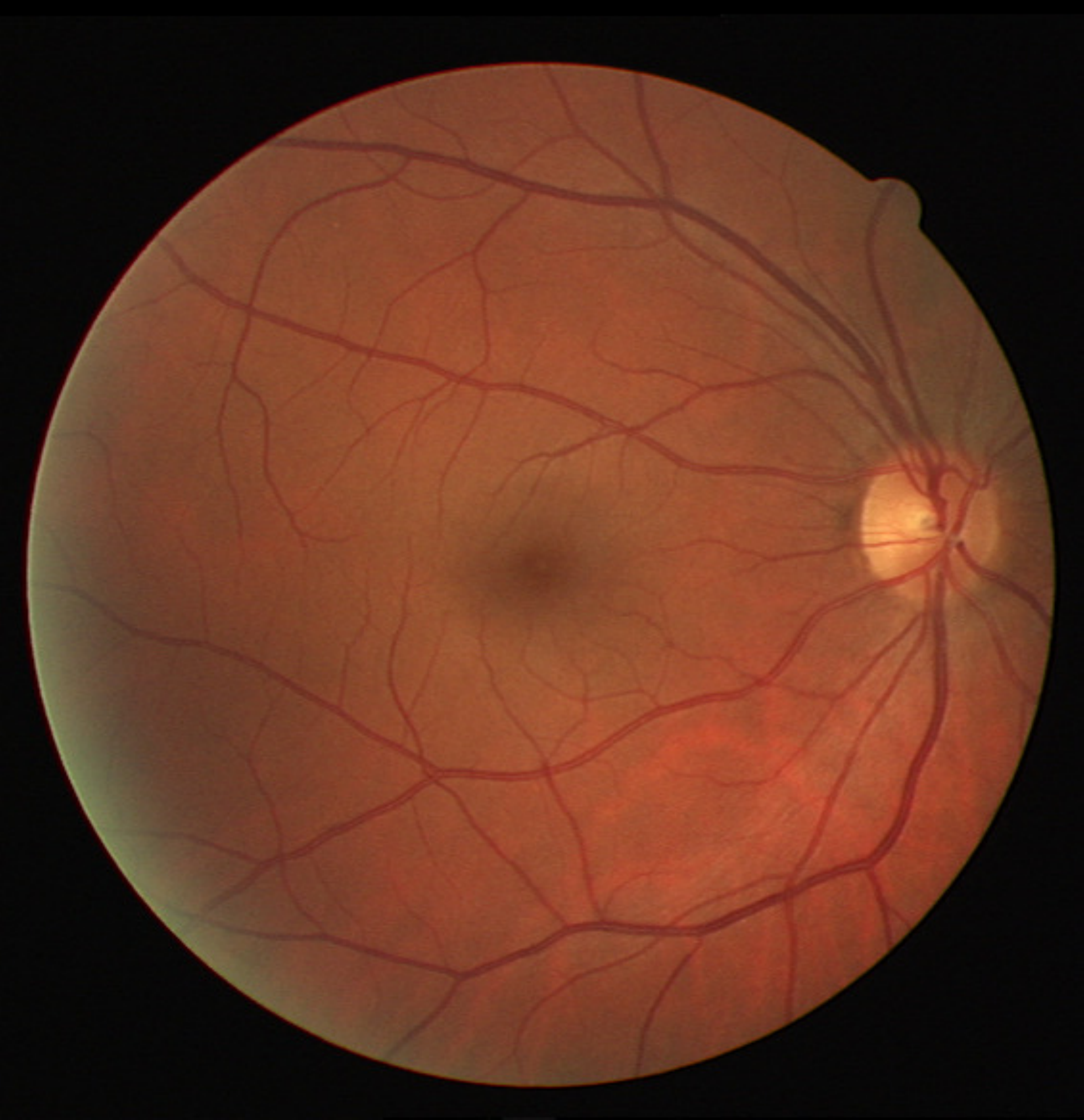}   & \includegraphics[width=0.14\textwidth]{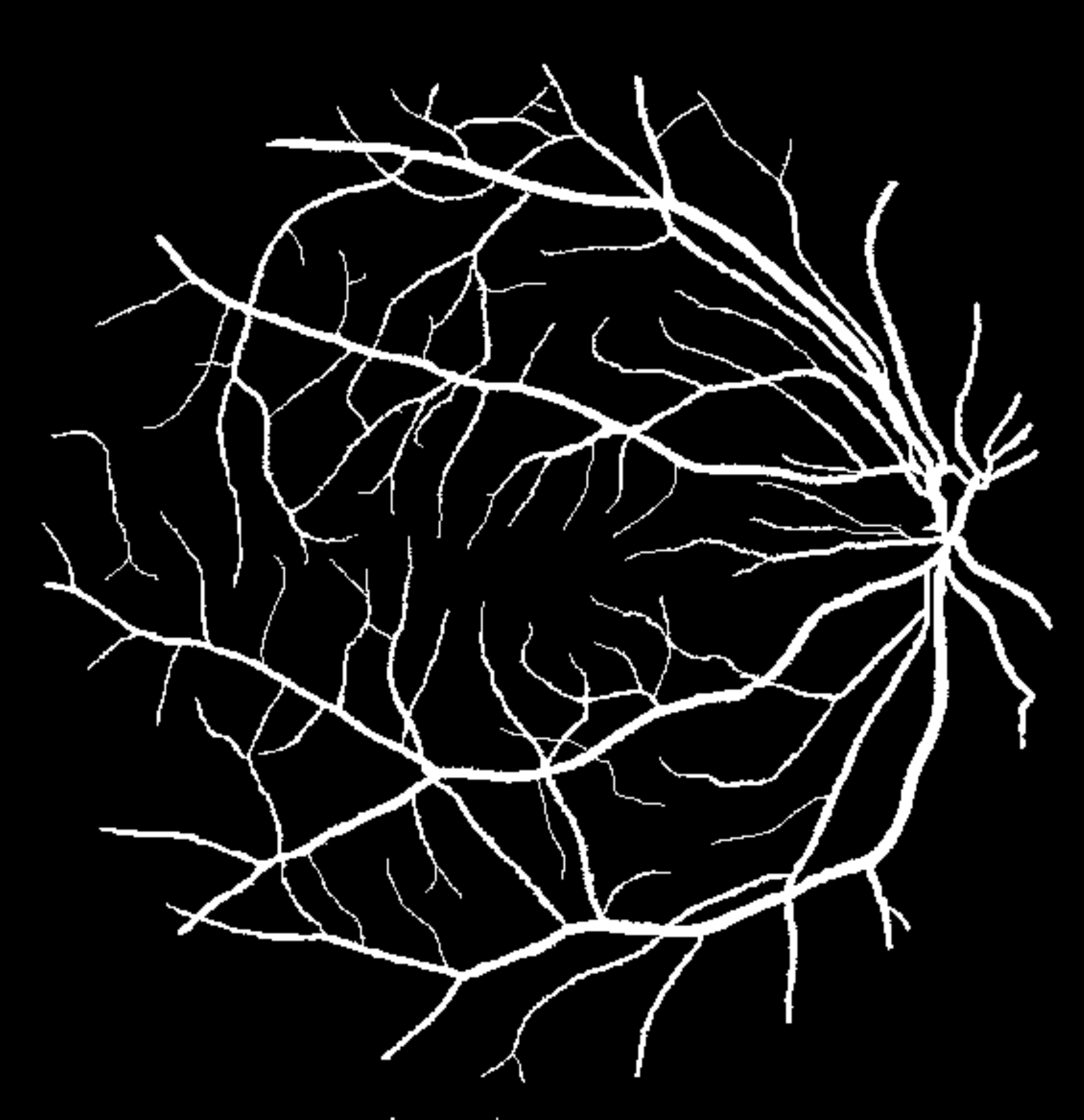}      &    \includegraphics[width=0.14\textwidth]{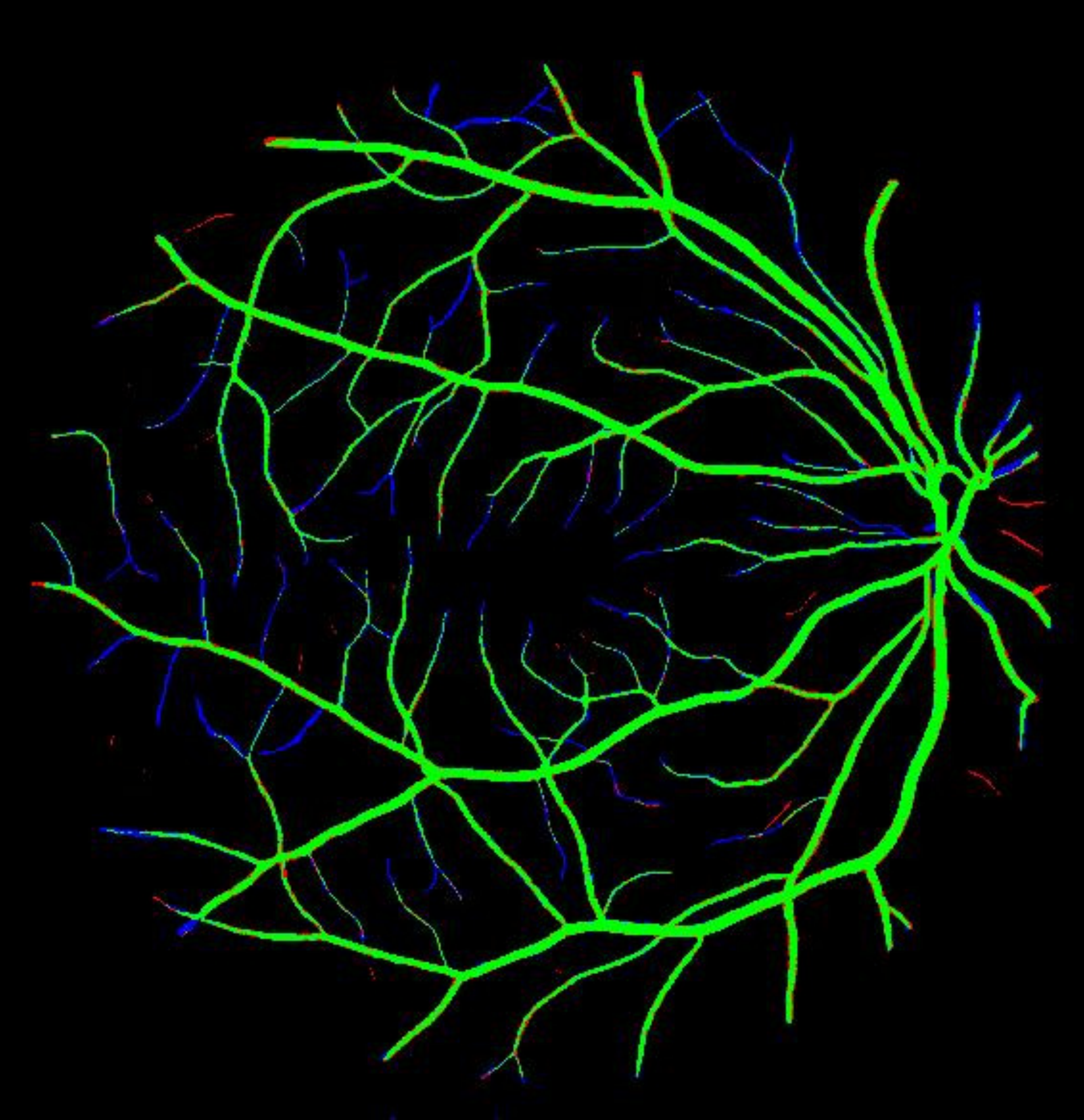}       & \includegraphics[width=0.14\textwidth]{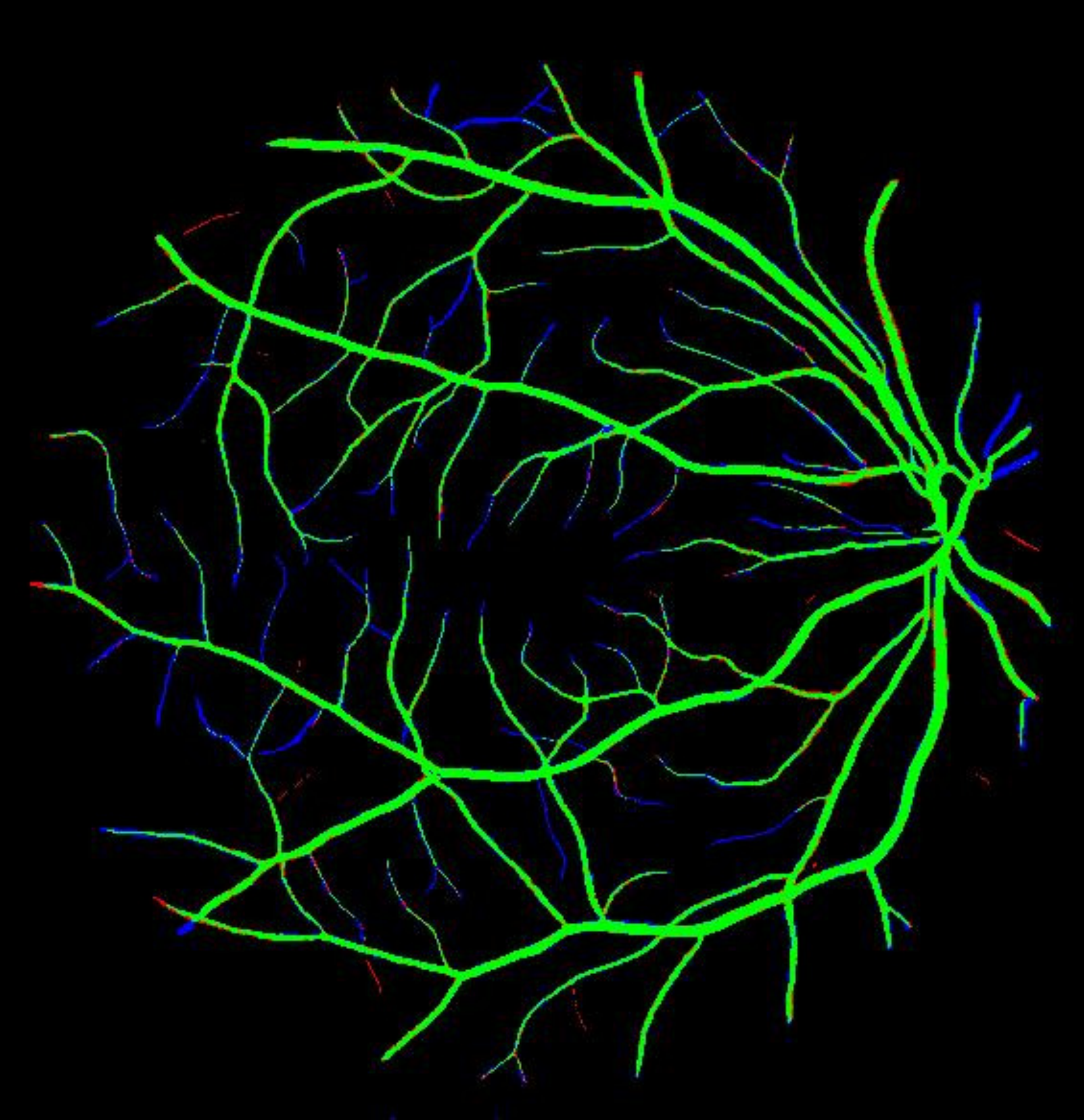}       & \includegraphics[width=0.14\textwidth]{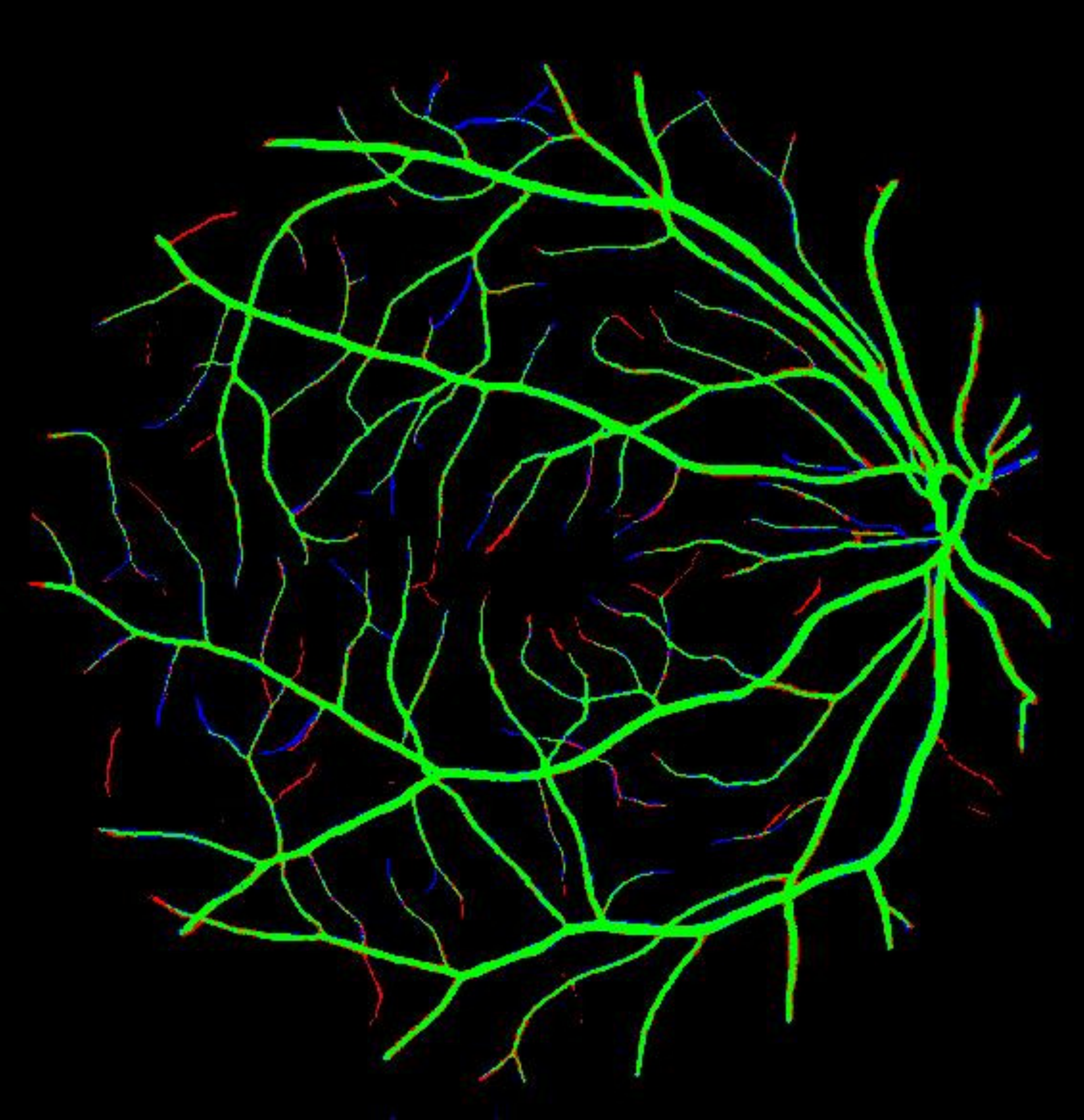}      & \includegraphics[width=0.14\textwidth]{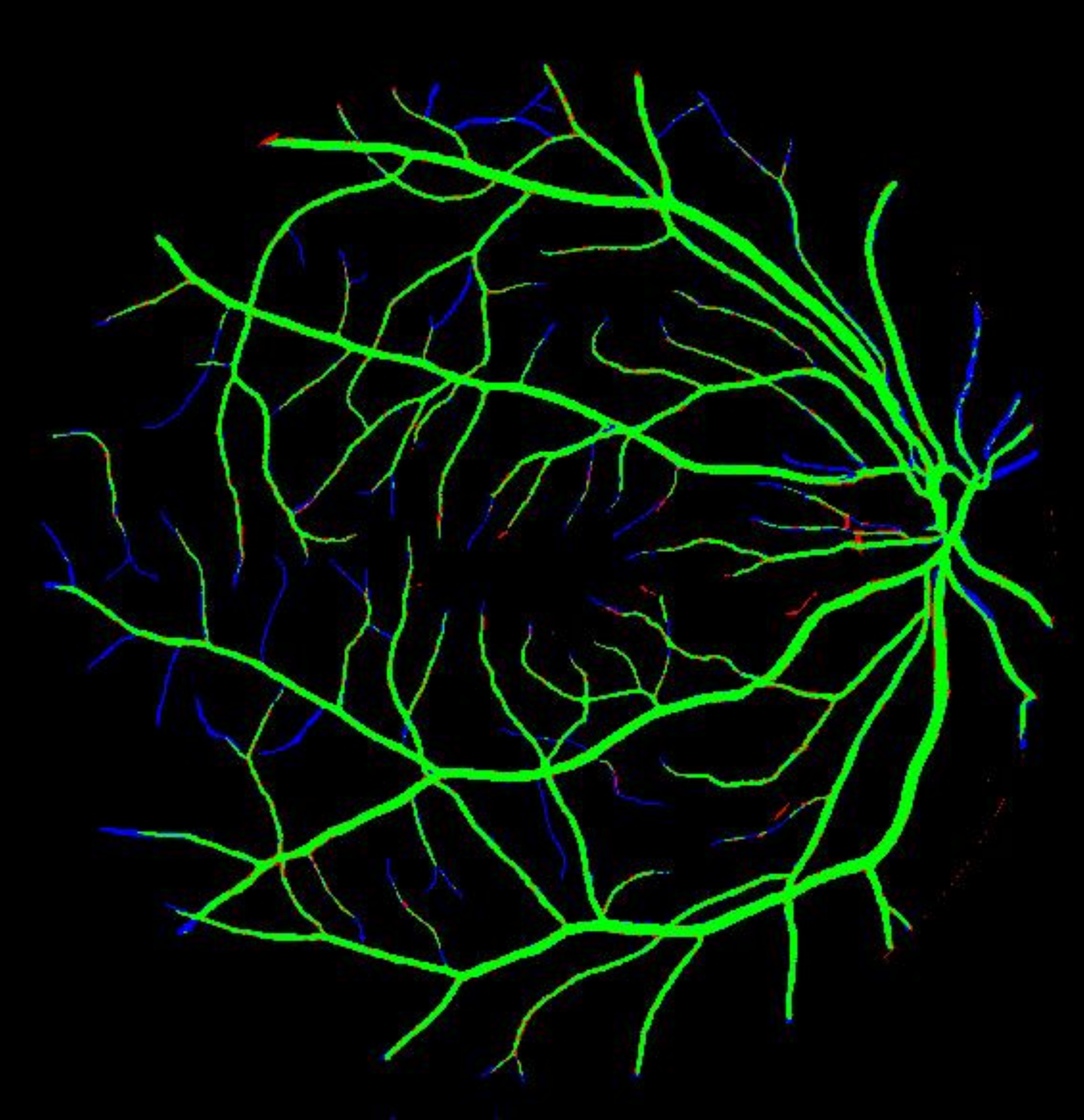}      & \includegraphics[width=0.14\textwidth]{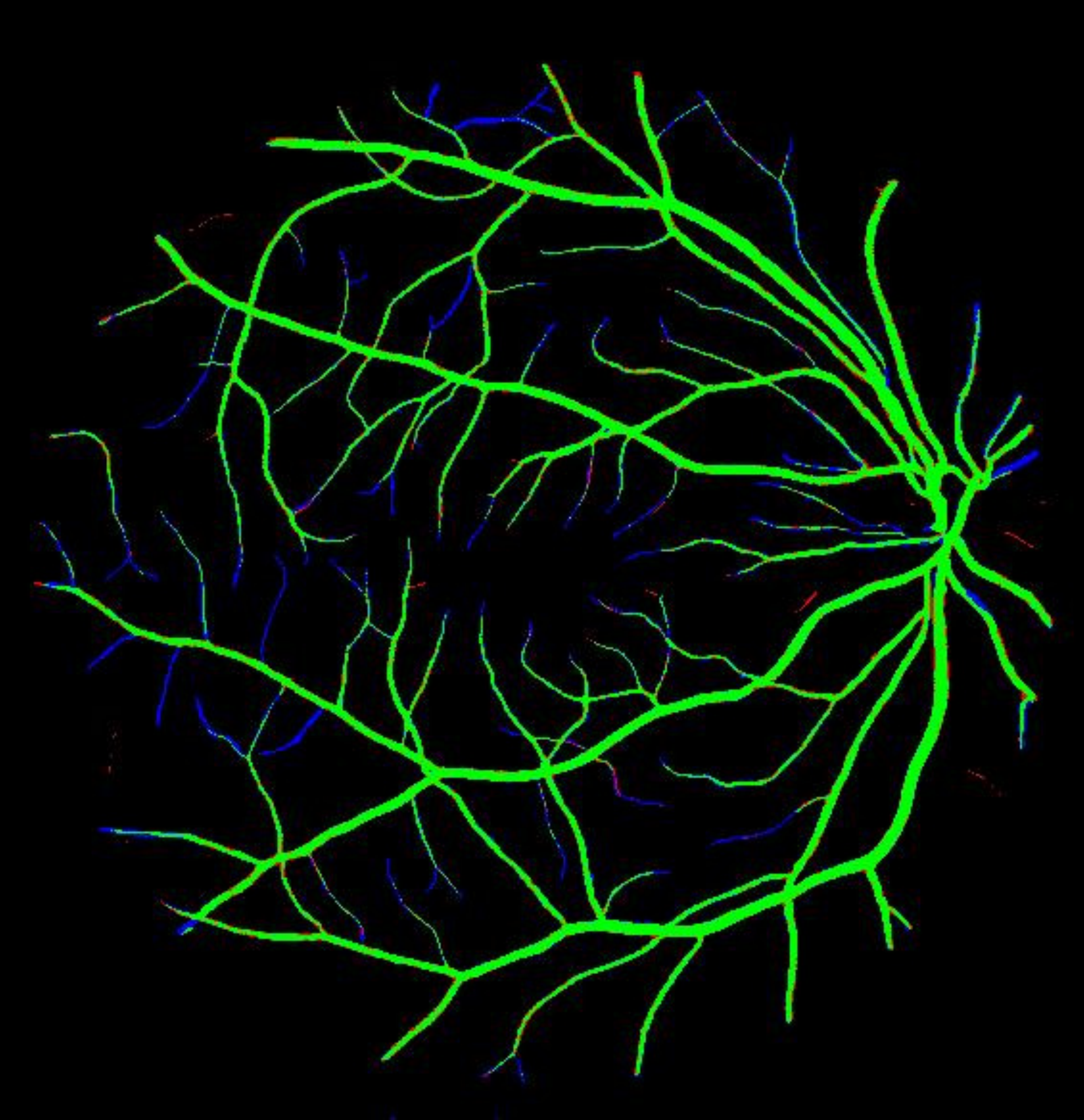} \\
		\end{tabular}}
	\caption{Segmentation results delivered by our MRC-Net method on representative test images, i.e. image numbers 1, 2, 4, 16, and 19 from the DRIVE dataset. From left to right, we show the input image, ground truth, and the results yielded by BCDUNet, MultiResUNet, SegNet, U-Net++, and our network. In the figure, false positives are shown in red, whereas blue pixels depict false negatives.}
	\label{visualDRIVE}%
\end{figure*}%

\begin{figure*}[h!]
	\centering
	\resizebox{0.85\textwidth}{!}{%
		\begin{tabular}{ccccccc}
        Image & Ground Truth & BCDUNet & {MultiResUNet} & {SegNet} & {U-Net++} & {MRC-Net} \\
			\includegraphics[width=0.14\textwidth]{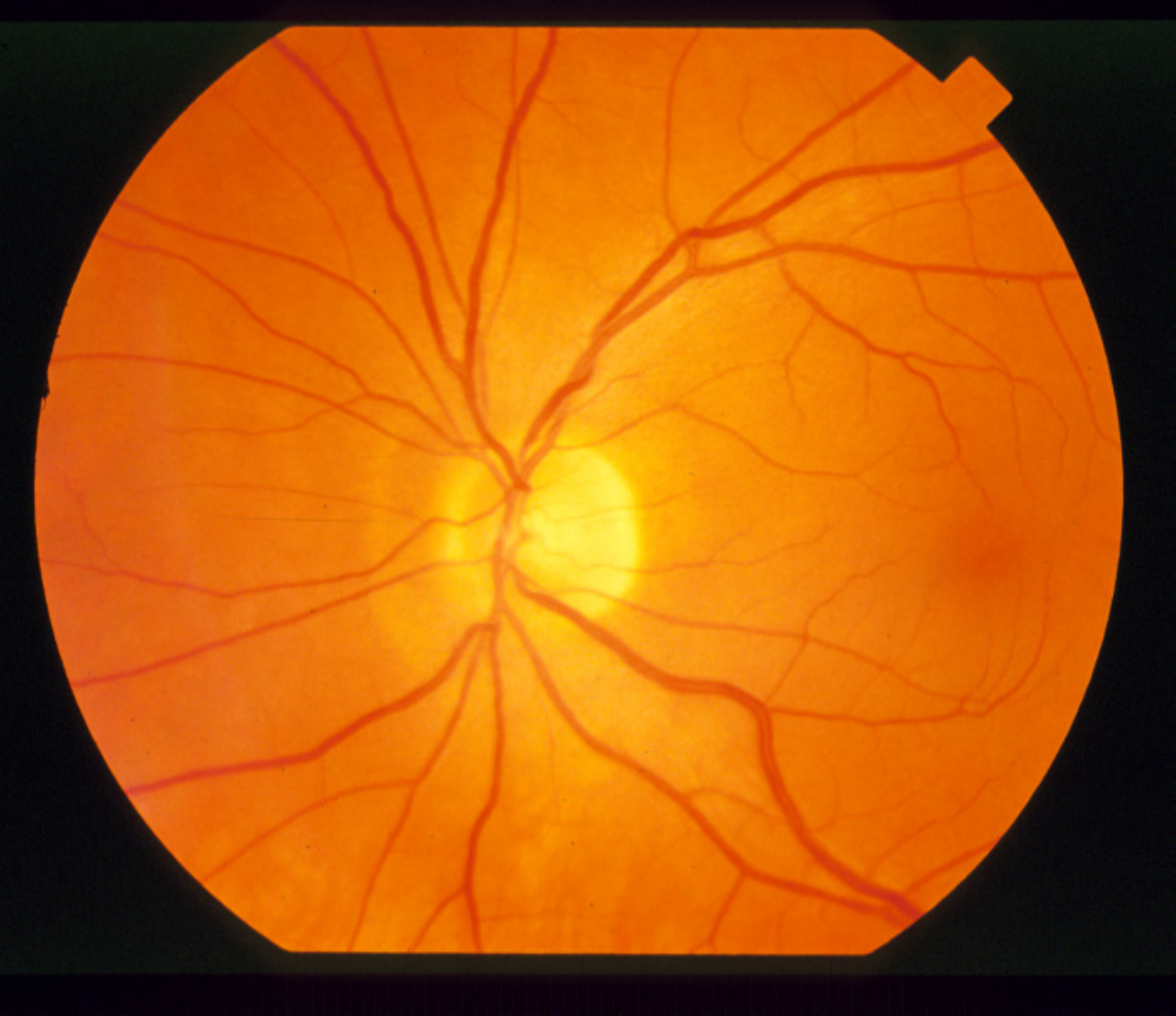}   & \includegraphics[width=0.14\textwidth]{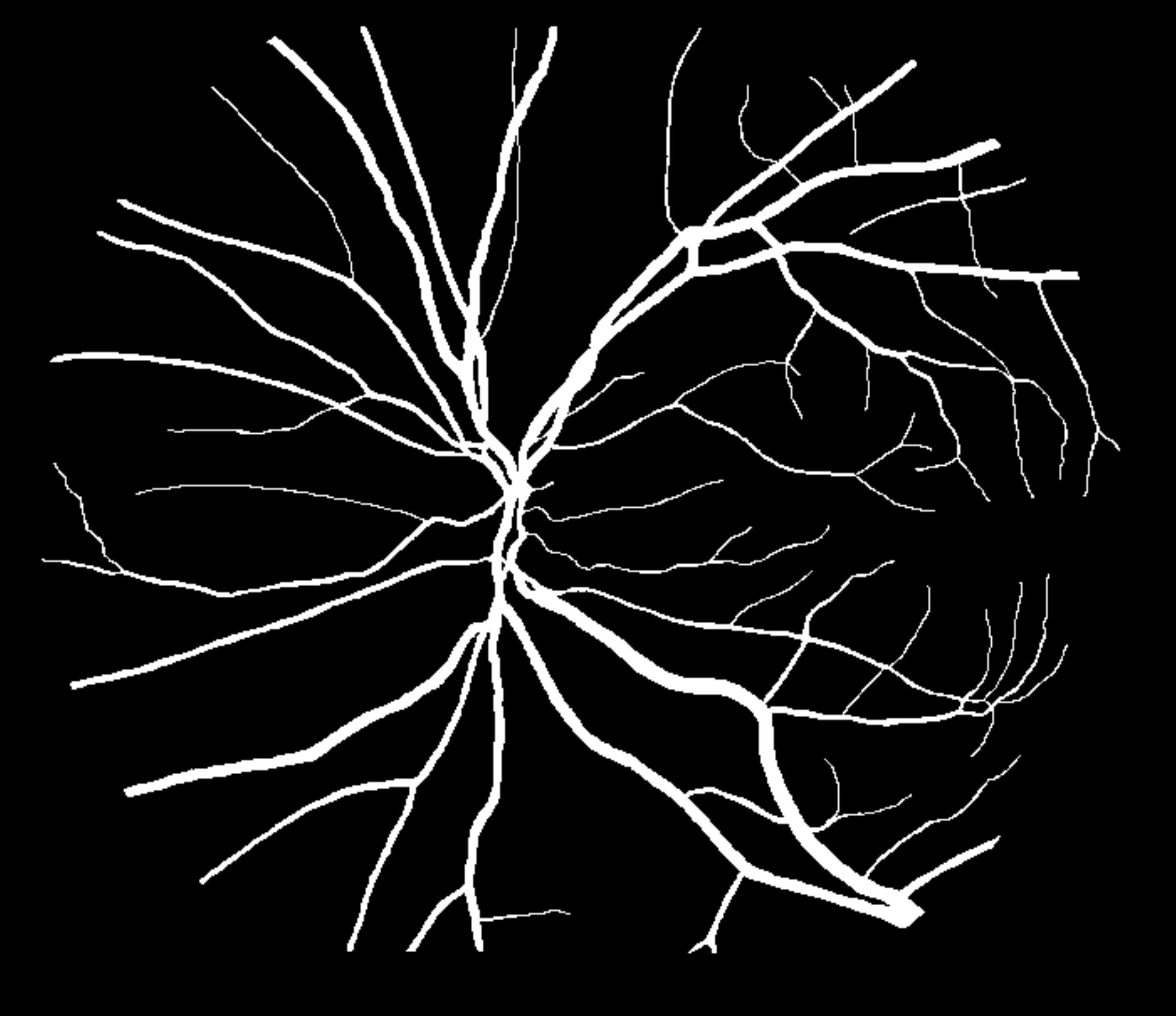}    & \includegraphics[width=0.14\textwidth]{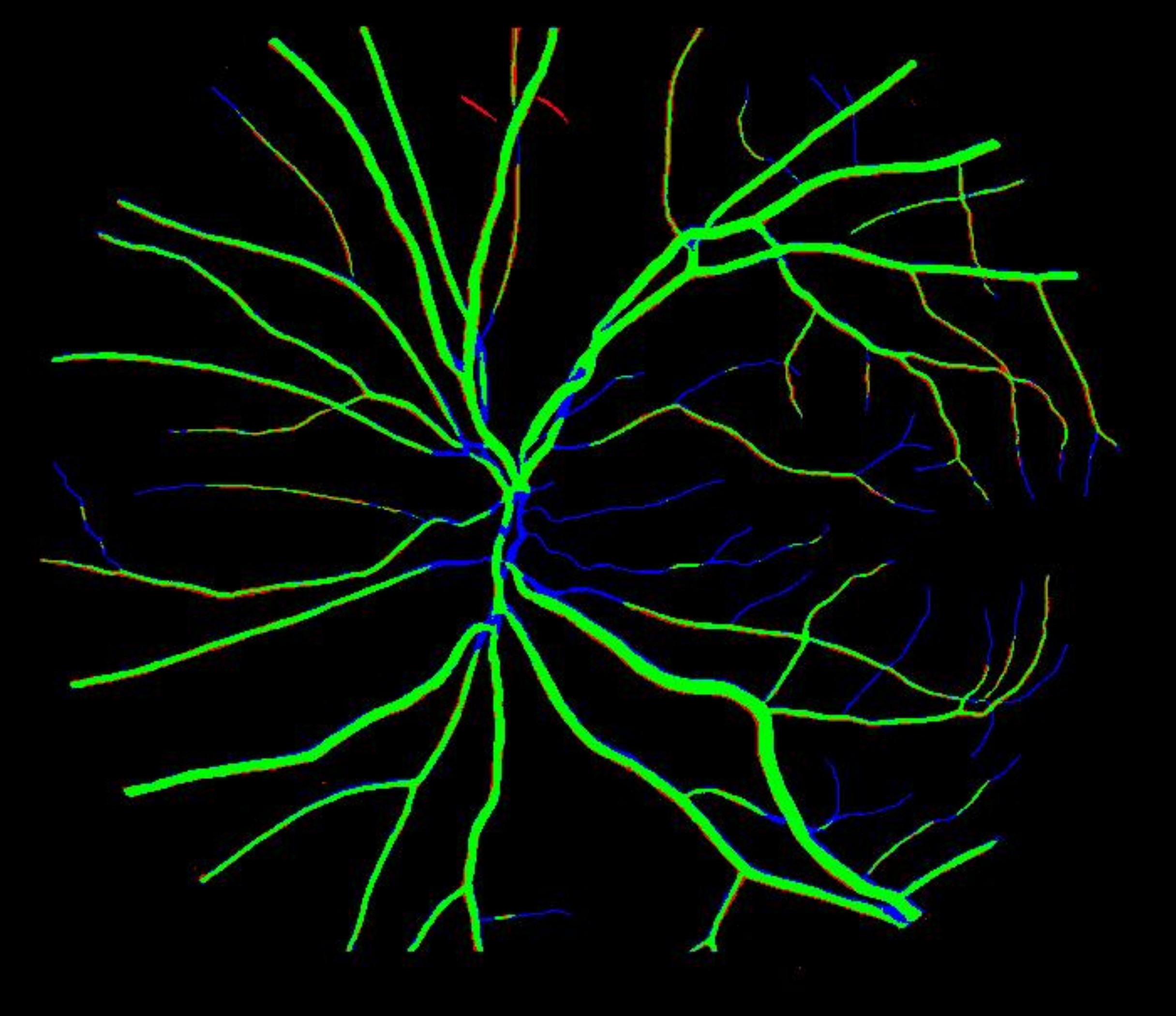}       & \includegraphics[width=0.14\textwidth]{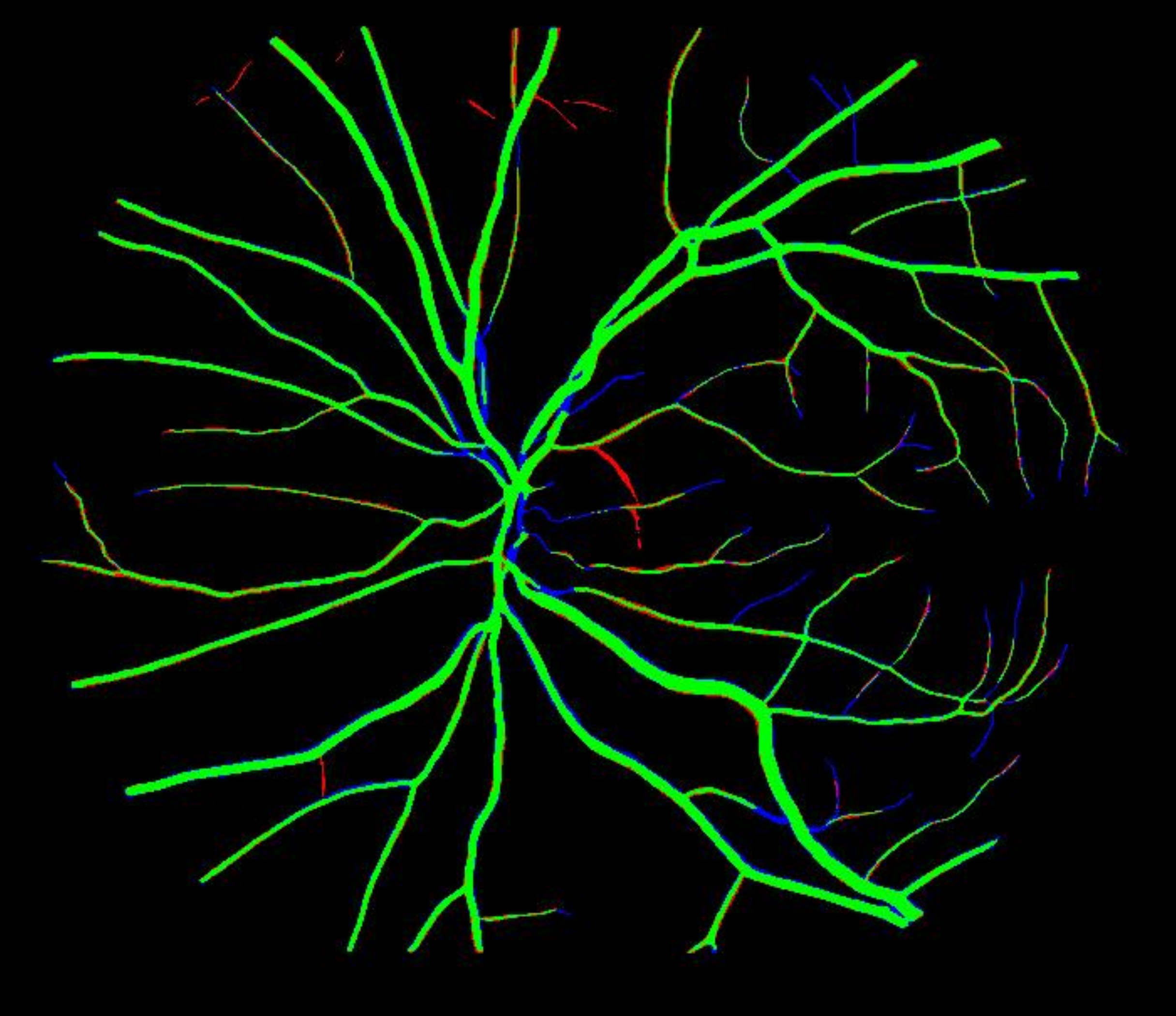}       & \includegraphics[width=0.14\textwidth]{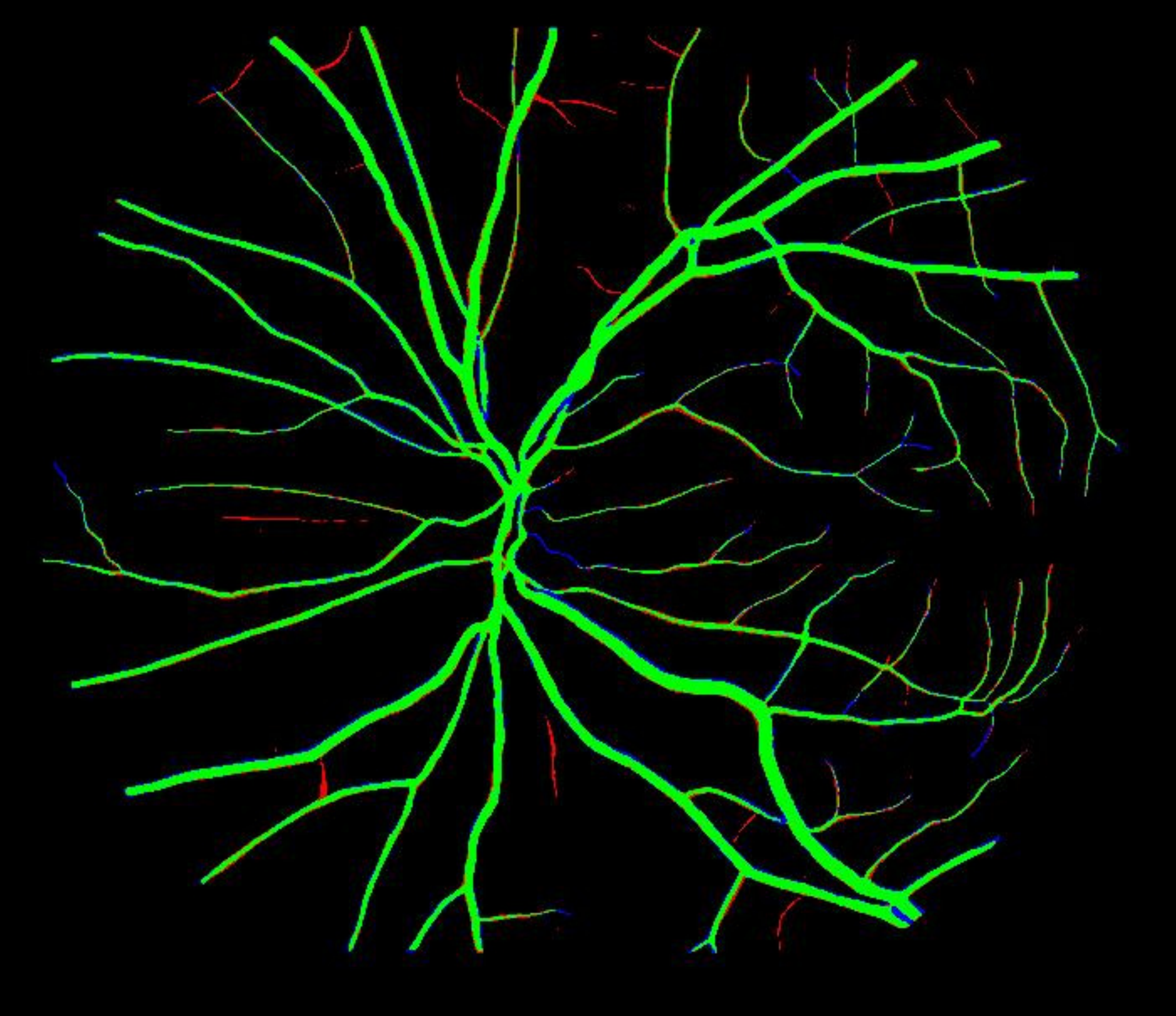}      & \includegraphics[width=0.14\textwidth]{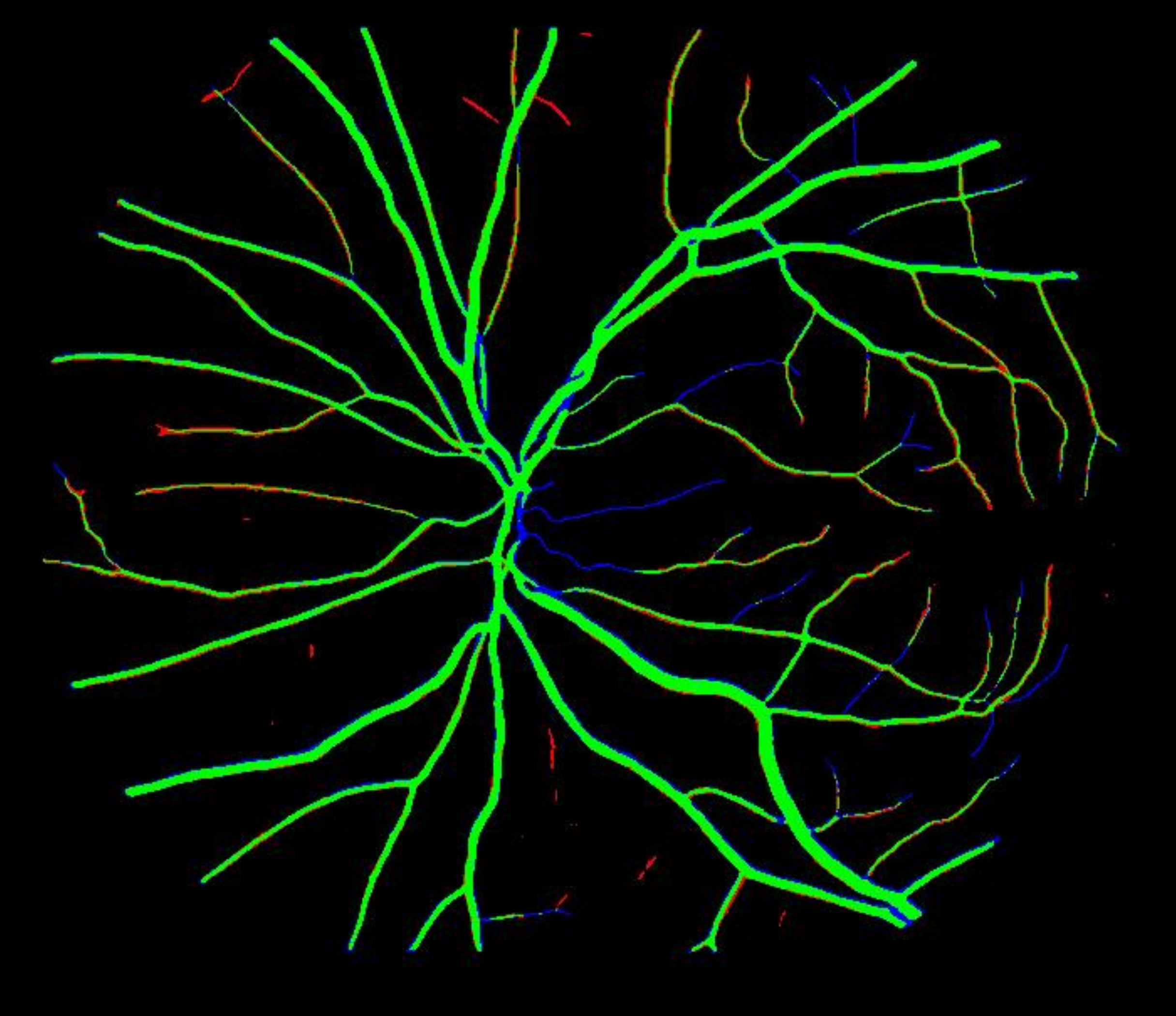}      & \includegraphics[width=0.14\textwidth]{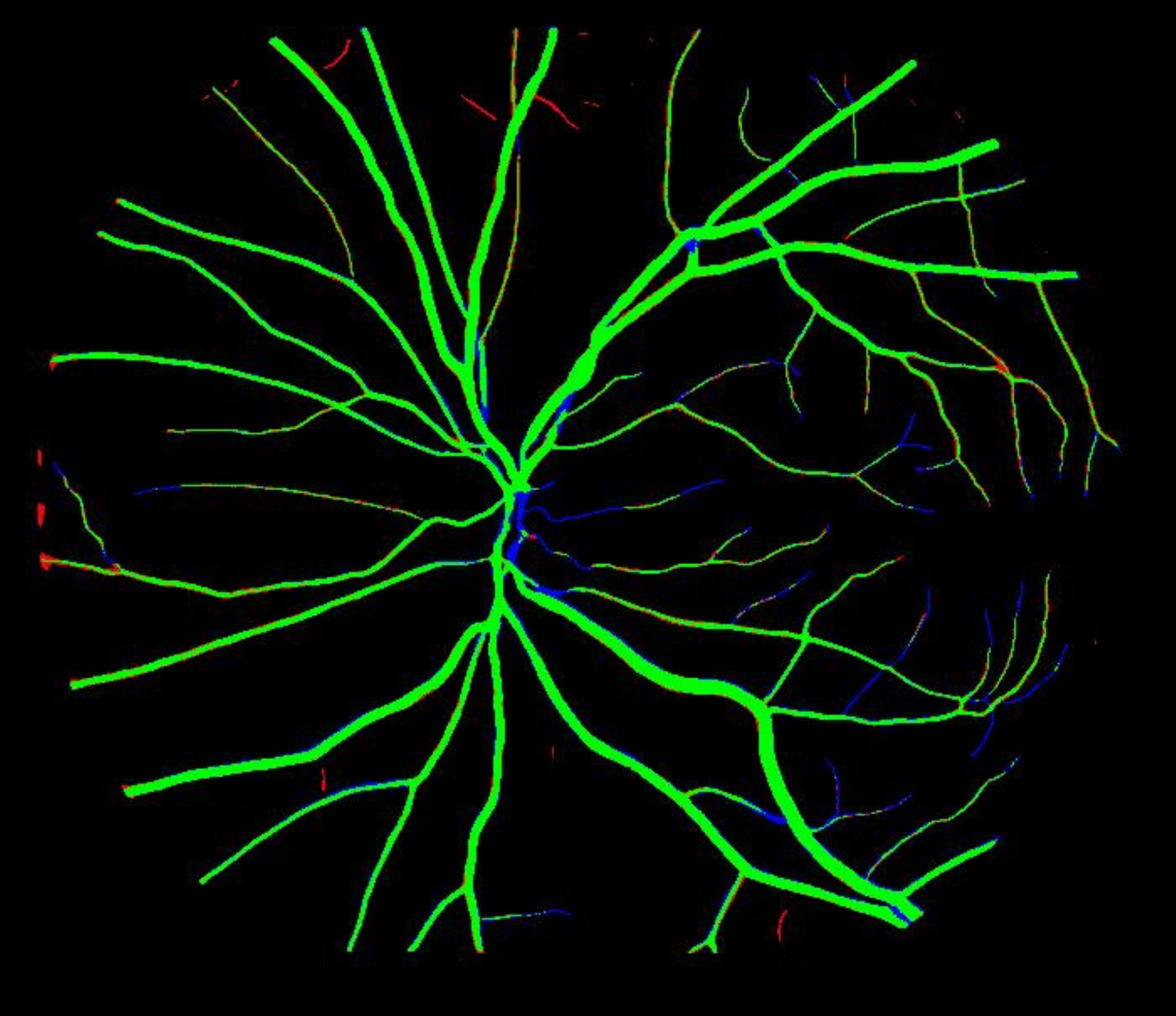} \\
			\includegraphics[width=0.14\textwidth]{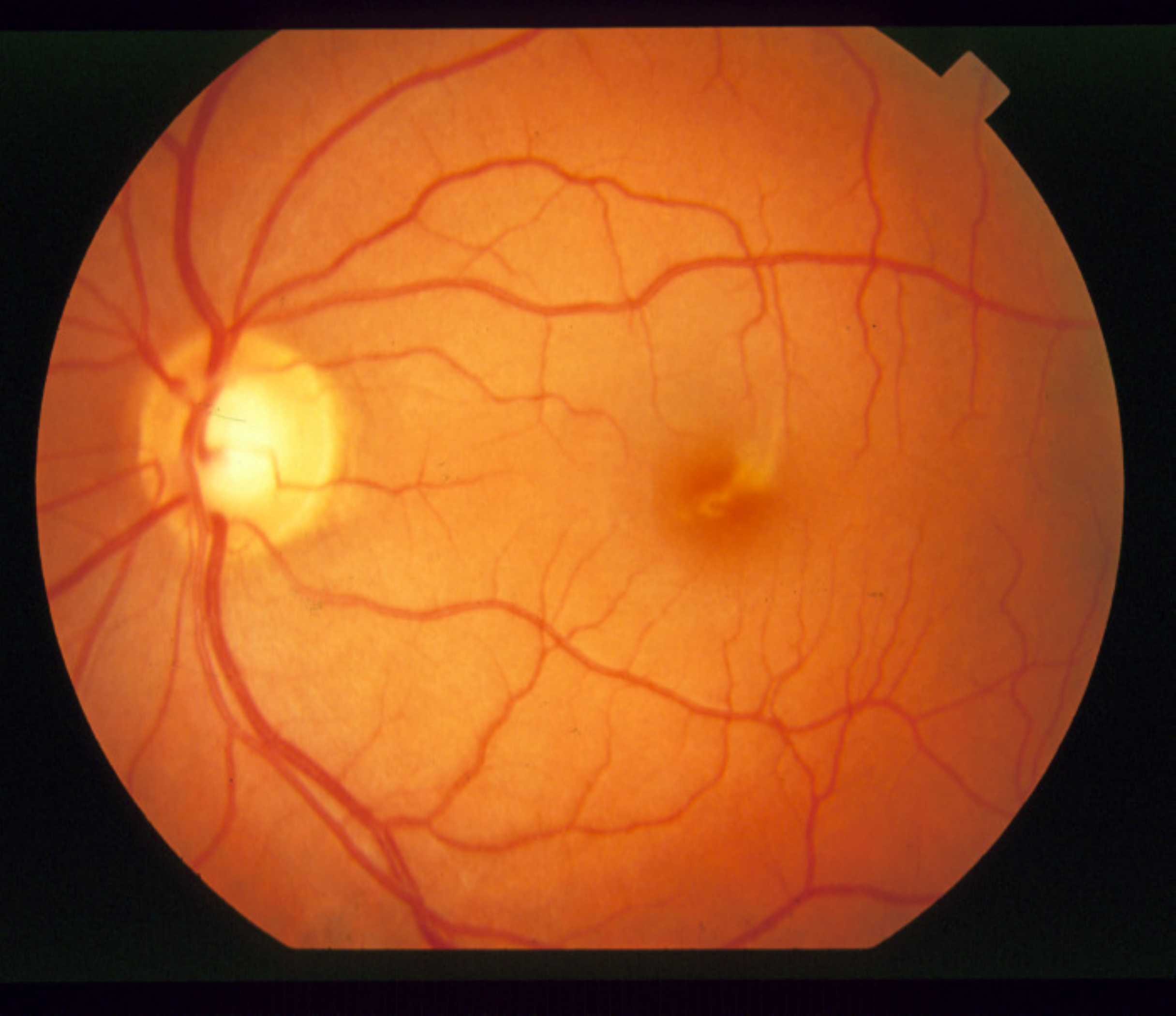}   & \includegraphics[width=0.14\textwidth]{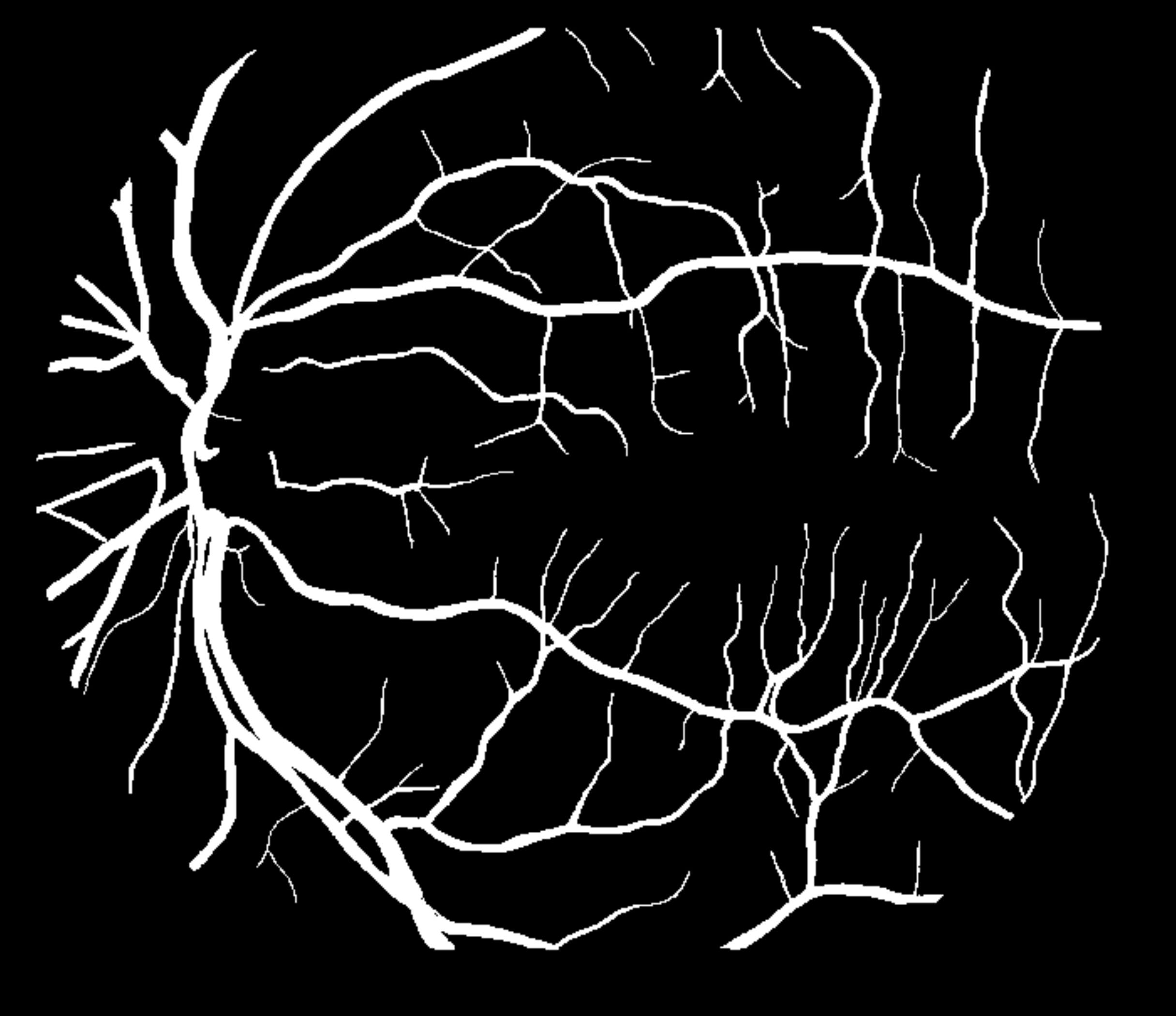}      &  \includegraphics[width=0.14\textwidth]{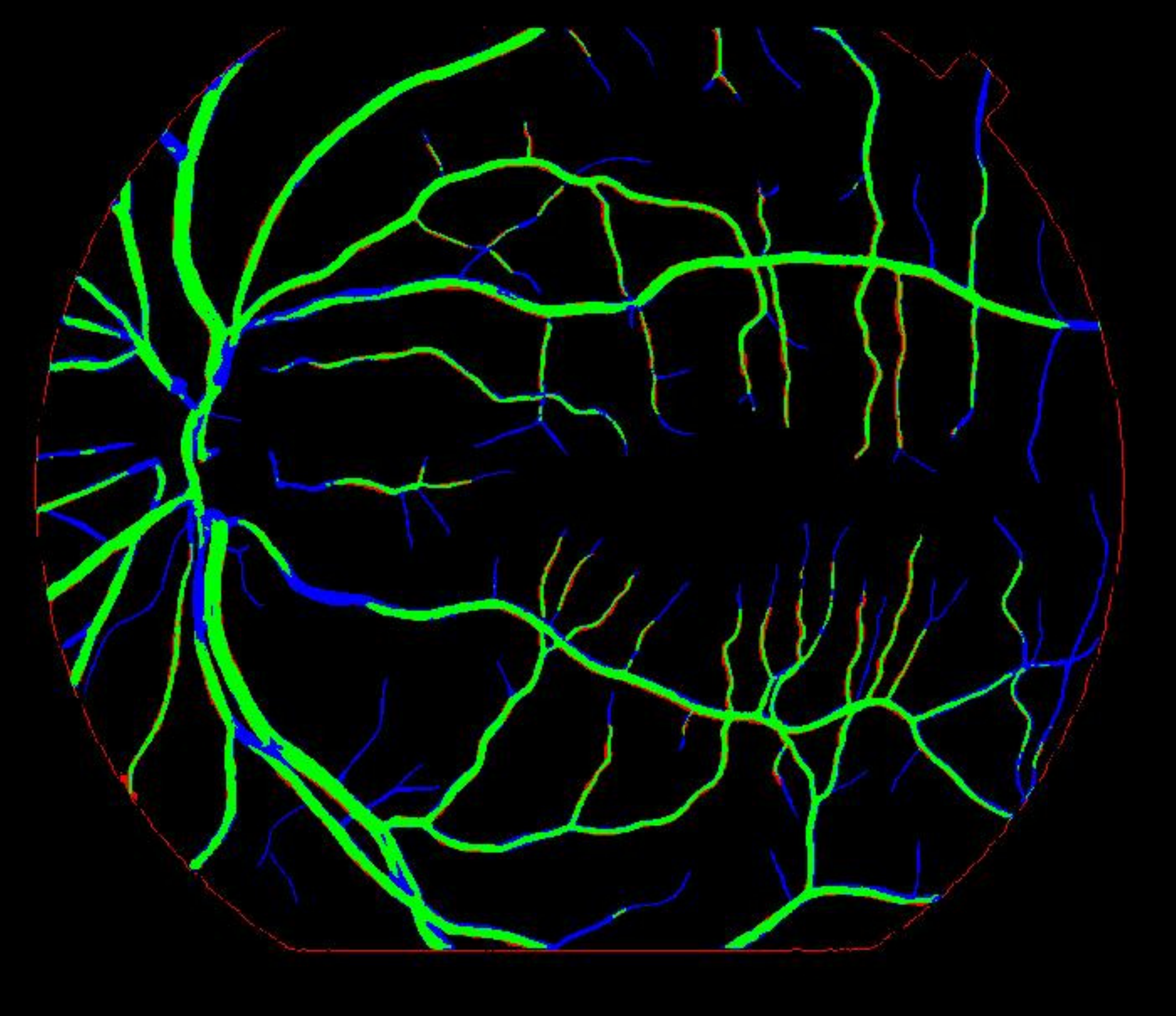}       & \includegraphics[width=0.14\textwidth]{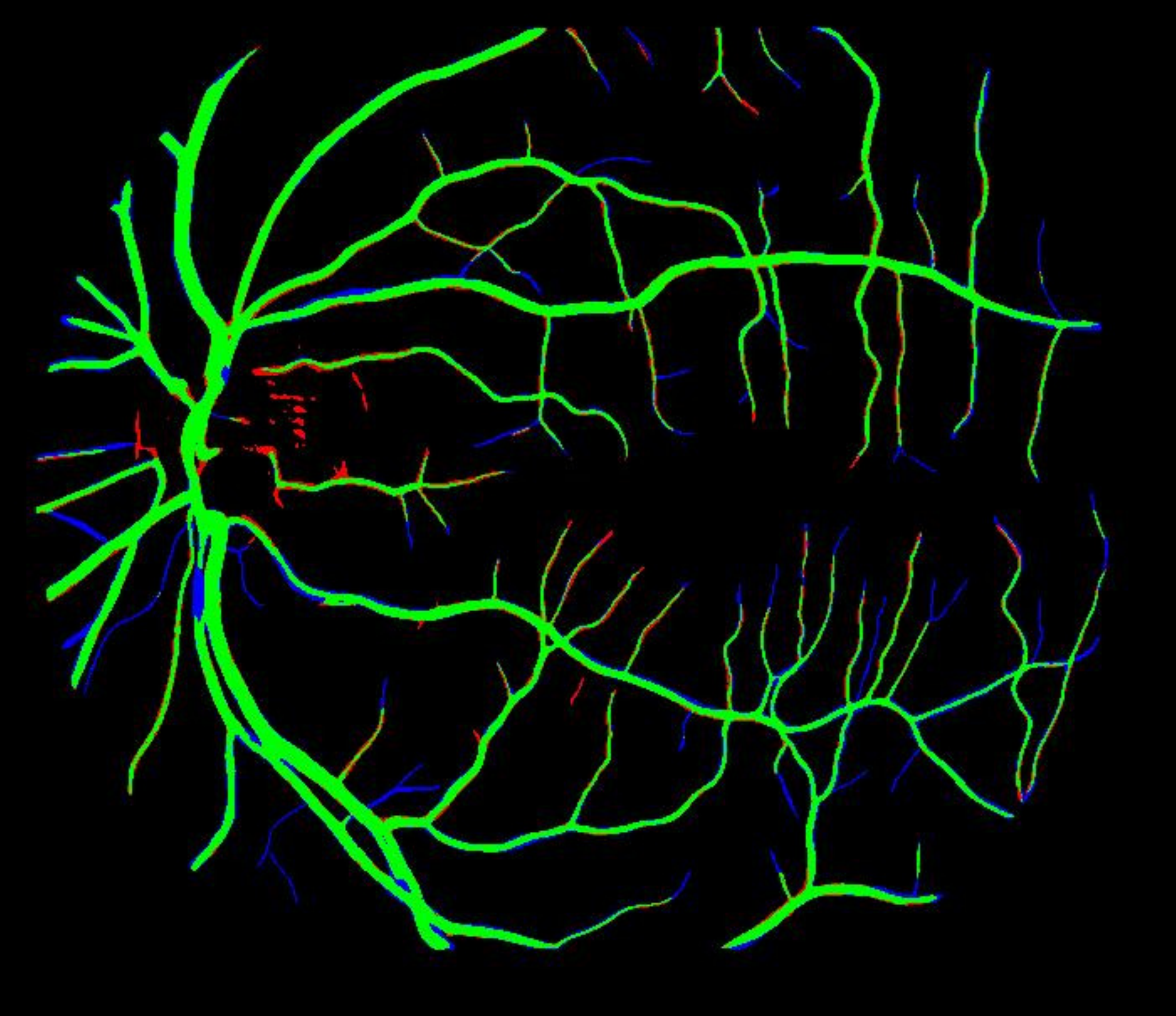}       & \includegraphics[width=0.14\textwidth]{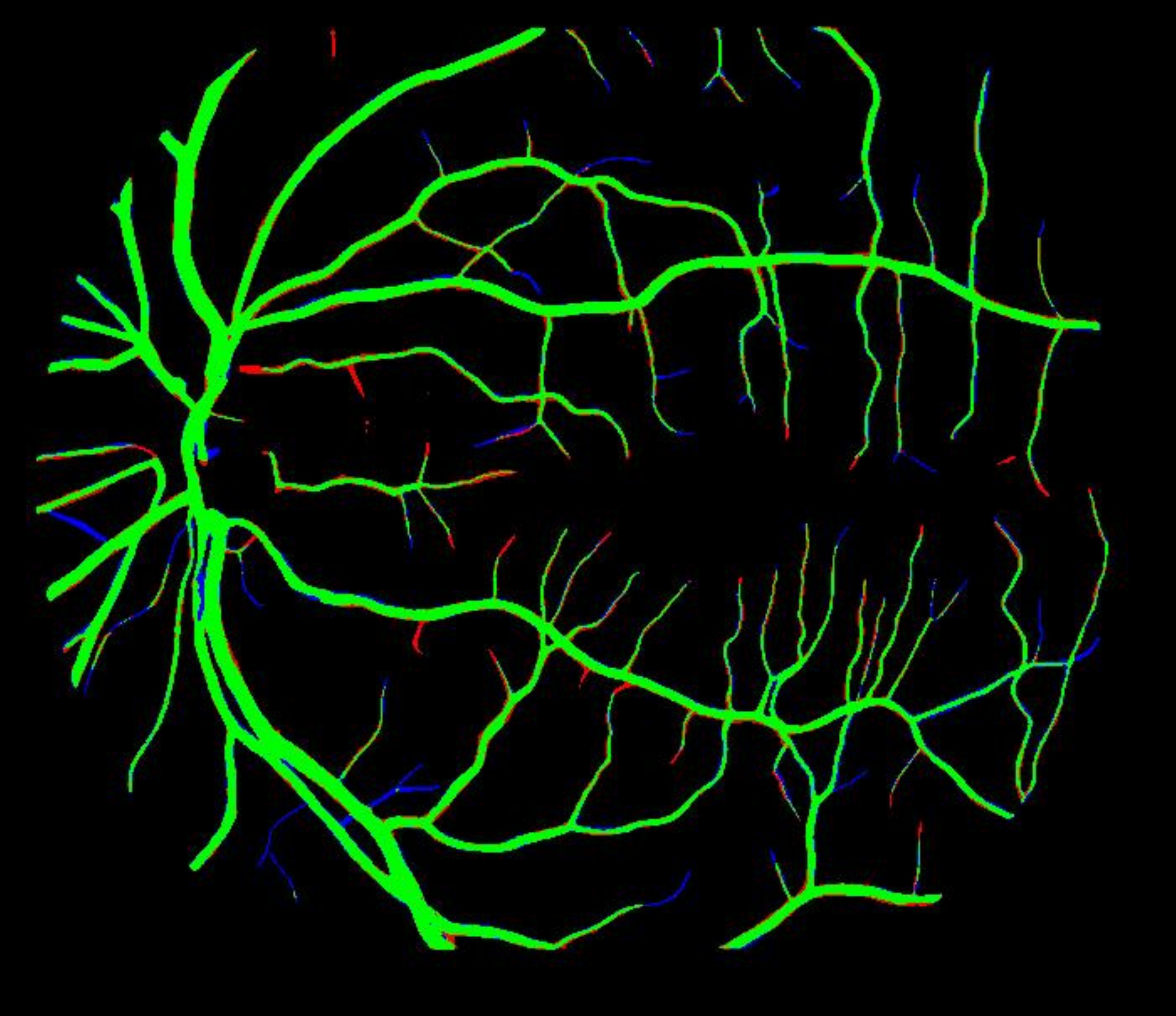}      & \includegraphics[width=0.14\textwidth]{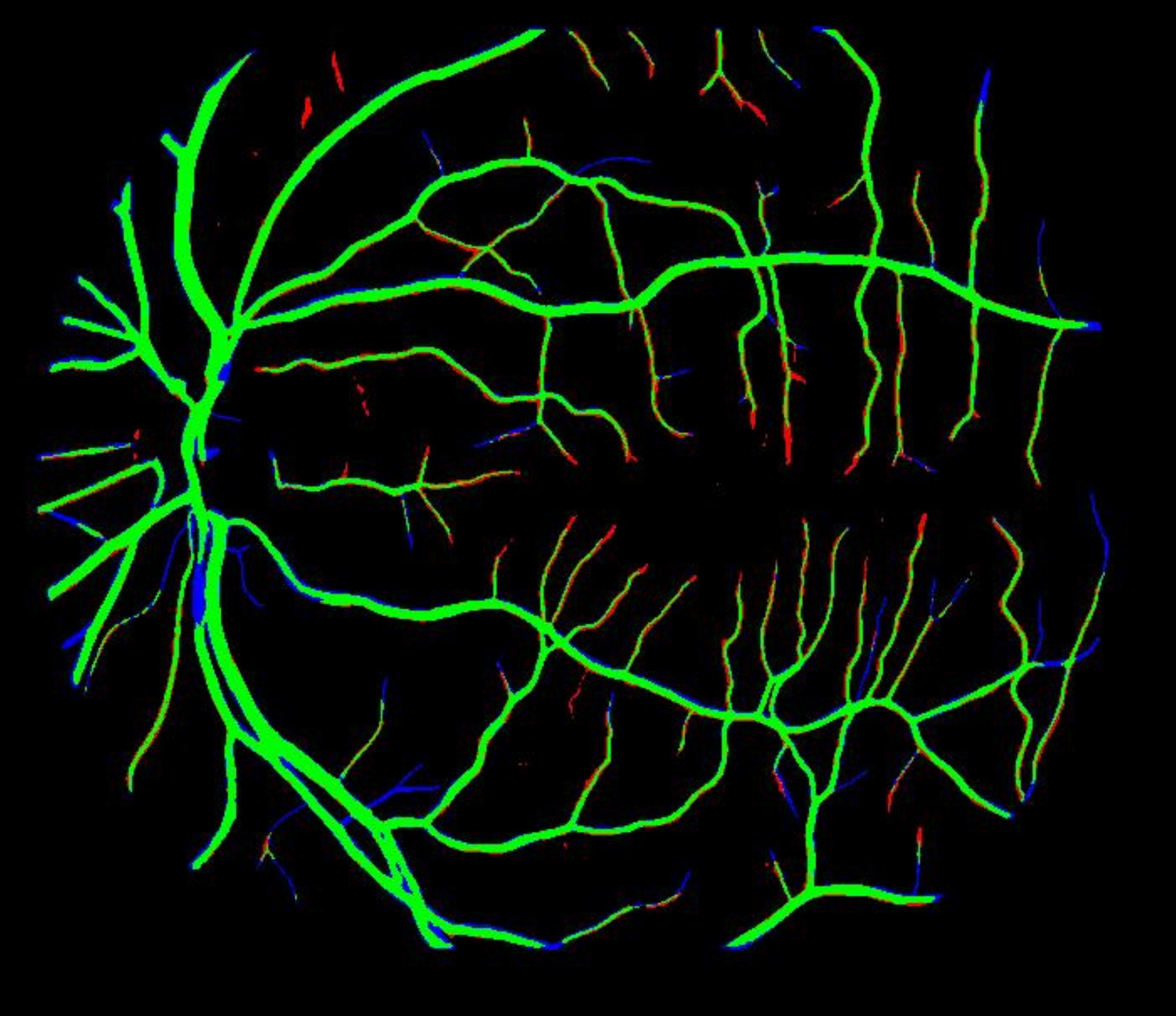}      & \includegraphics[width=0.14\textwidth]{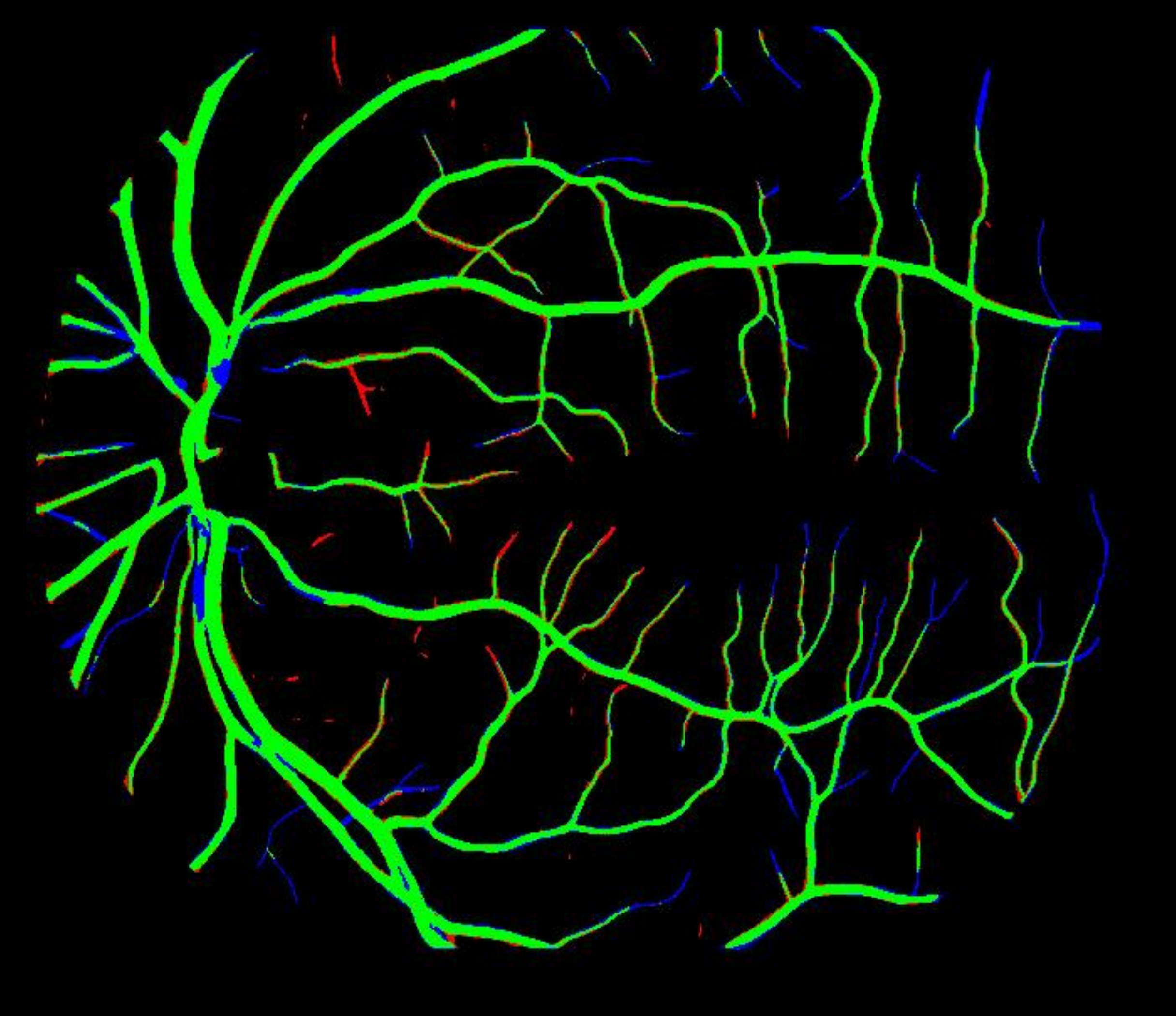} \\
			\includegraphics[width=0.14\textwidth]{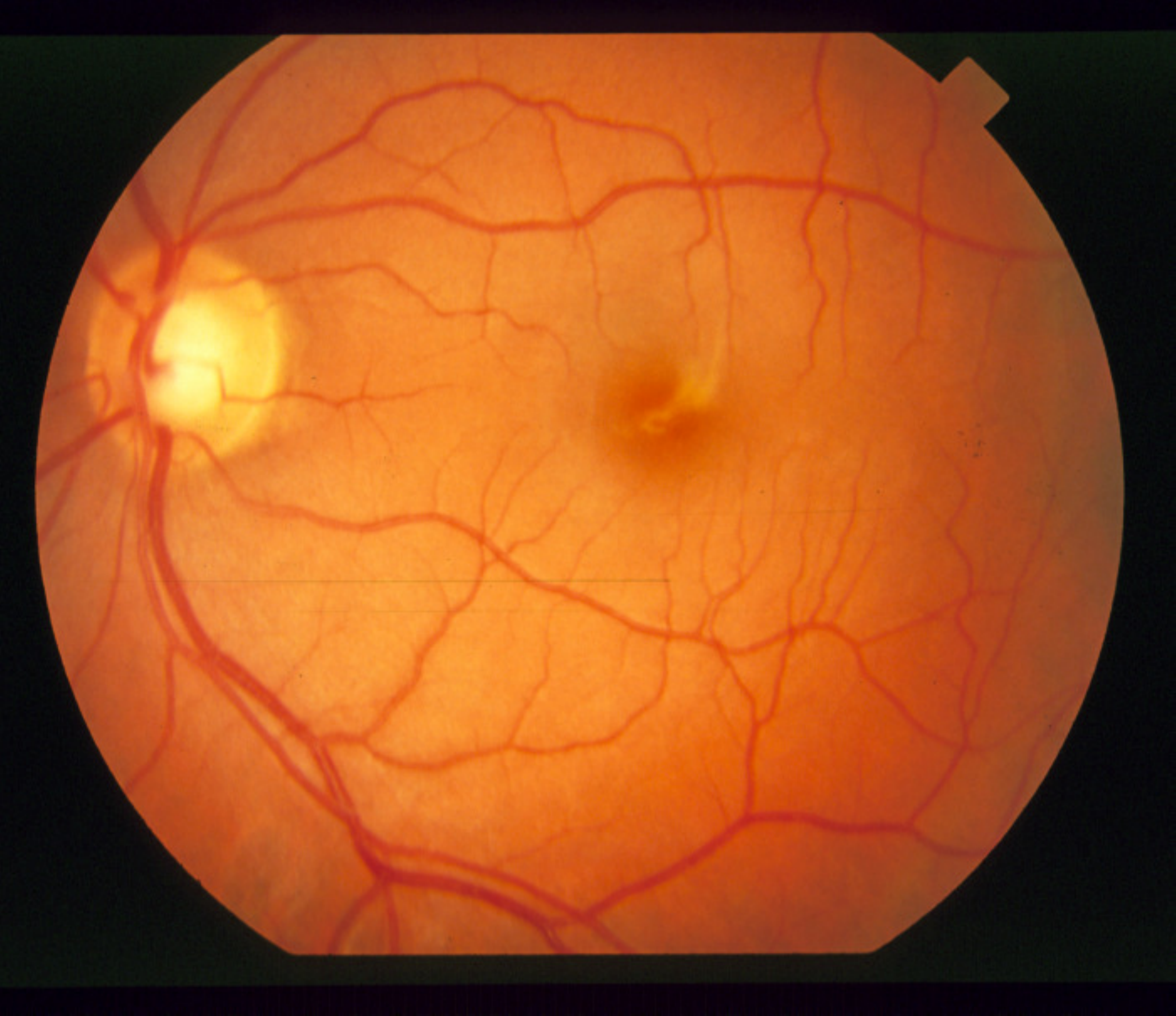}   & \includegraphics[width=0.14\textwidth]{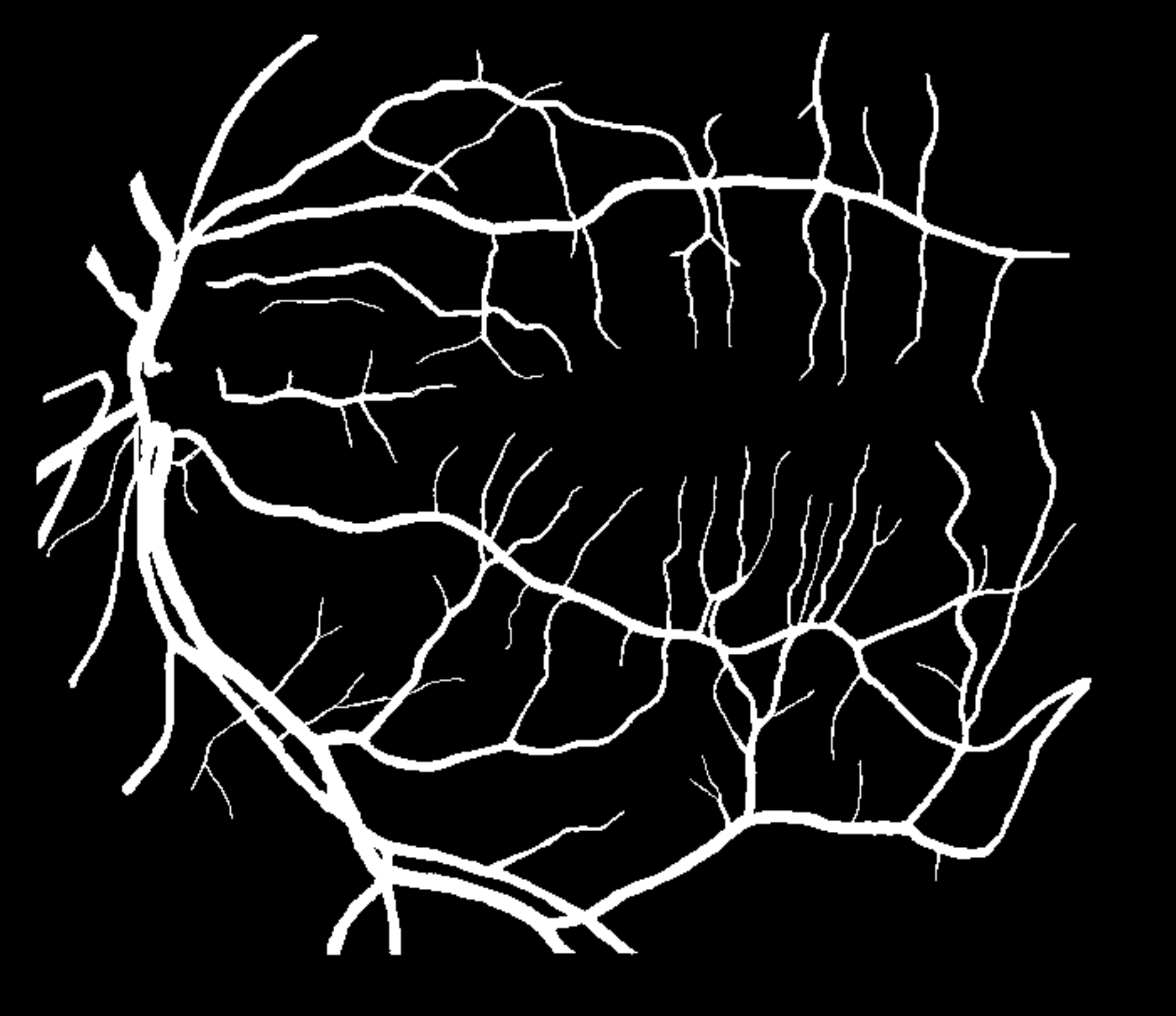}      &     \includegraphics[width=0.14\textwidth]{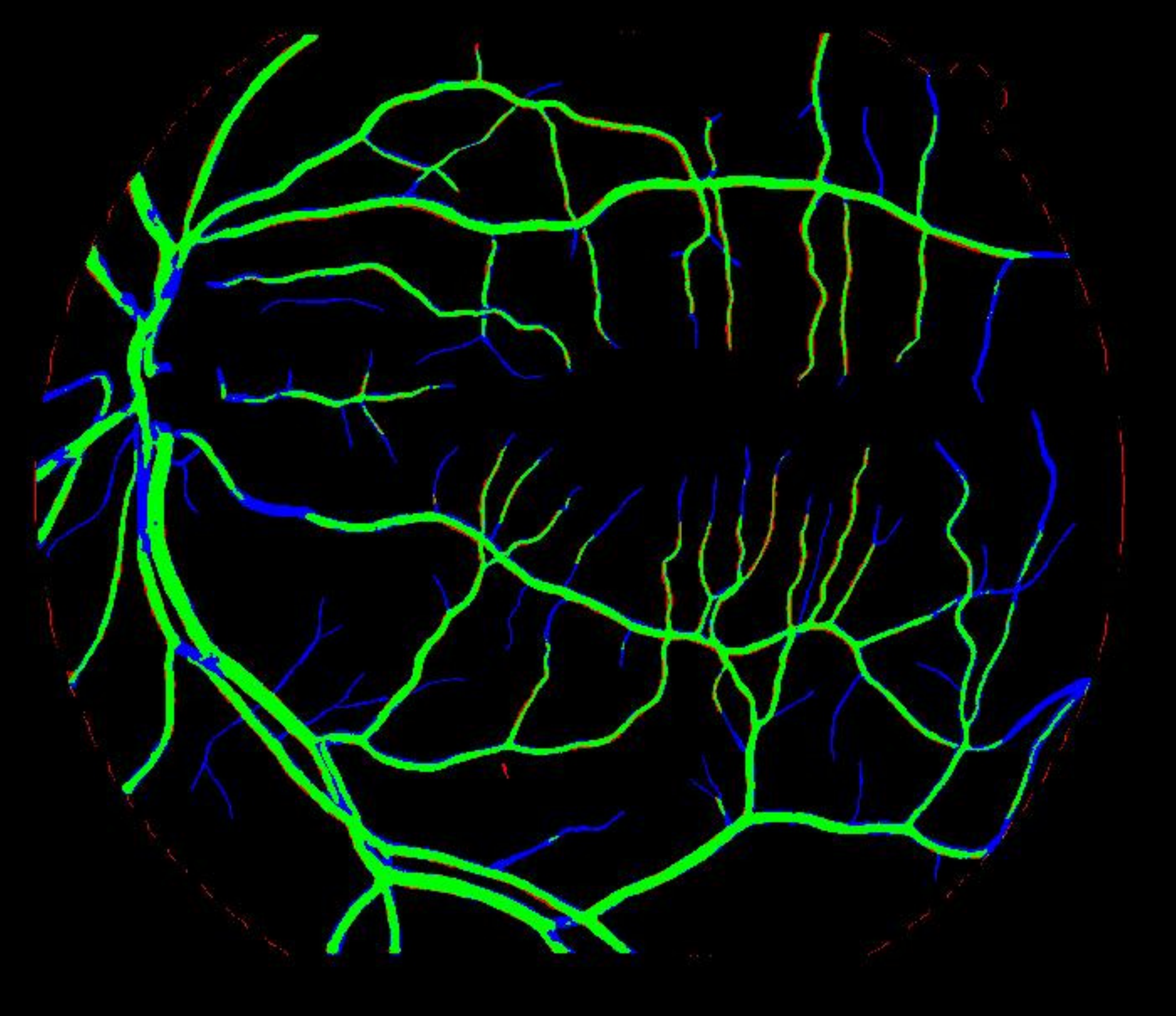}       & \includegraphics[width=0.14\textwidth]{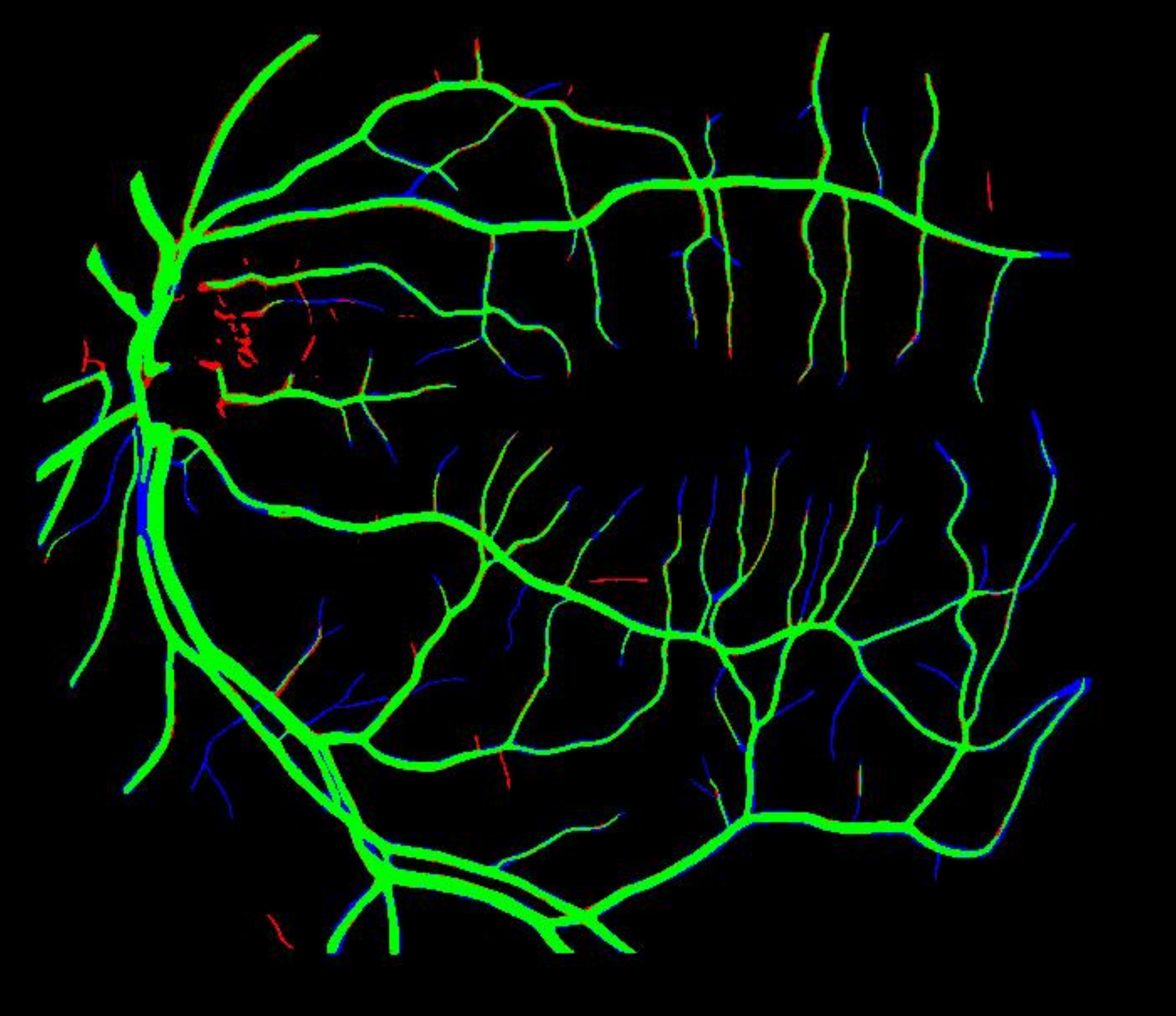}       & \includegraphics[width=0.14\textwidth]{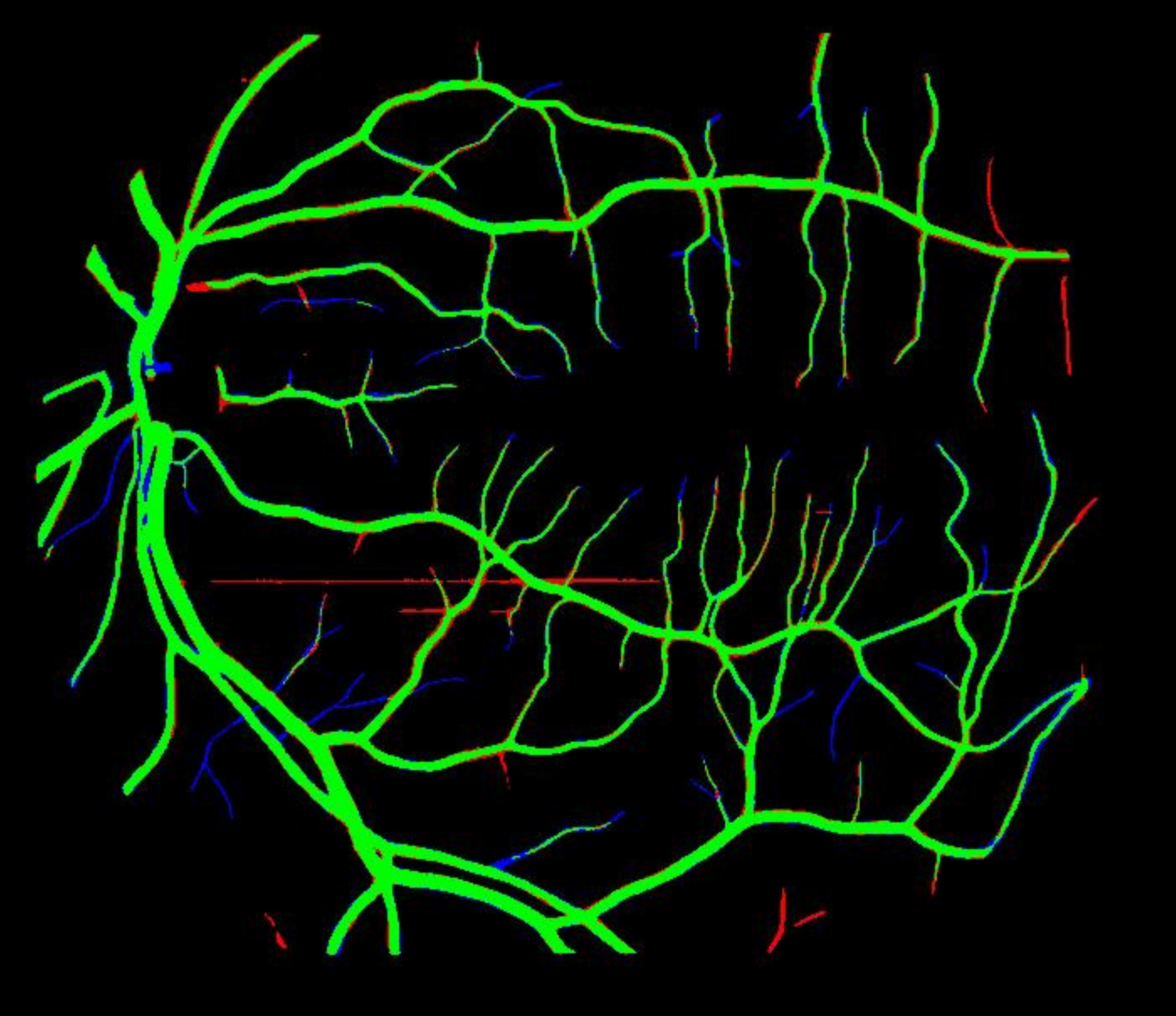}      & \includegraphics[width=0.14\textwidth]{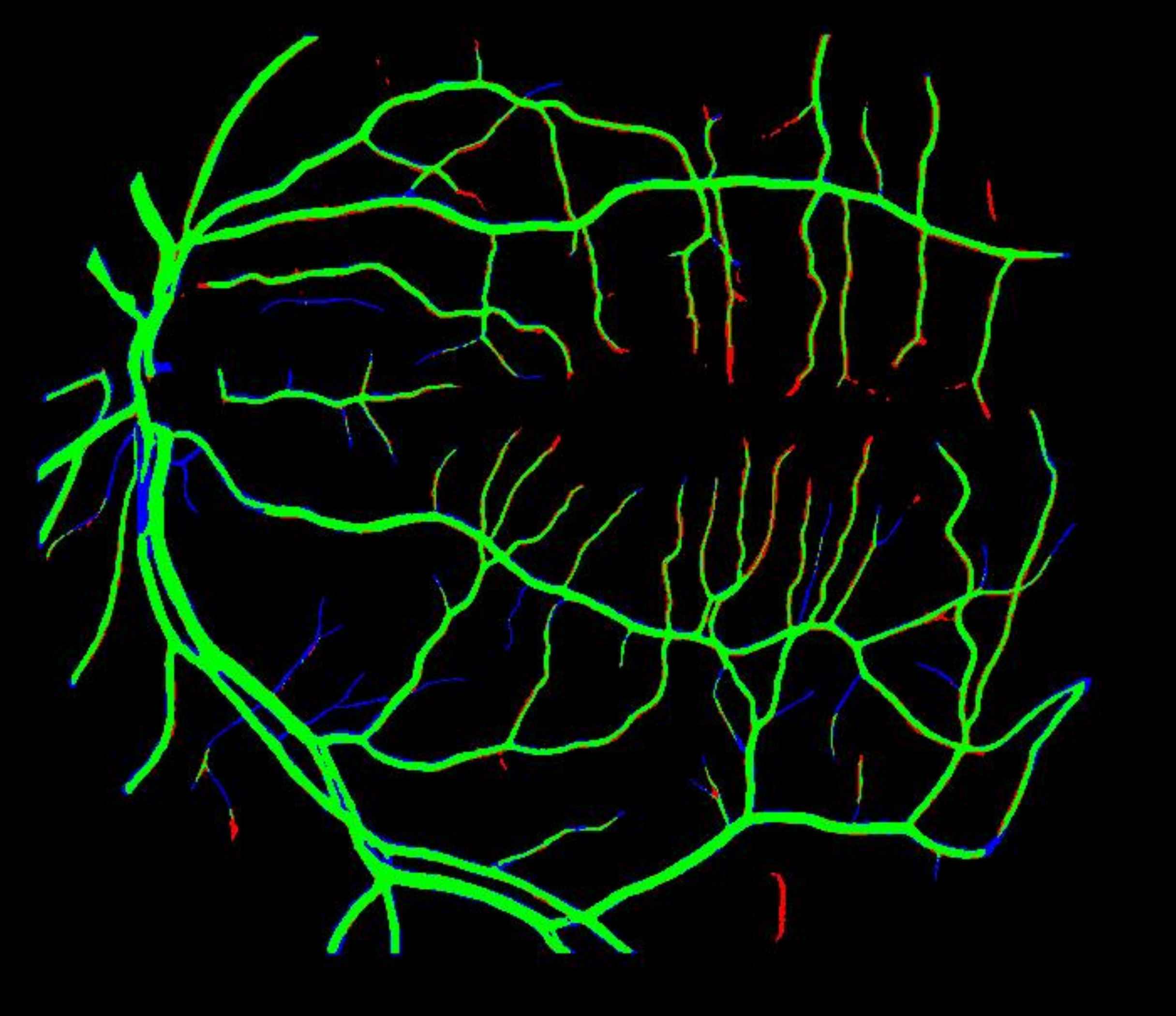}      & \includegraphics[width=0.14\textwidth]{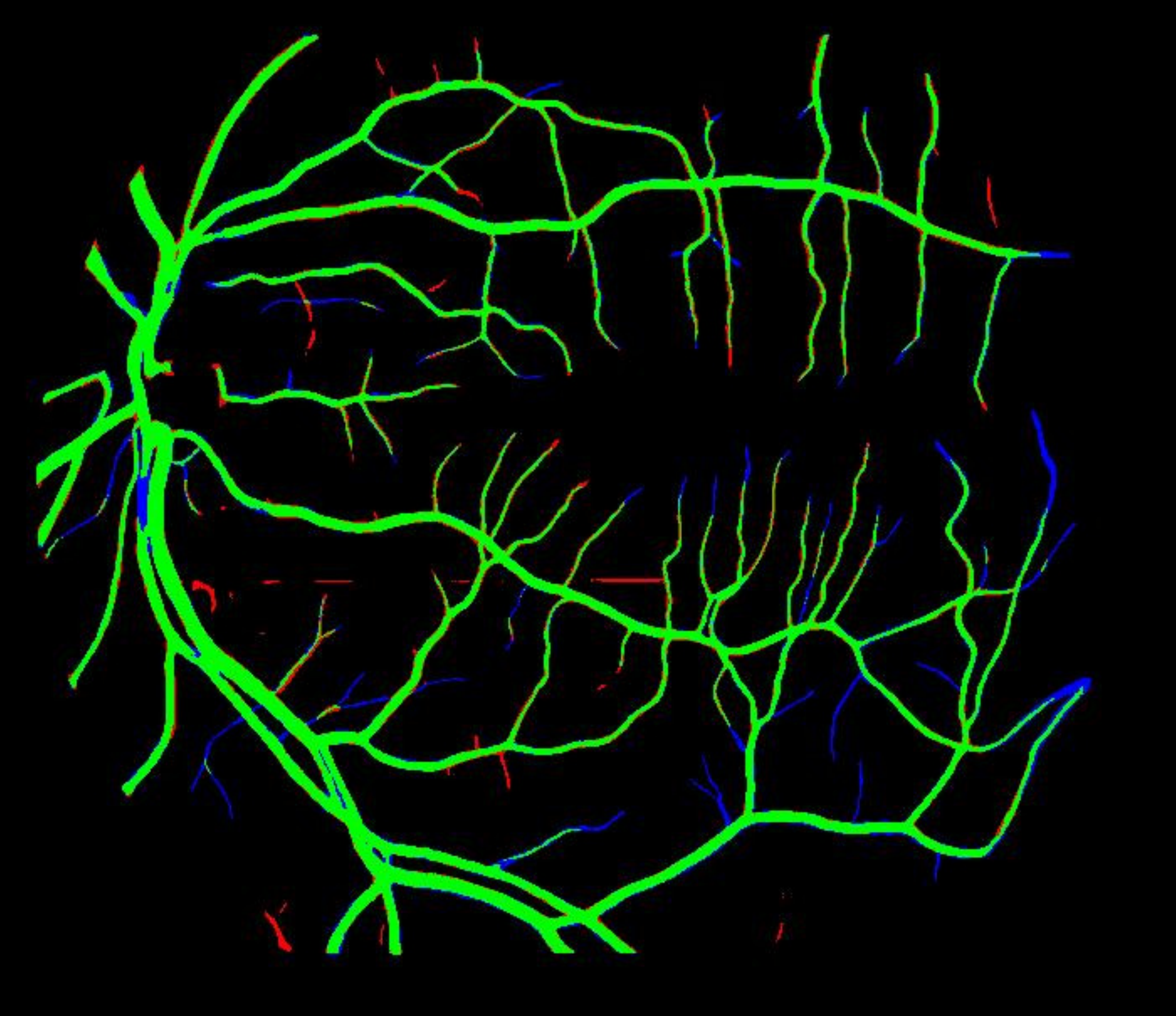} \\
			\includegraphics[width=0.14\textwidth]{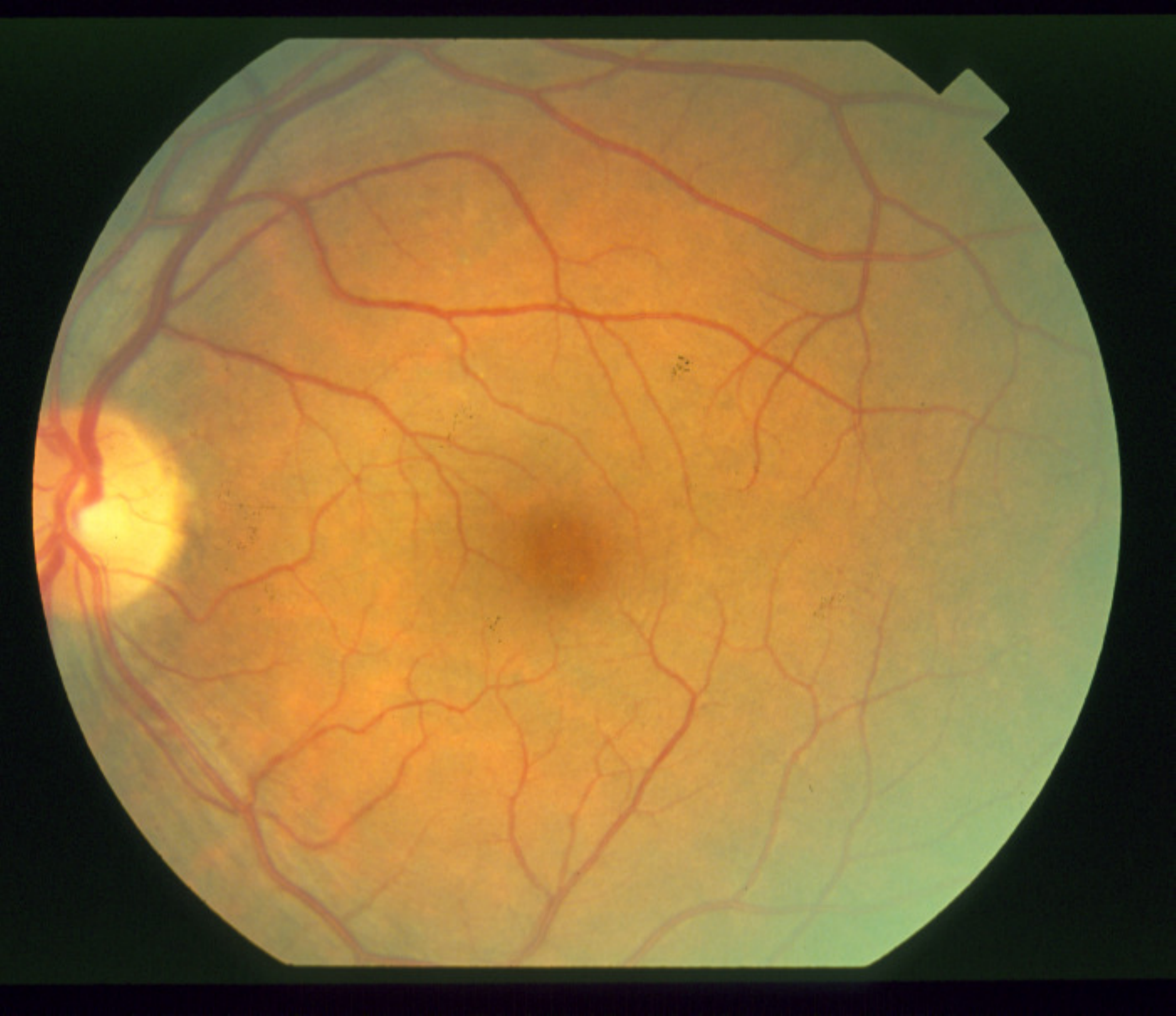}   & \includegraphics[width=0.14\textwidth]{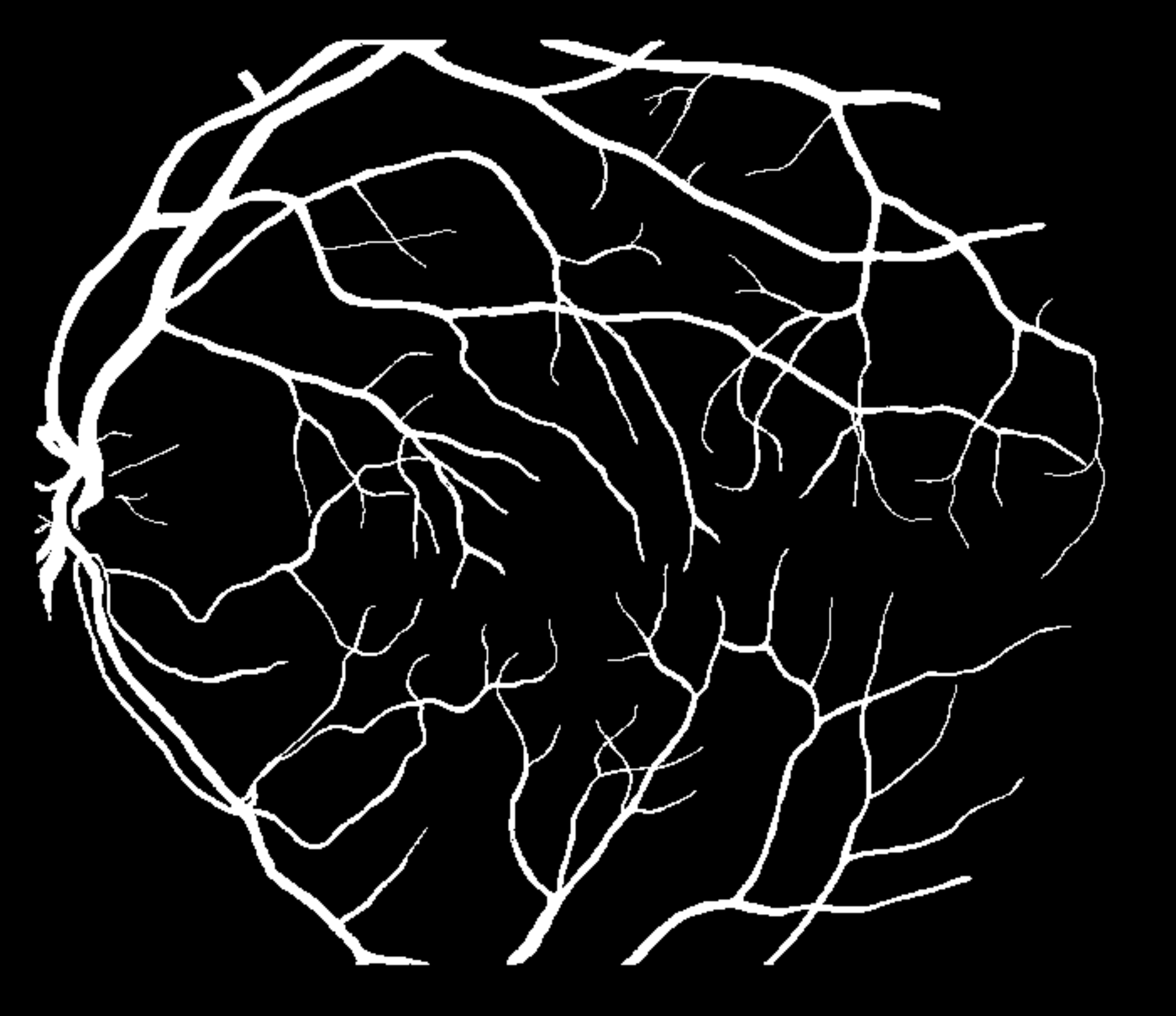}      &     \includegraphics[width=0.14\textwidth]{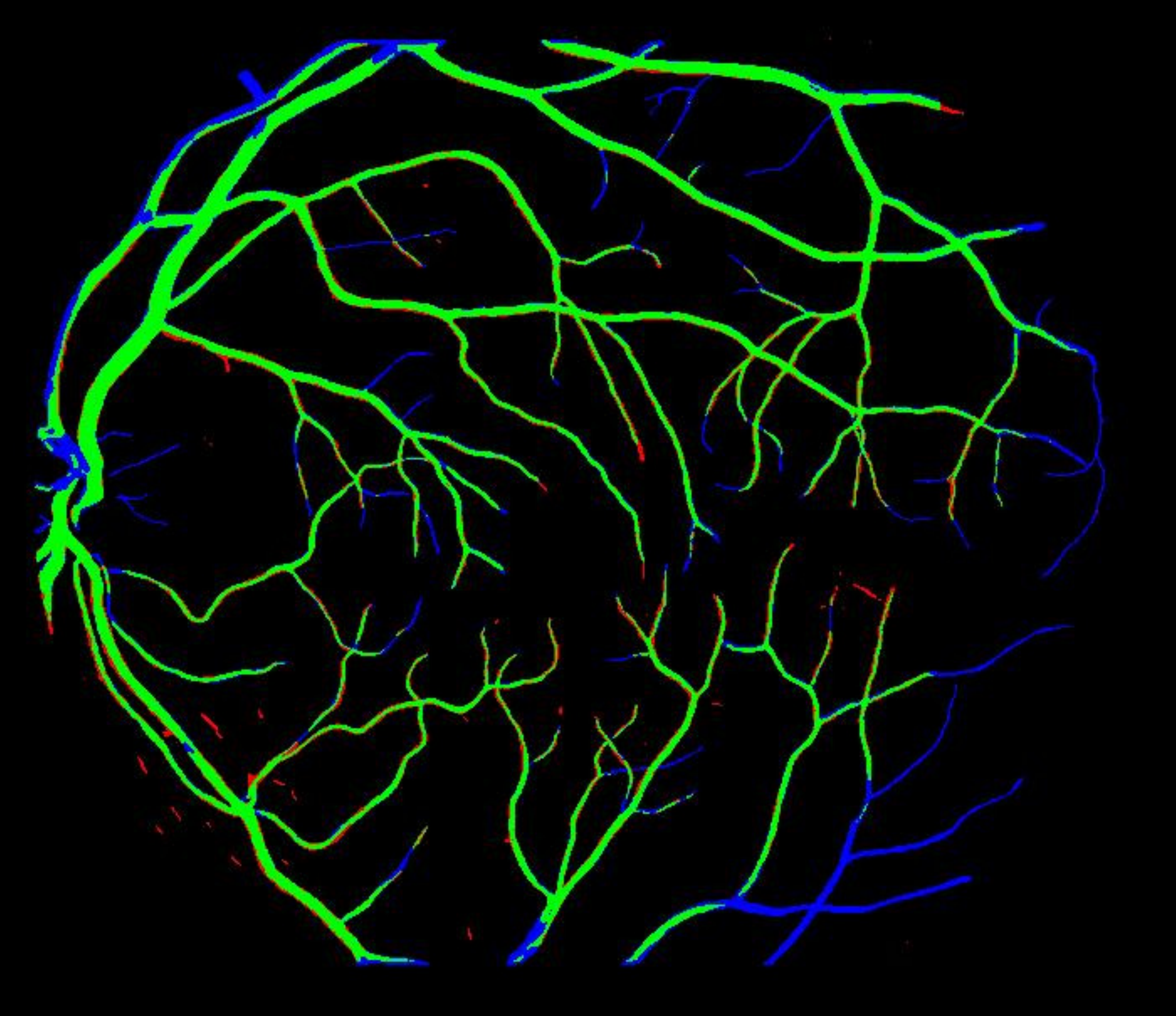}       & \includegraphics[width=0.14\textwidth]{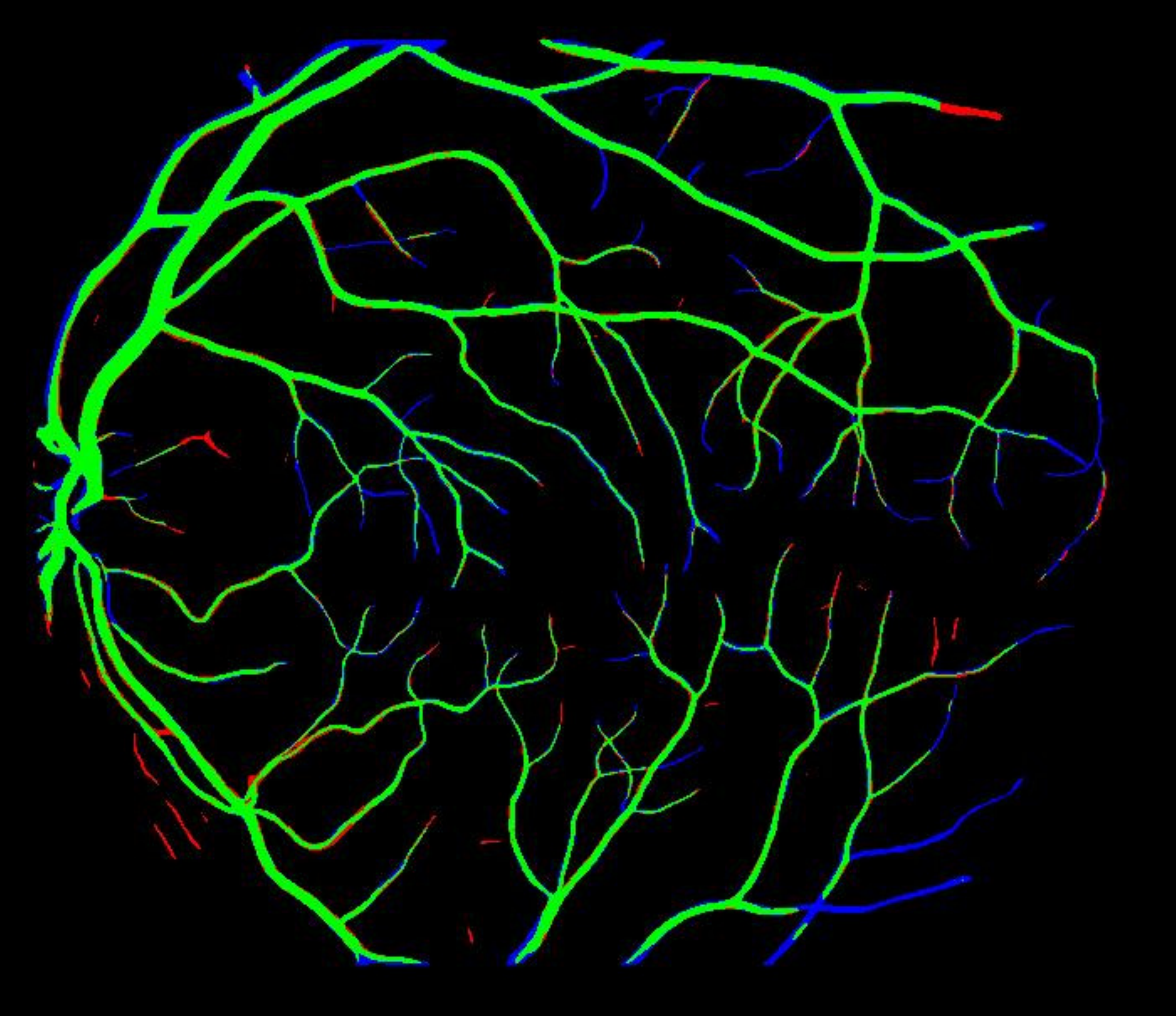}       & \includegraphics[width=0.14\textwidth]{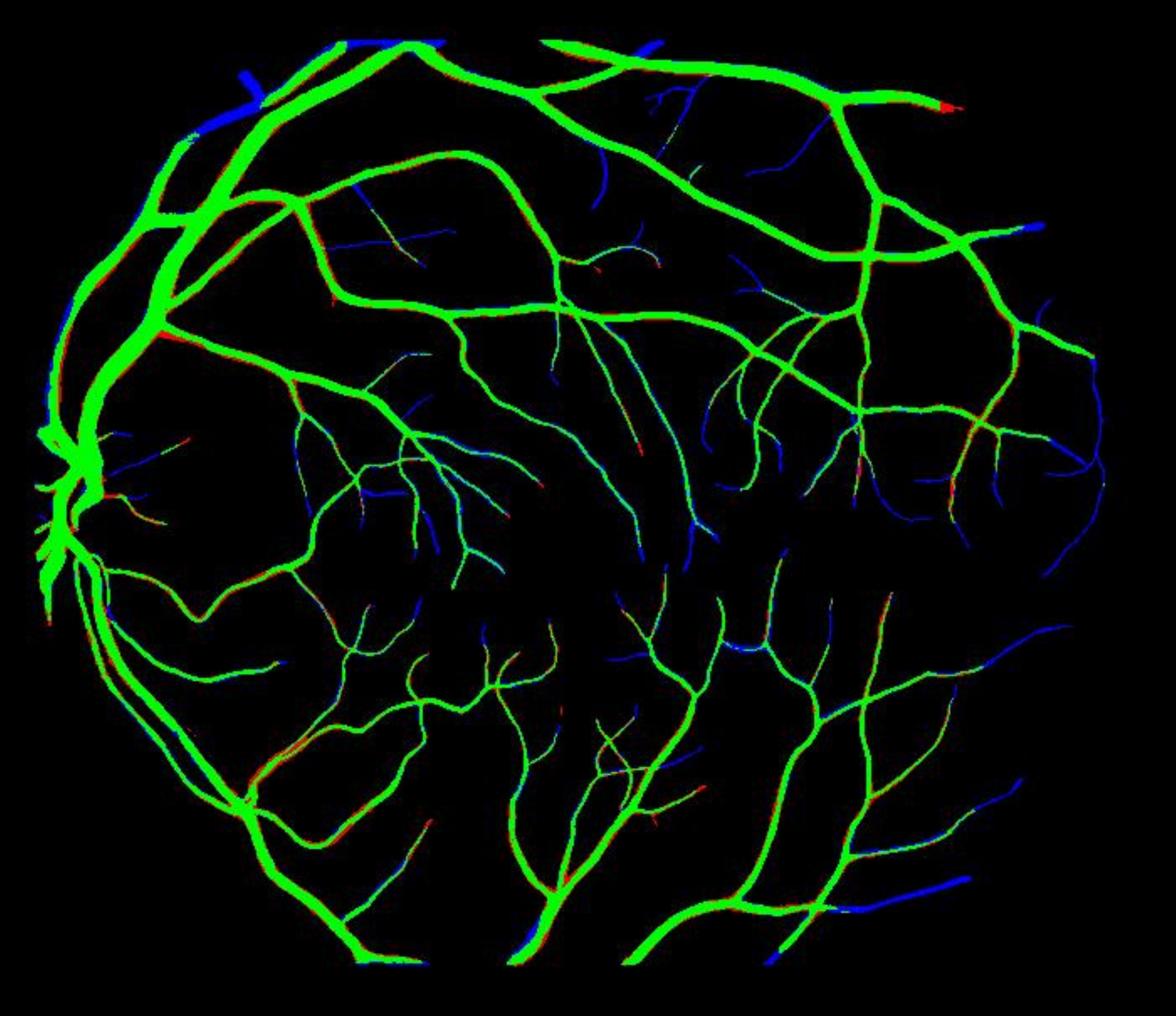}      & \includegraphics[width=0.14\textwidth]{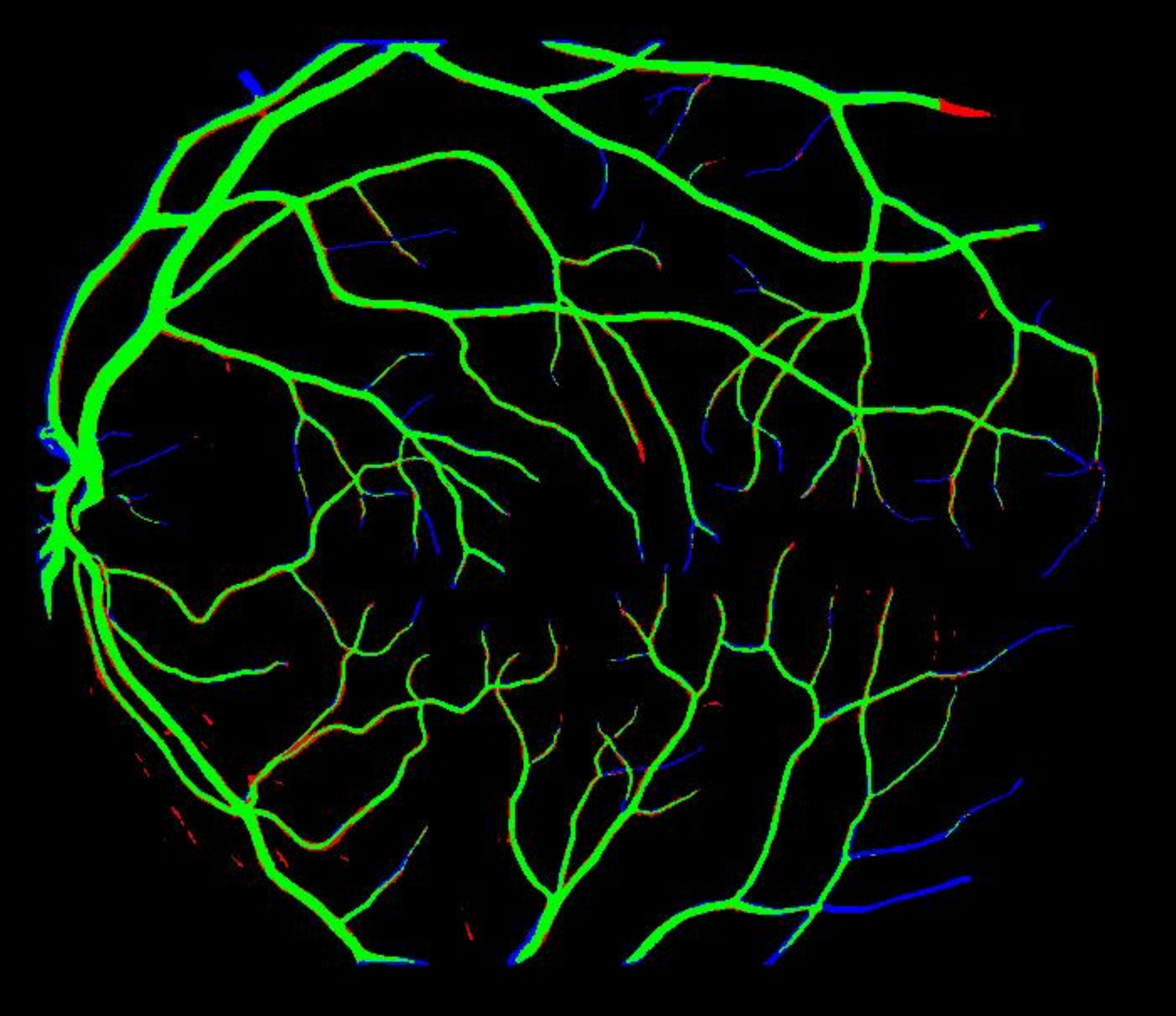}      & \includegraphics[width=0.14\textwidth]{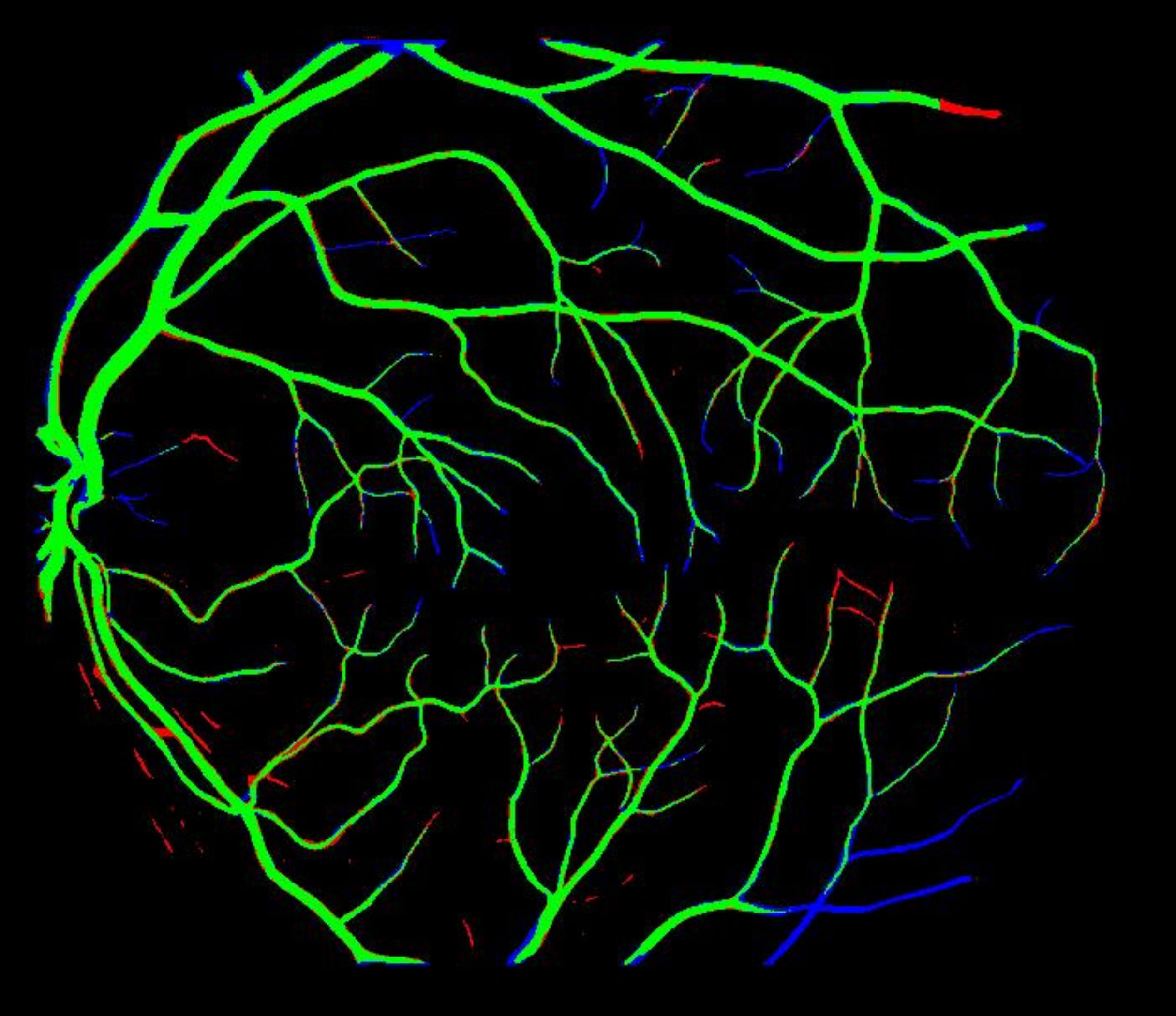} \\
			\includegraphics[width=0.14\textwidth]{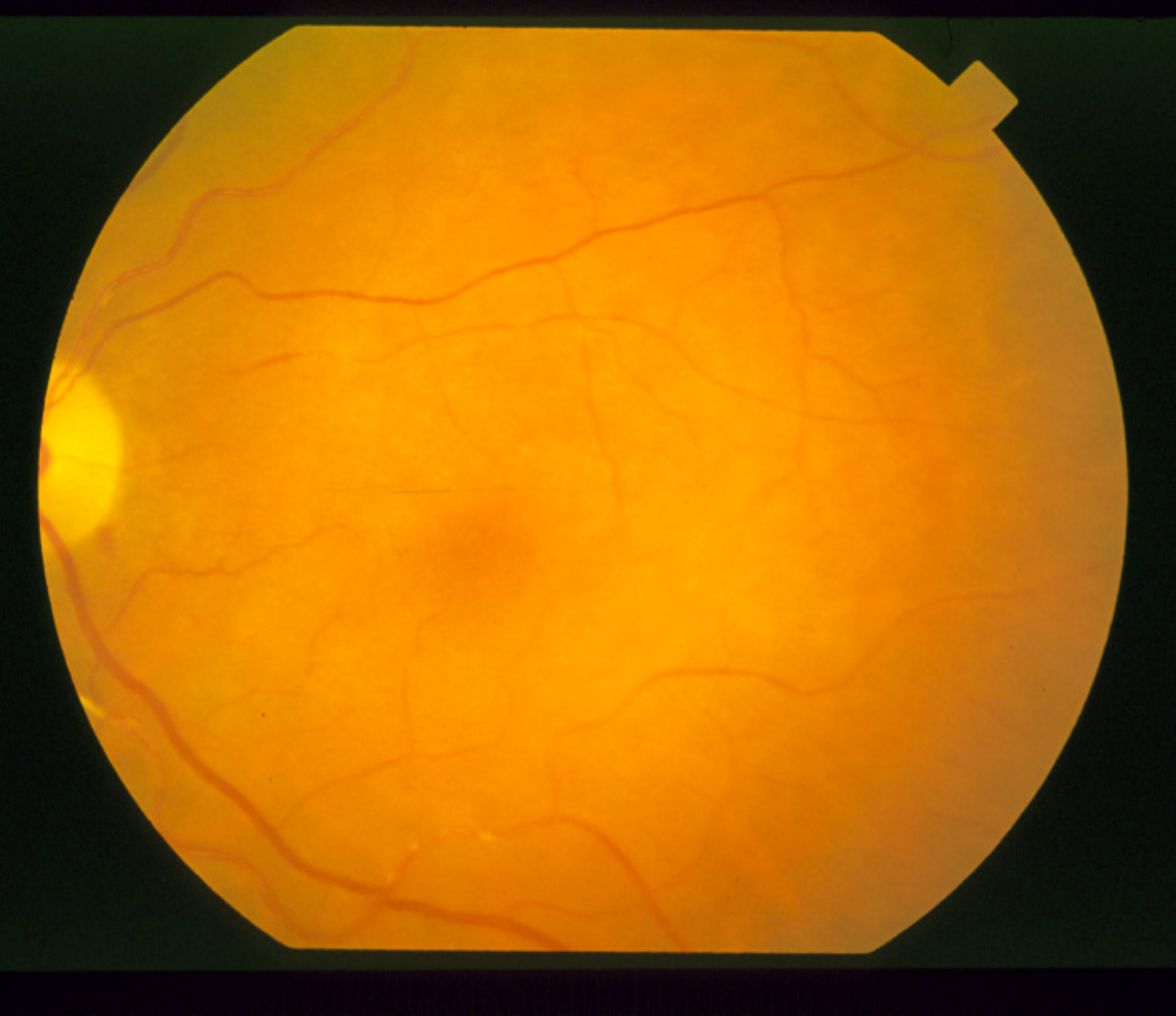}   & \includegraphics[width=0.14\textwidth]{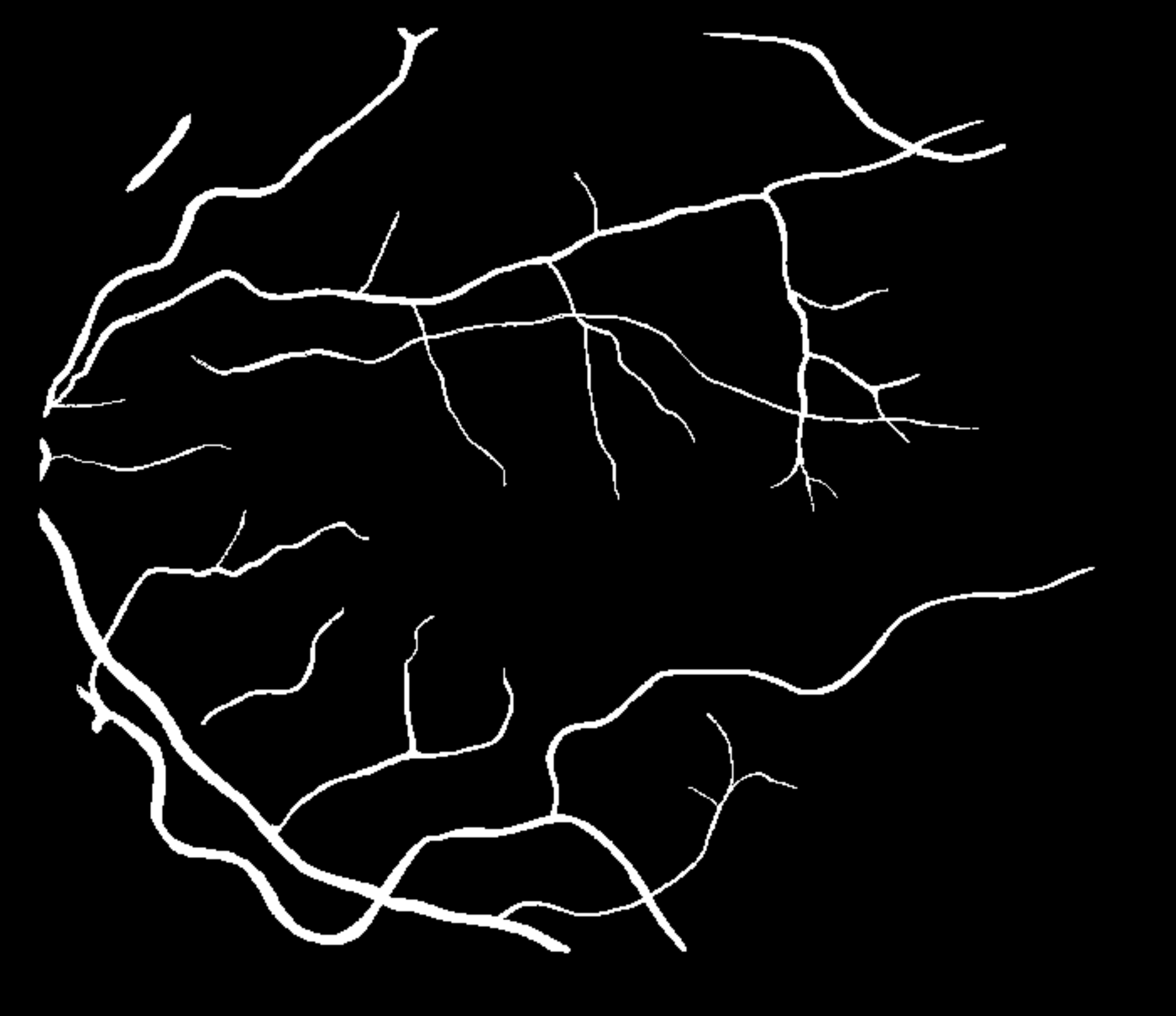}      &    \includegraphics[width=0.14\textwidth]{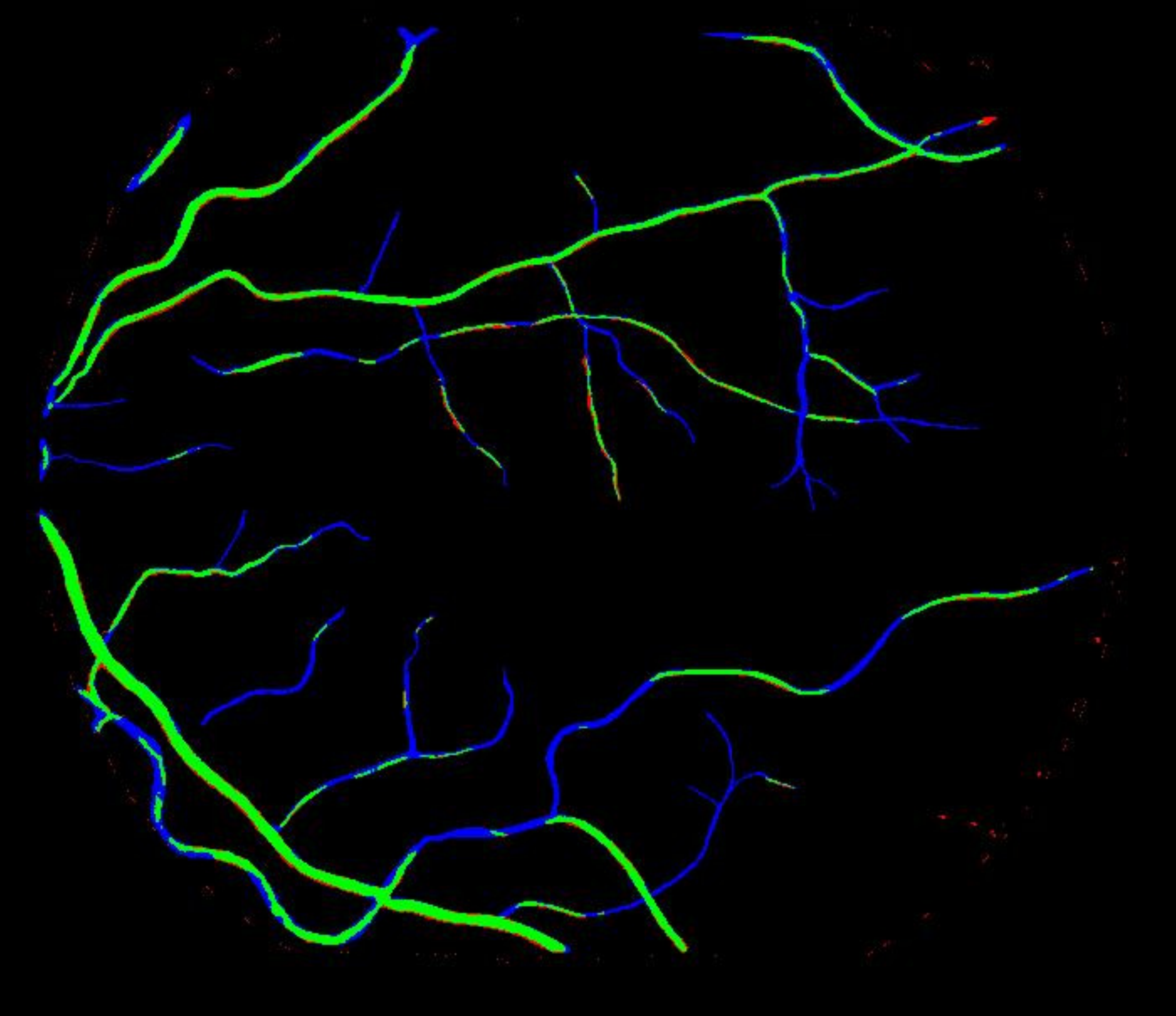}       & \includegraphics[width=0.14\textwidth]{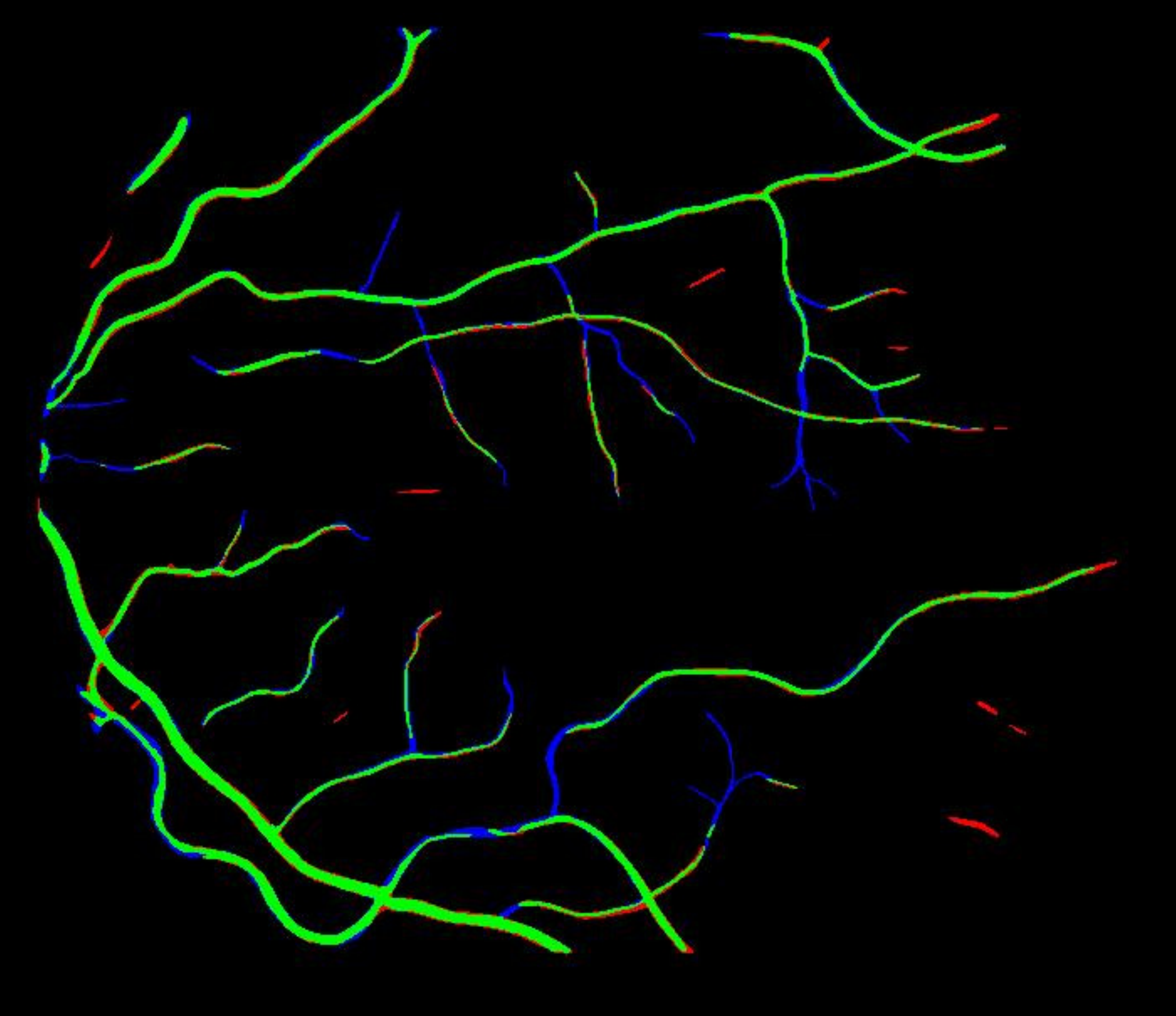}       & \includegraphics[width=0.14\textwidth]{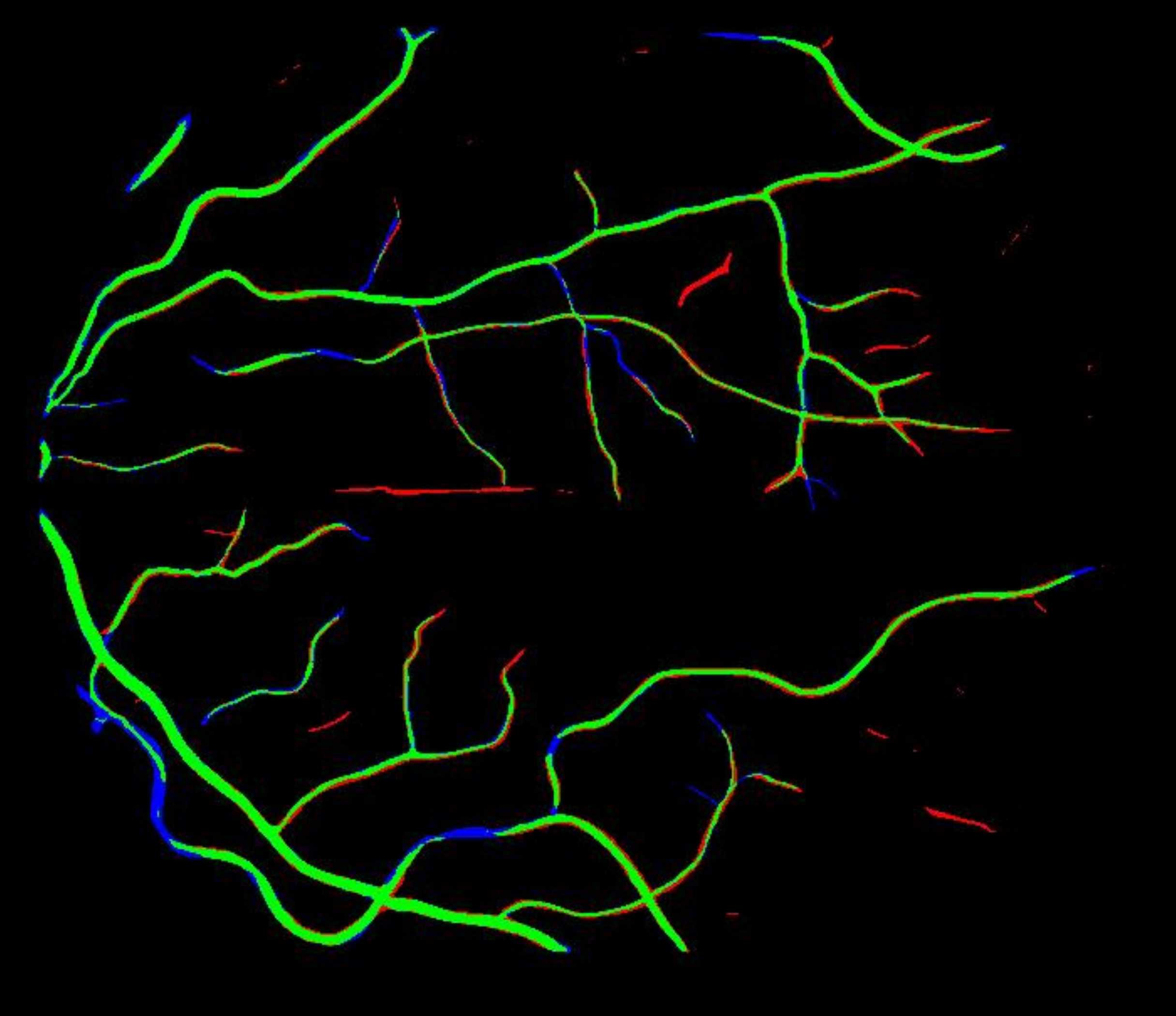}      & \includegraphics[width=0.14\textwidth]{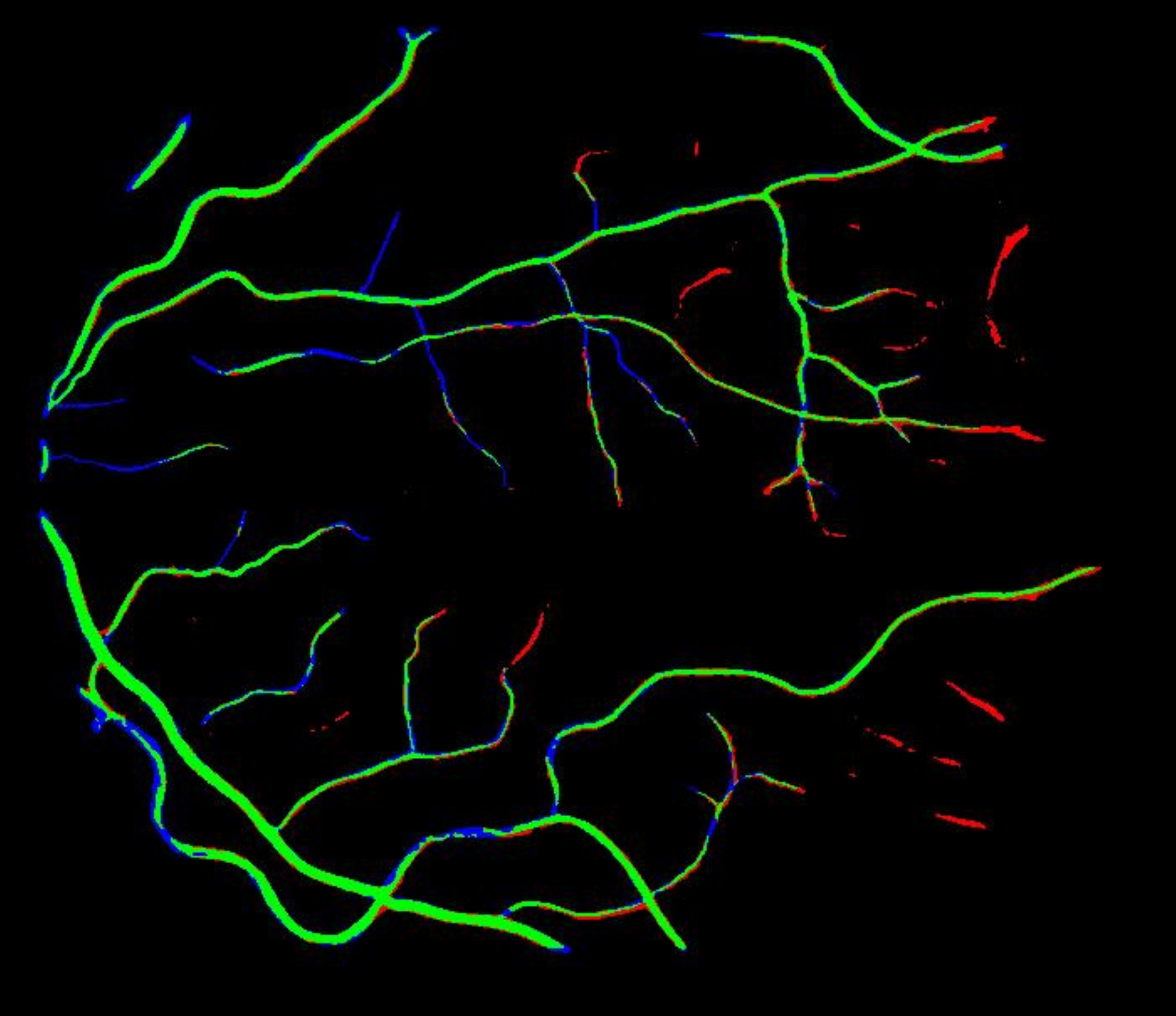}      & \includegraphics[width=0.14\textwidth]{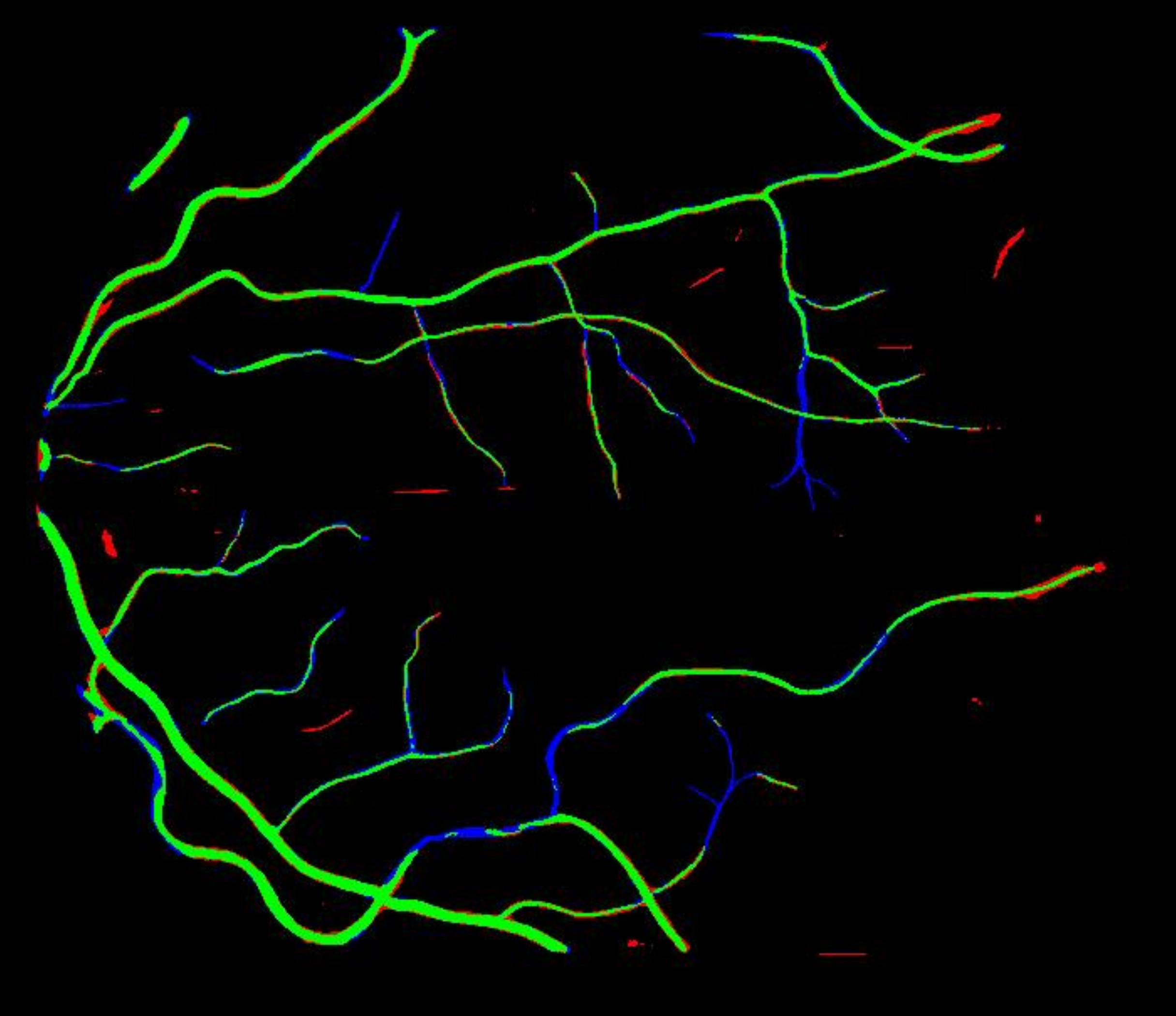} \\
		\end{tabular}%
	}
	\caption{Segmentation results delivered by our MRC-Net method on representative test images, i.e. image numbers 2, 3, 4, 5, and 7, from the STARE dataset. From left to right, we show the input image, ground truth, and the results yielded by BCDUNet, MultiResUNet, SegNet, U-Net++, and our network.}
	\label{visualSTARE}%
\end{figure*}%

\begin{figure*}[h!]
	\centering
	\resizebox{0.7\textwidth}{!}{%
		\begin{tabular}{ccccc}
            Image & Ground Truth& SegNet & vVessSeg & MRC-Net \\
			\includegraphics[width=0.14\textwidth]{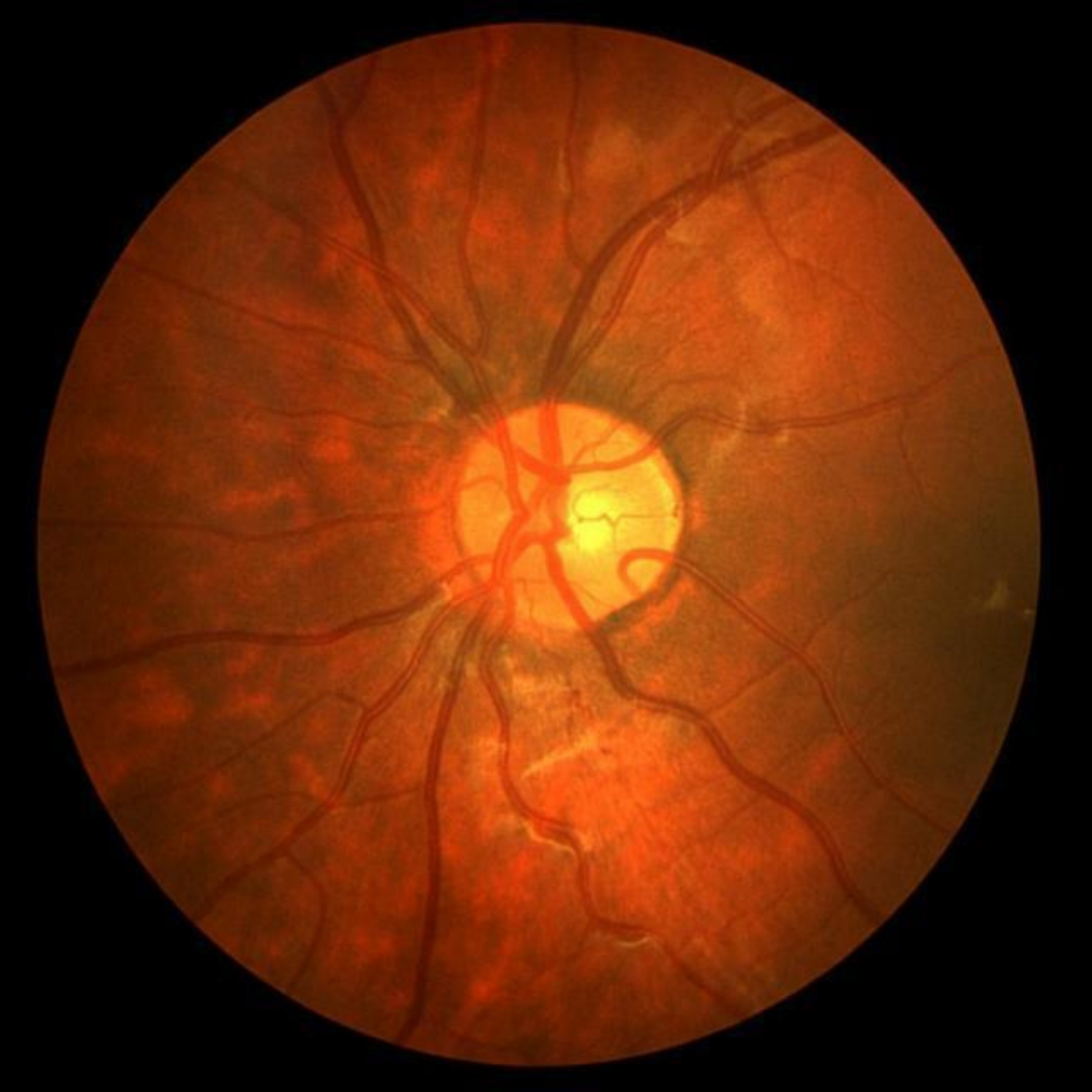}   & \includegraphics[width=0.14\textwidth]{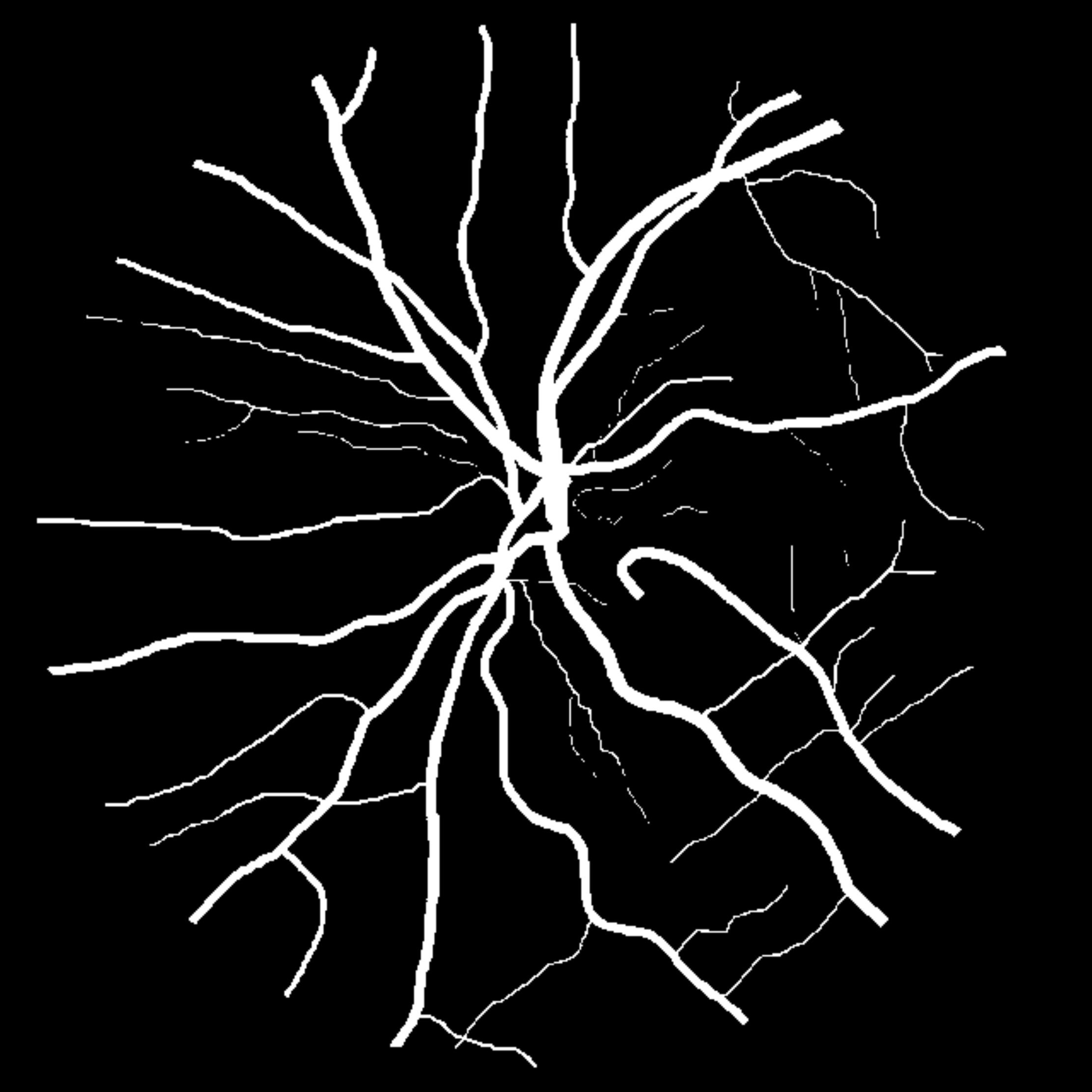}    &  \includegraphics[width=0.14\textwidth]{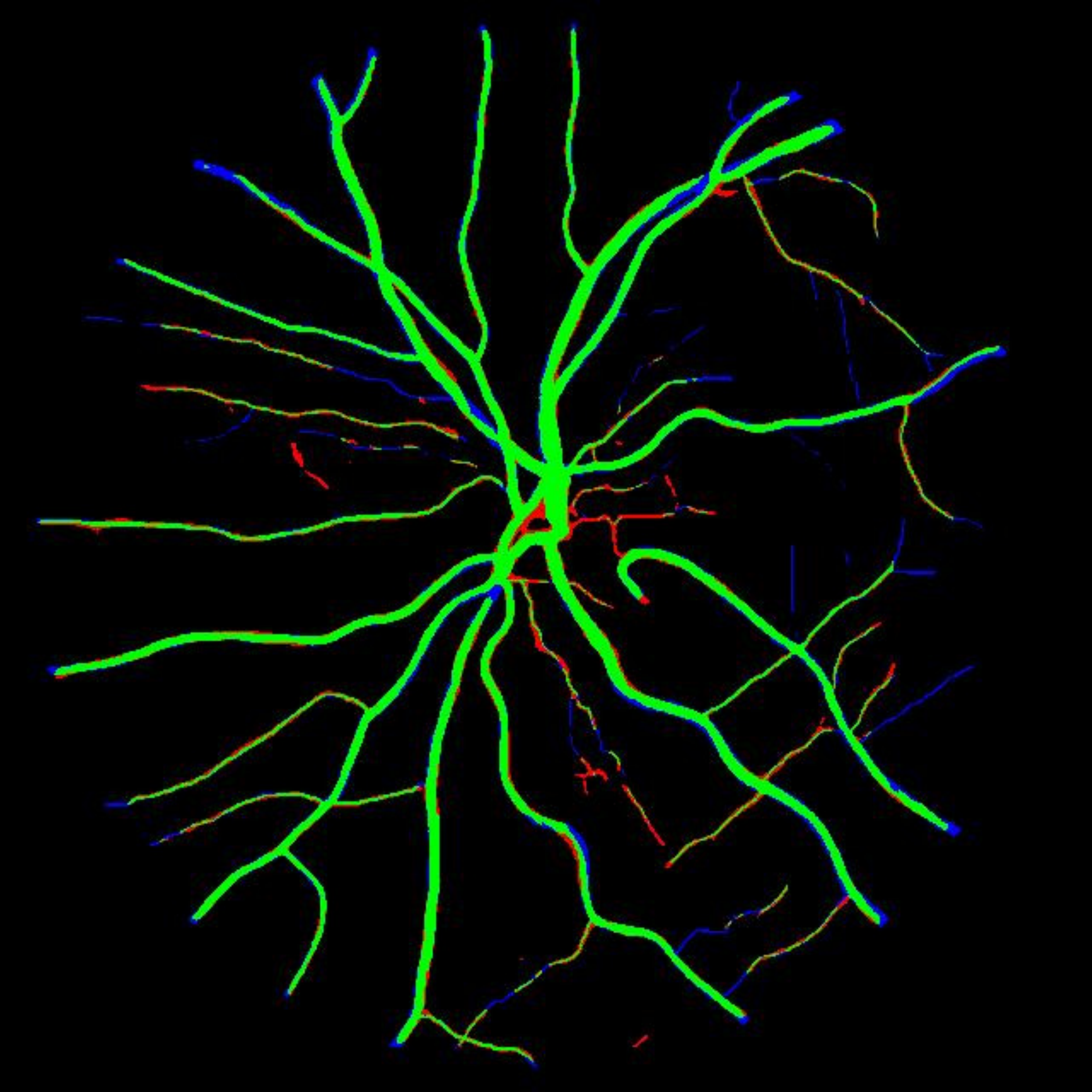}      & \includegraphics[width=0.14\textwidth]{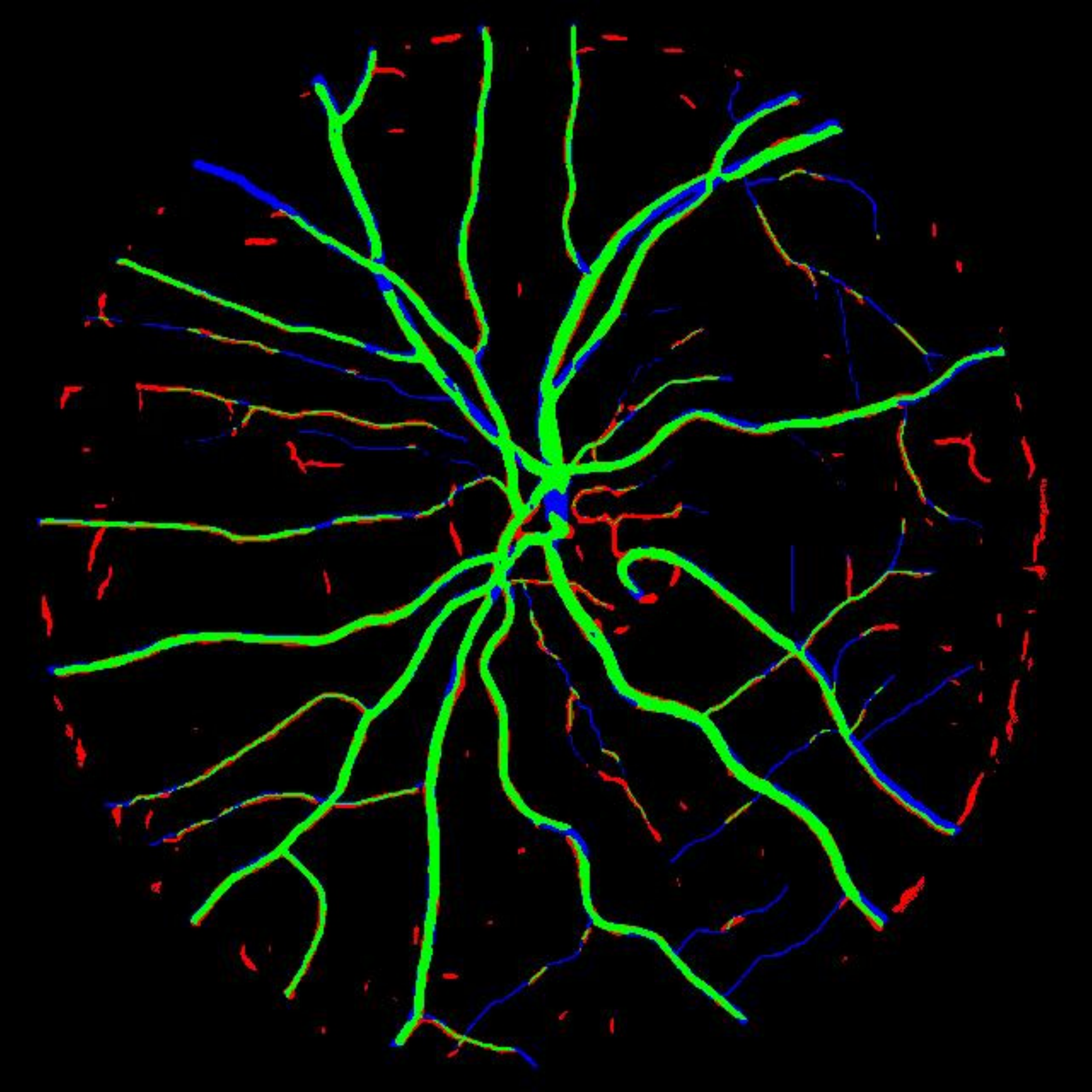}      & \includegraphics[width=0.14\textwidth]{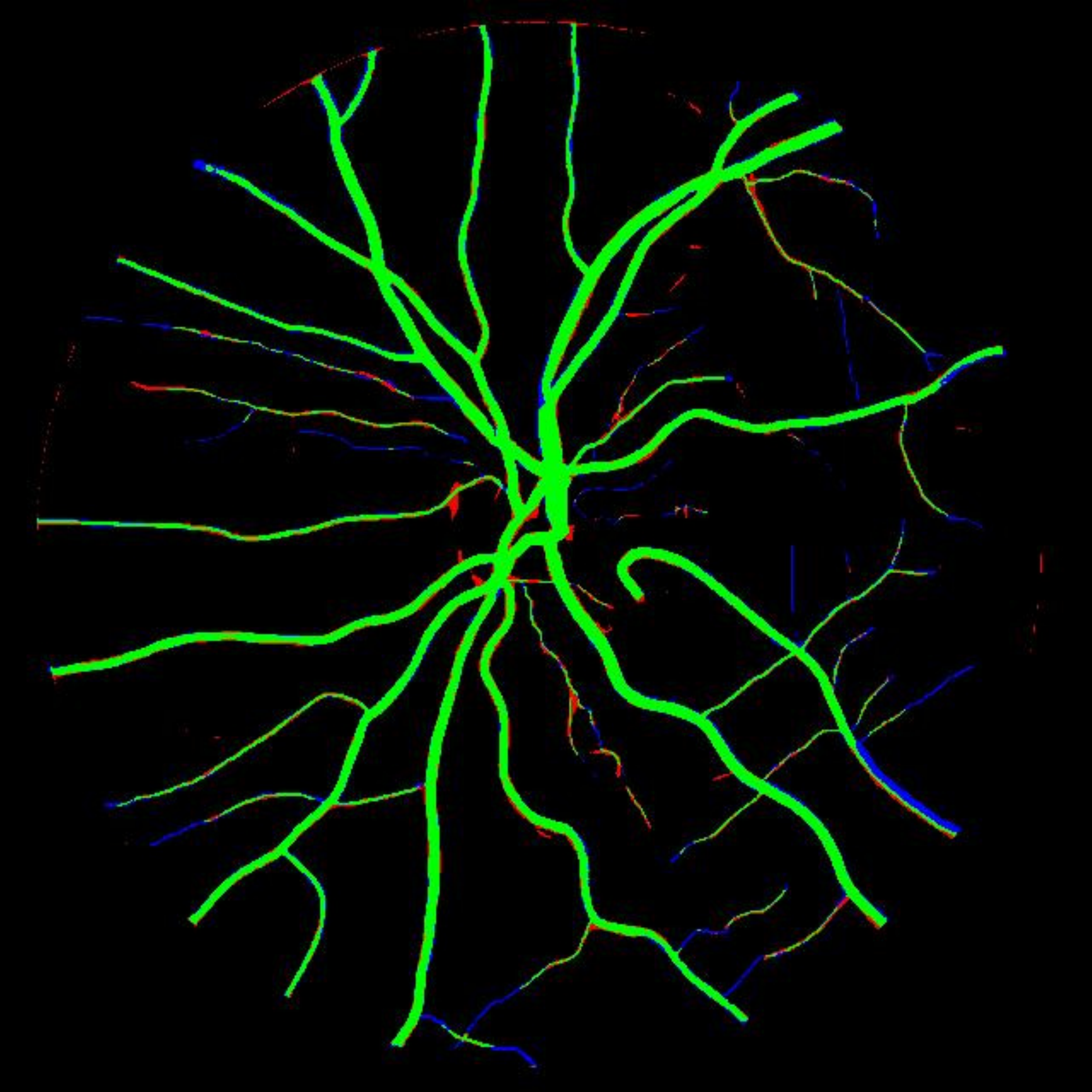} \\
			\includegraphics[width=0.14\textwidth]{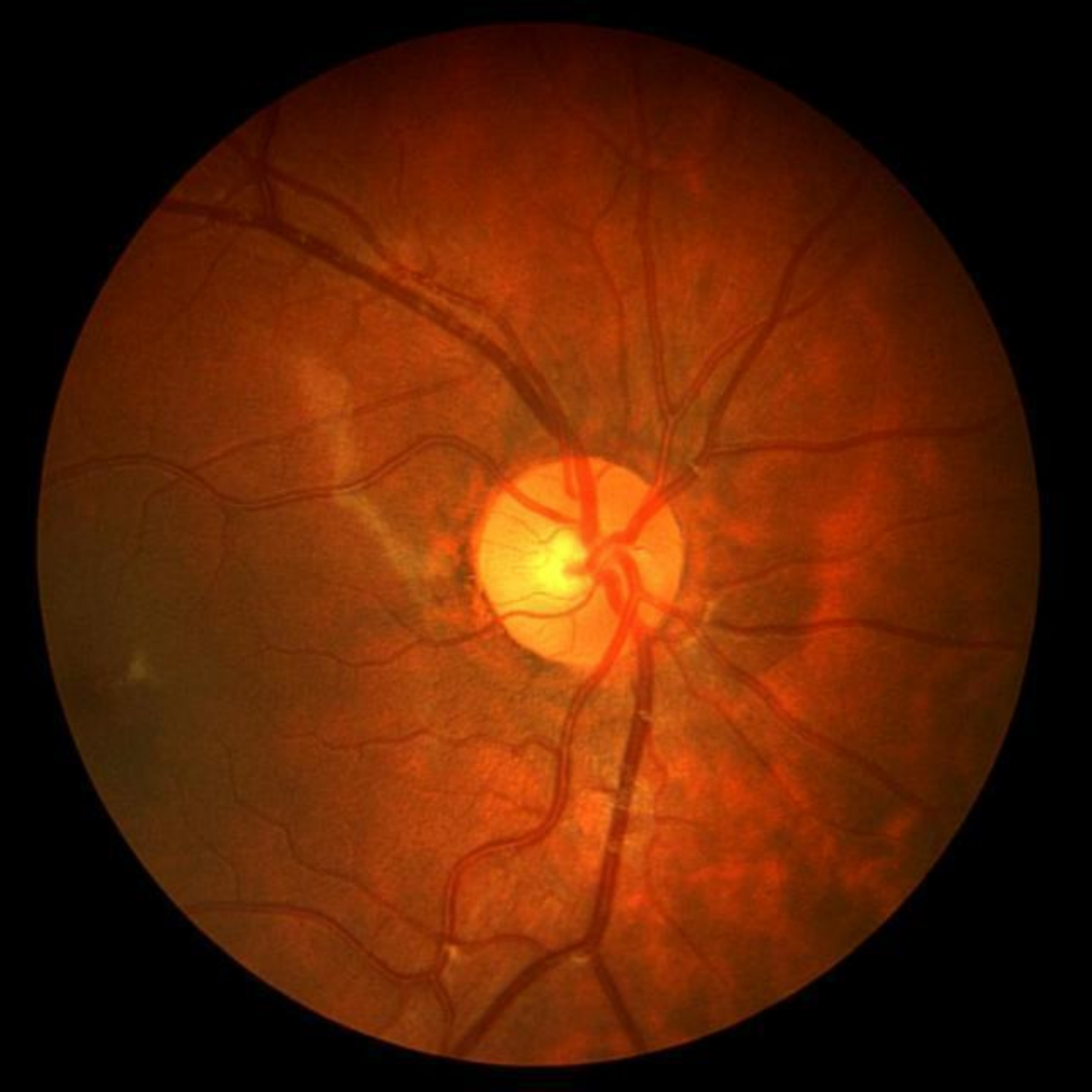}   & \includegraphics[width=0.14\textwidth]{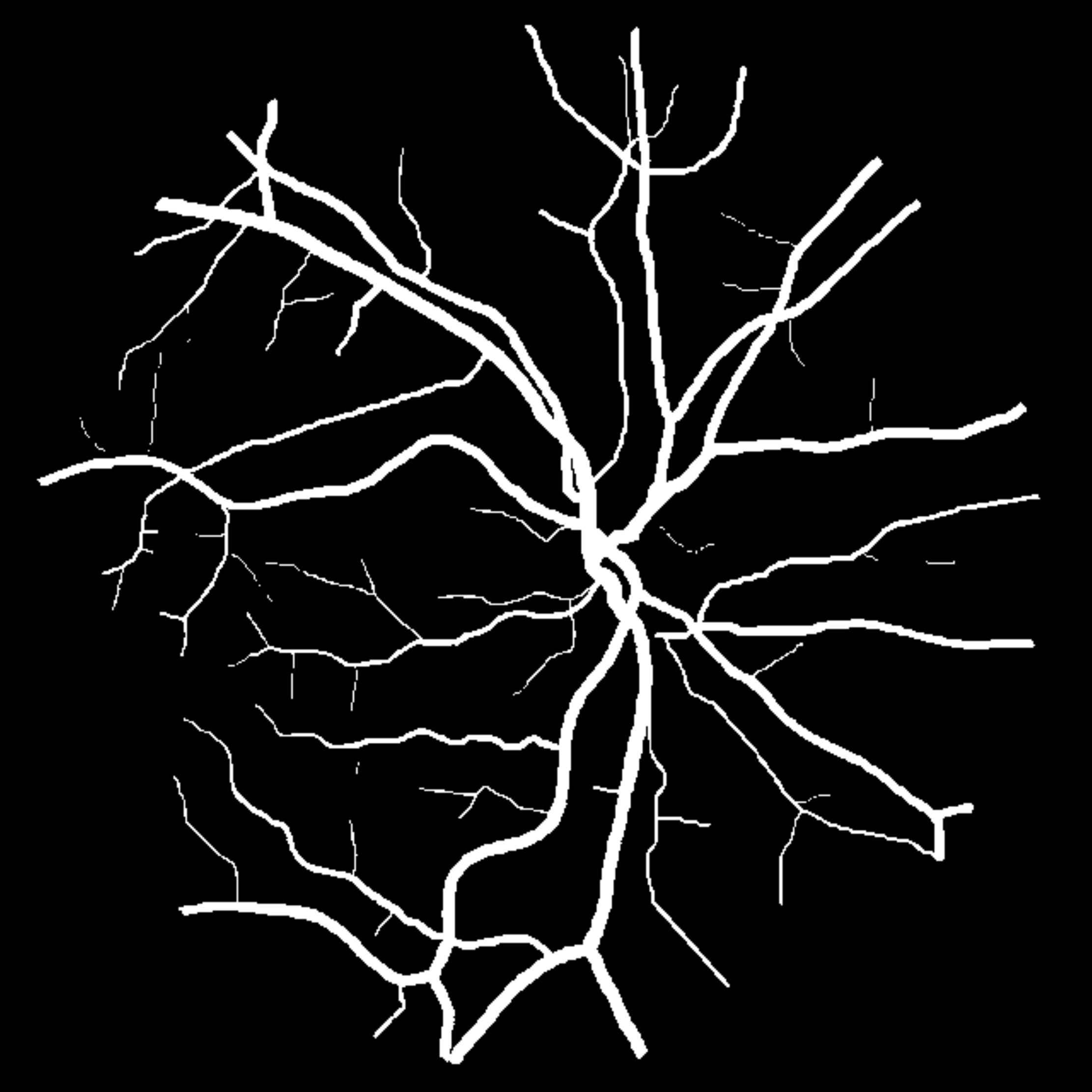}      &   \includegraphics[width=0.14\textwidth]{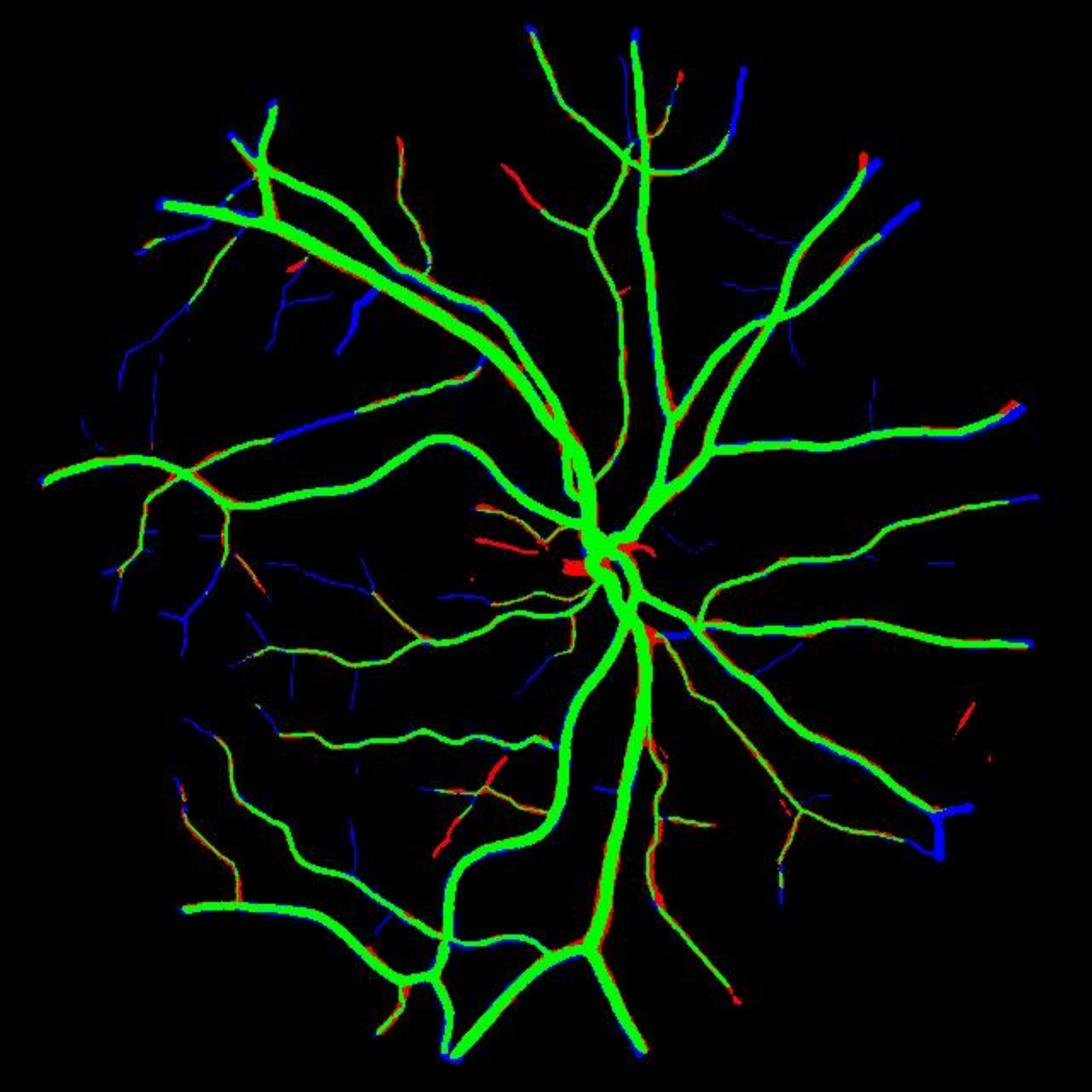}      & \includegraphics[width=0.14\textwidth]{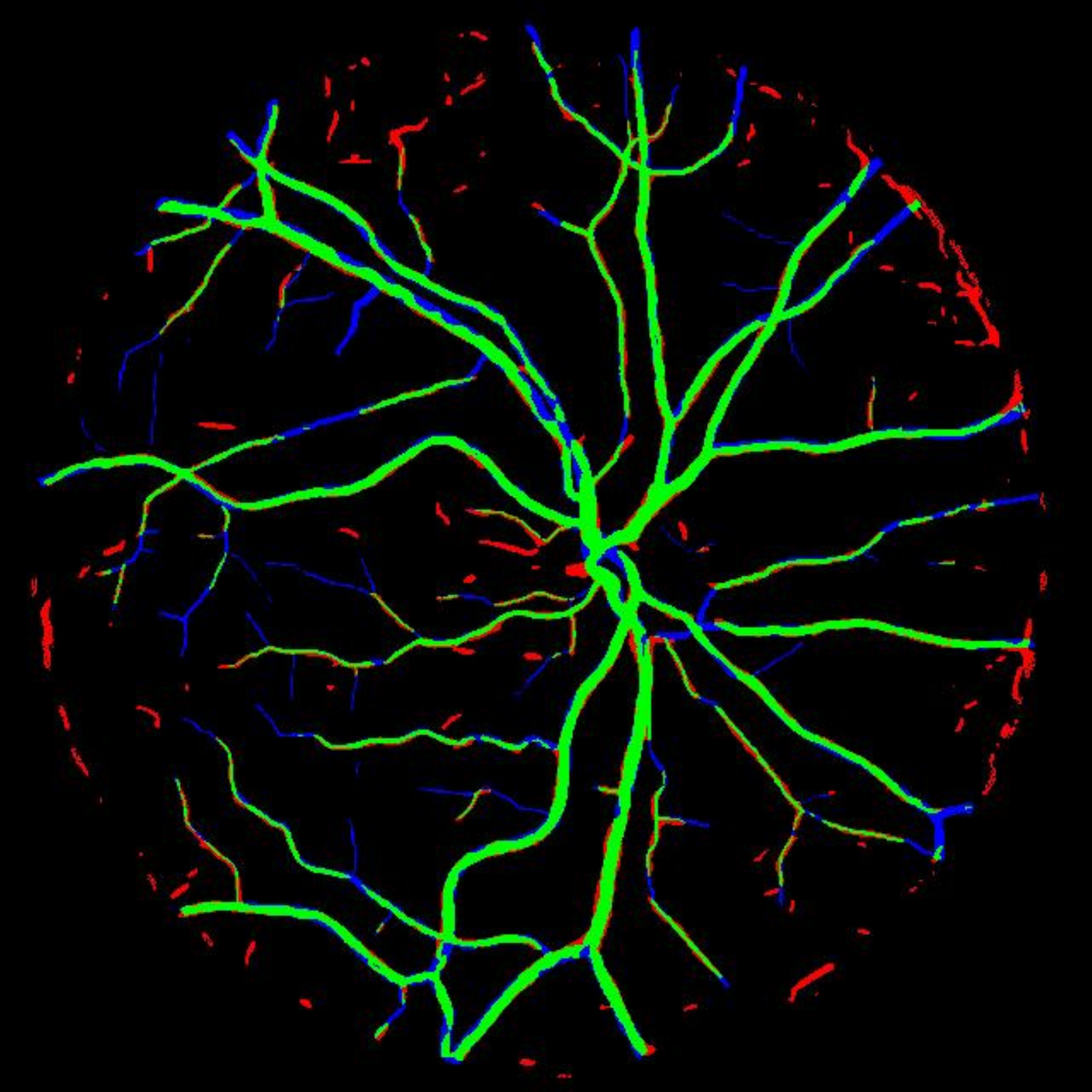}      & \includegraphics[width=0.14\textwidth]{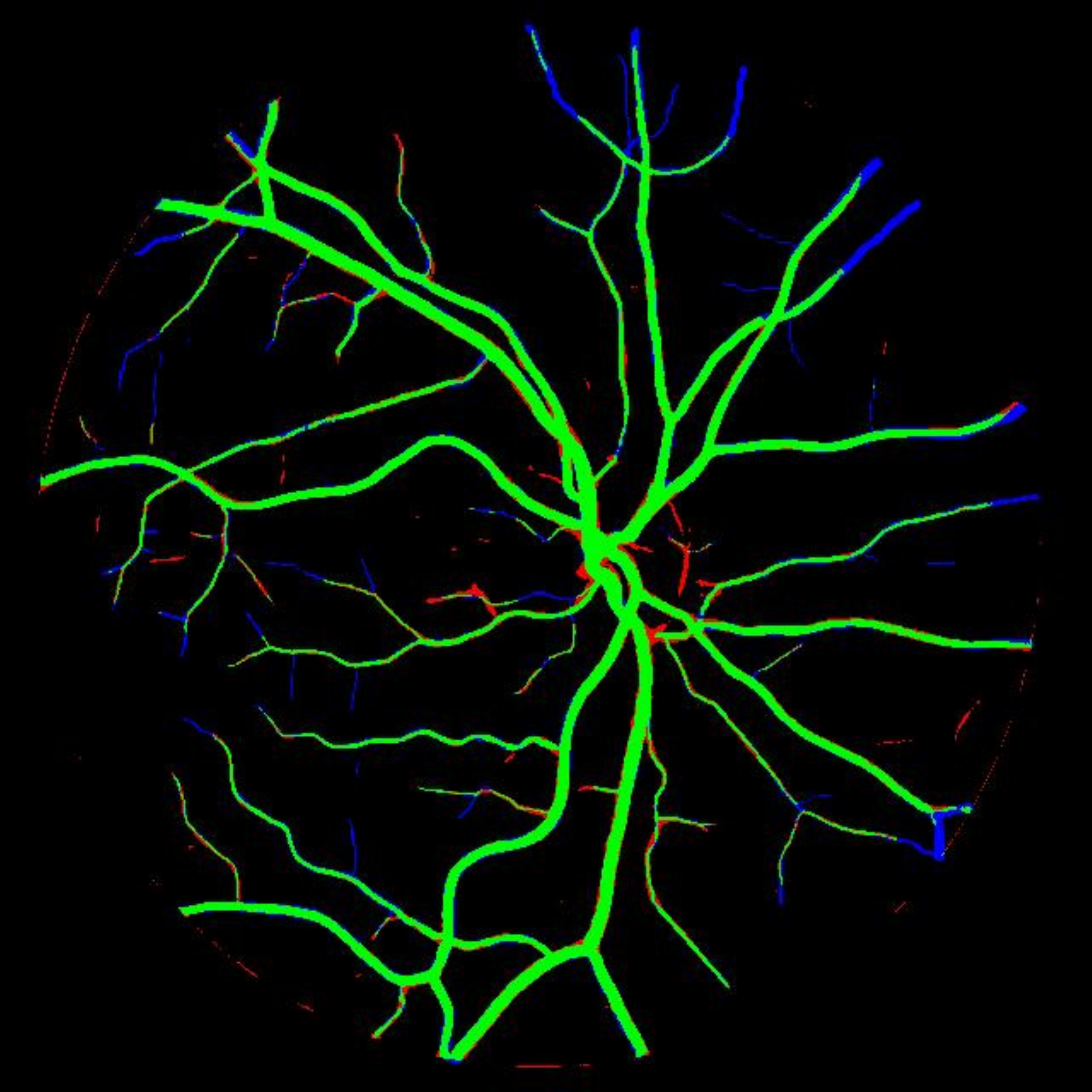} \\
			\includegraphics[width=0.14\textwidth]{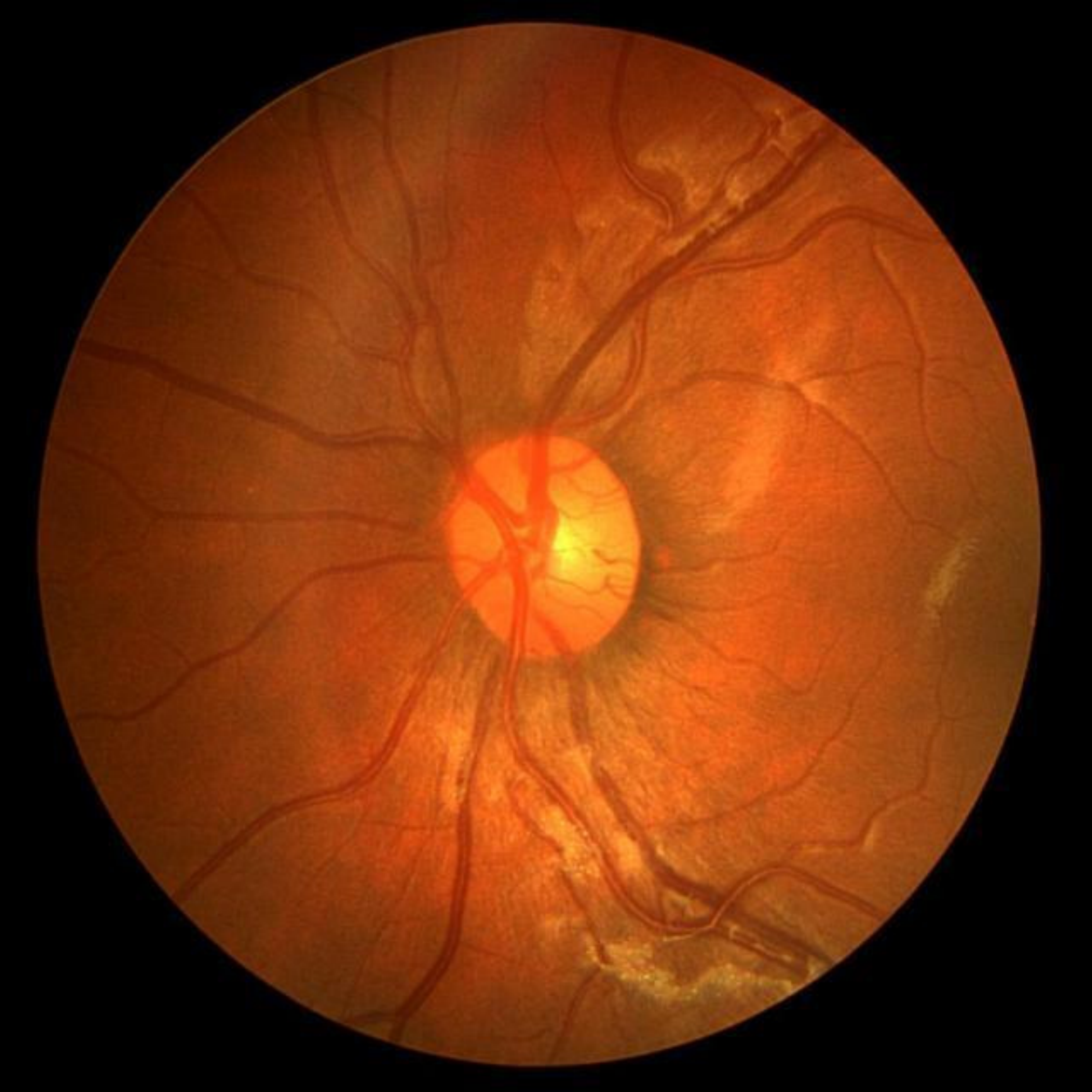}   & \includegraphics[width=0.14\textwidth]{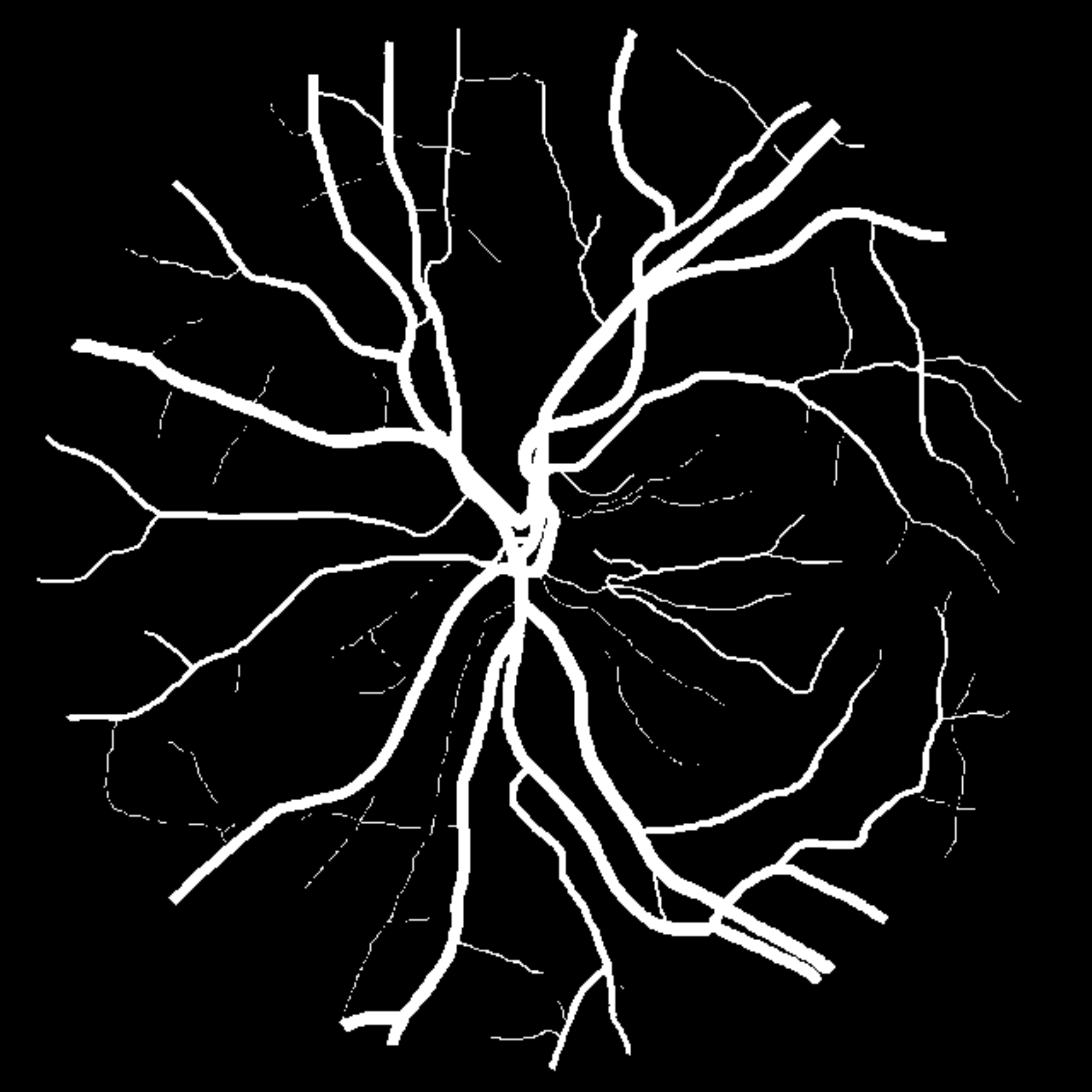}      &      \includegraphics[width=0.14\textwidth]{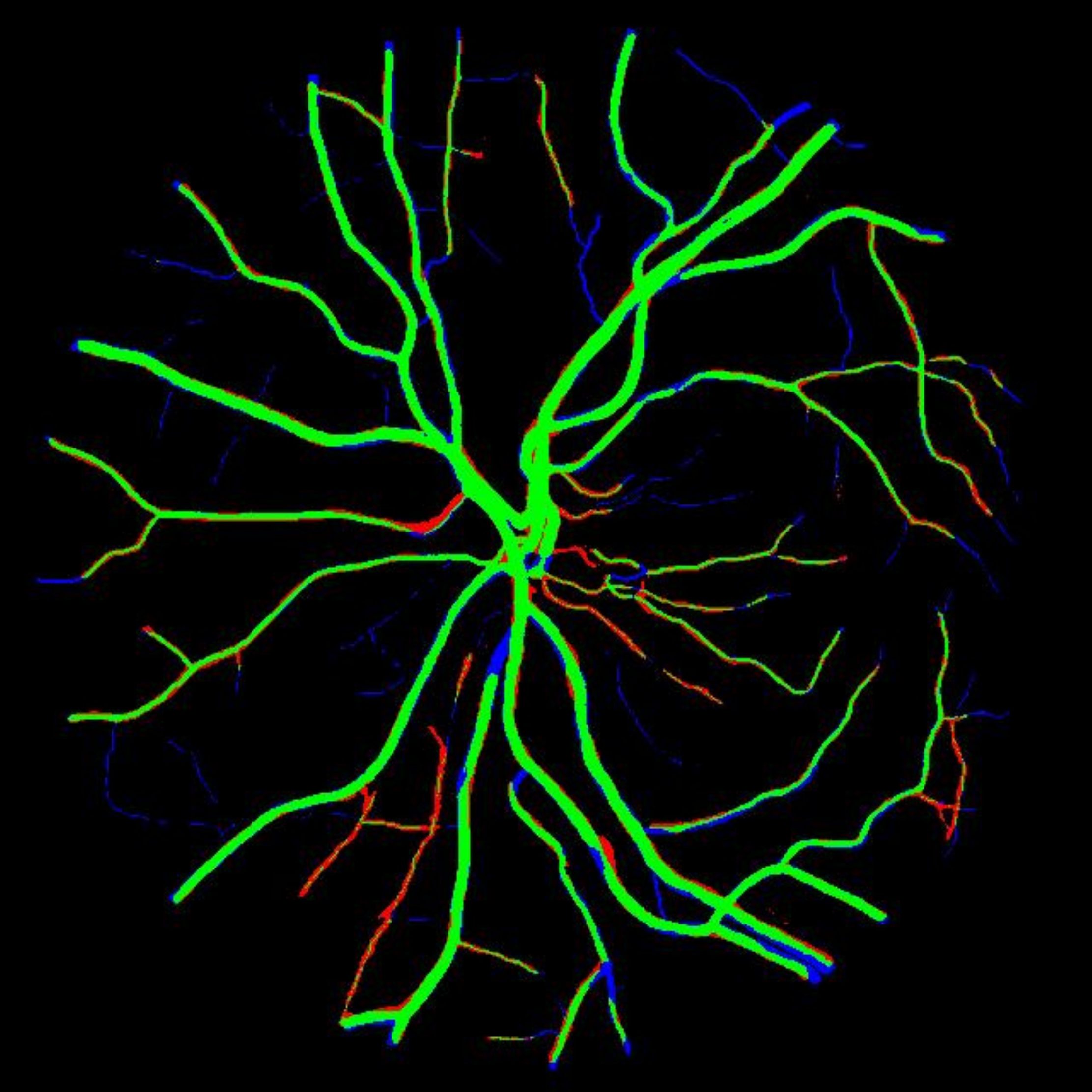}      & \includegraphics[width=0.14\textwidth]{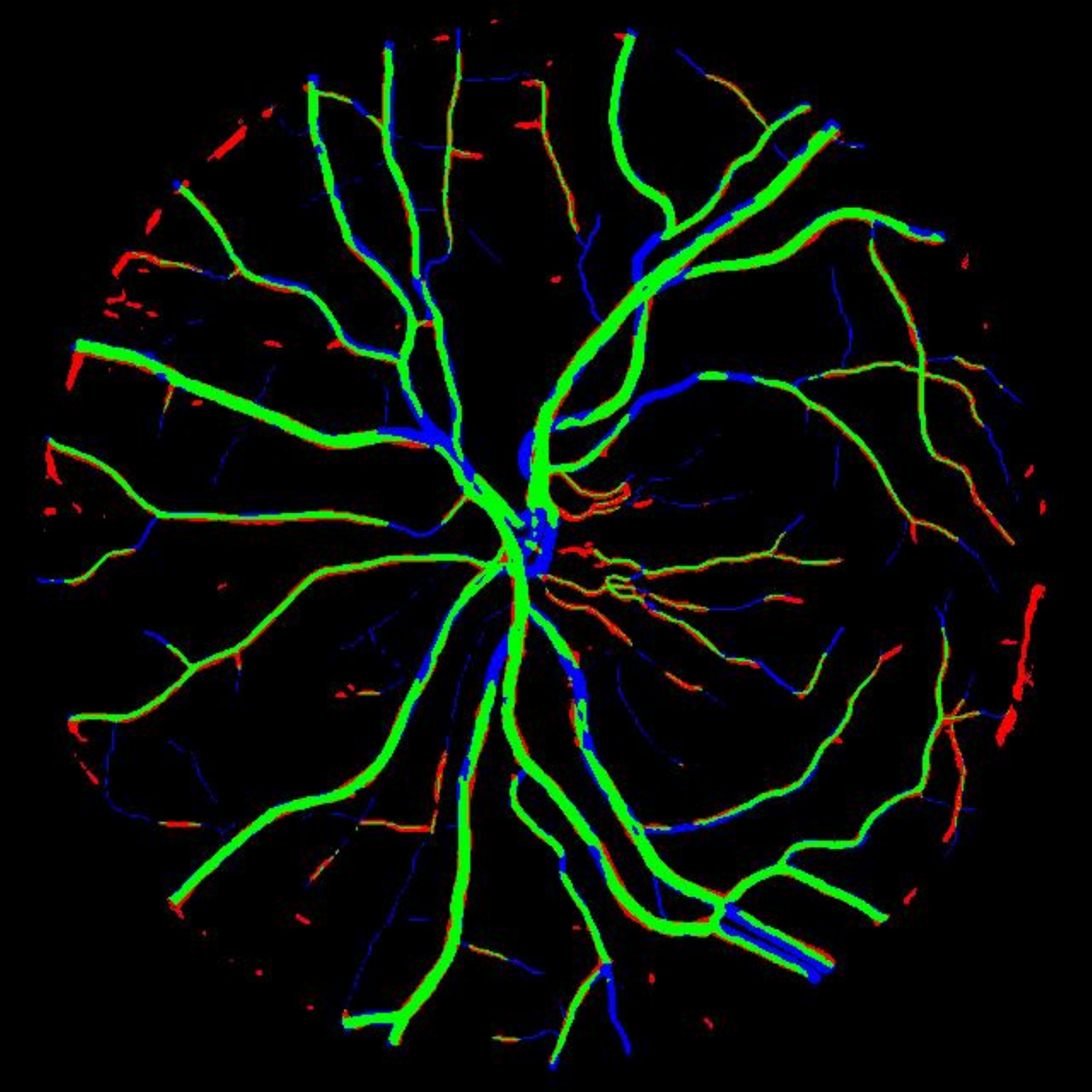}      & \includegraphics[width=0.14\textwidth]{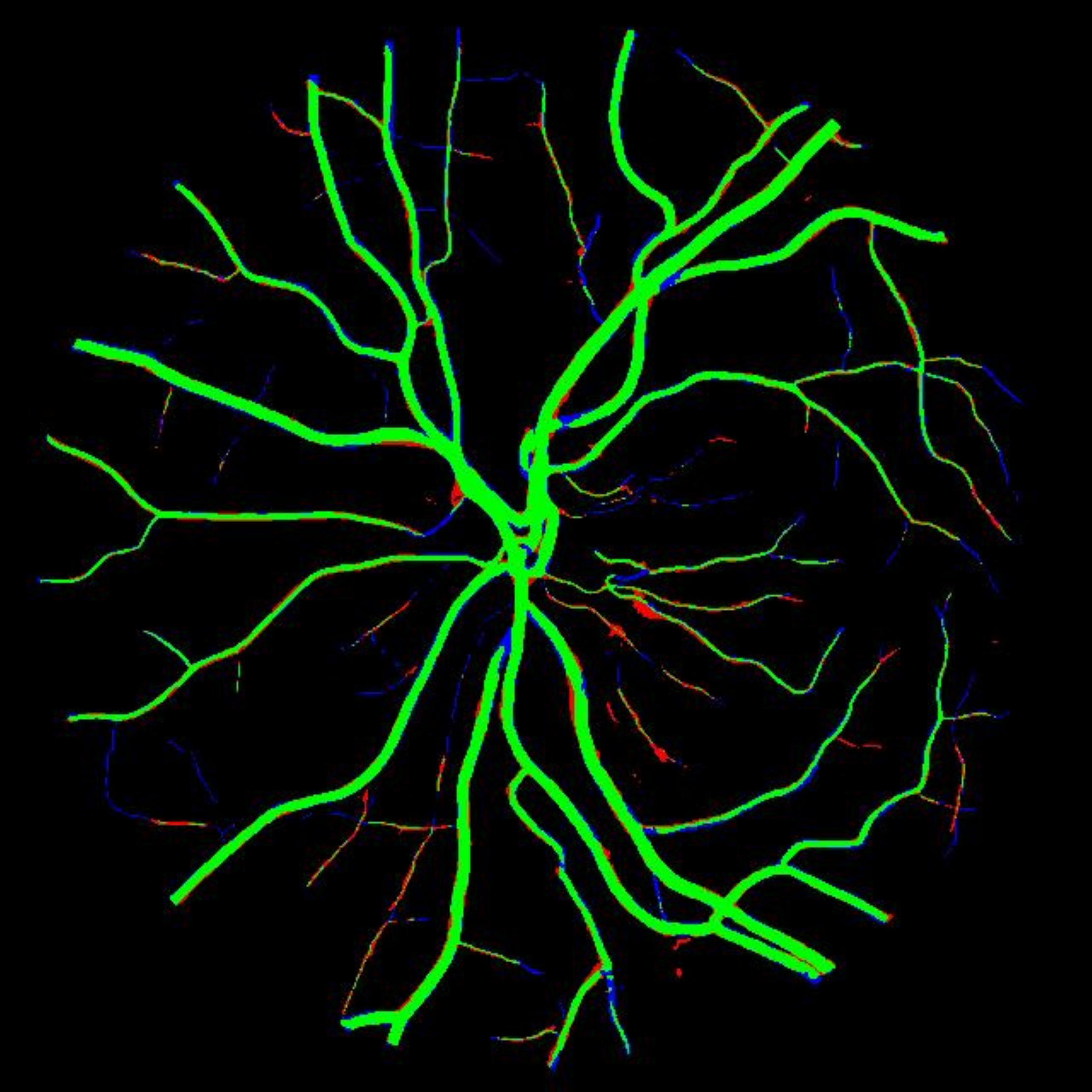} \\
			\includegraphics[width=0.14\textwidth]{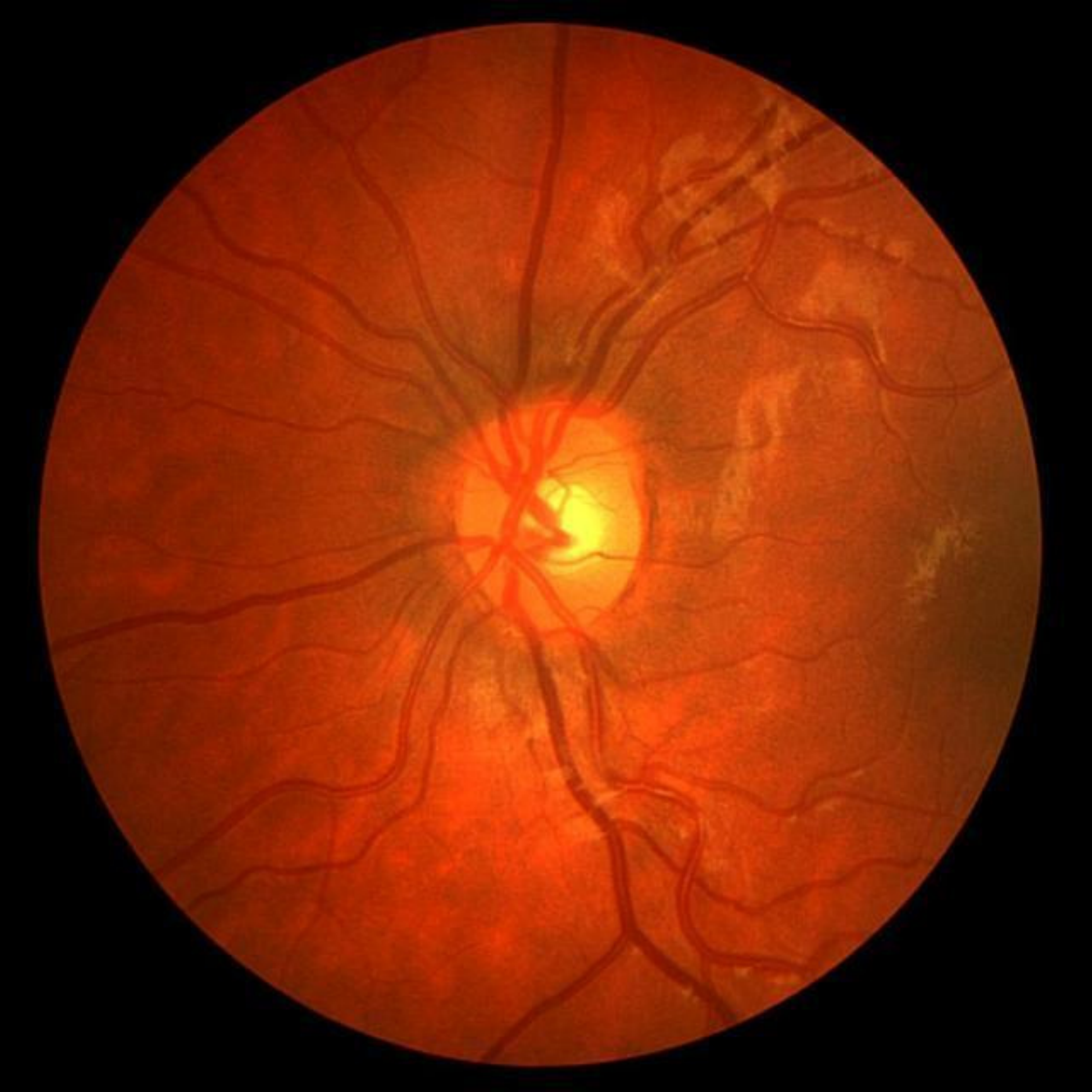}   & \includegraphics[width=0.14\textwidth]{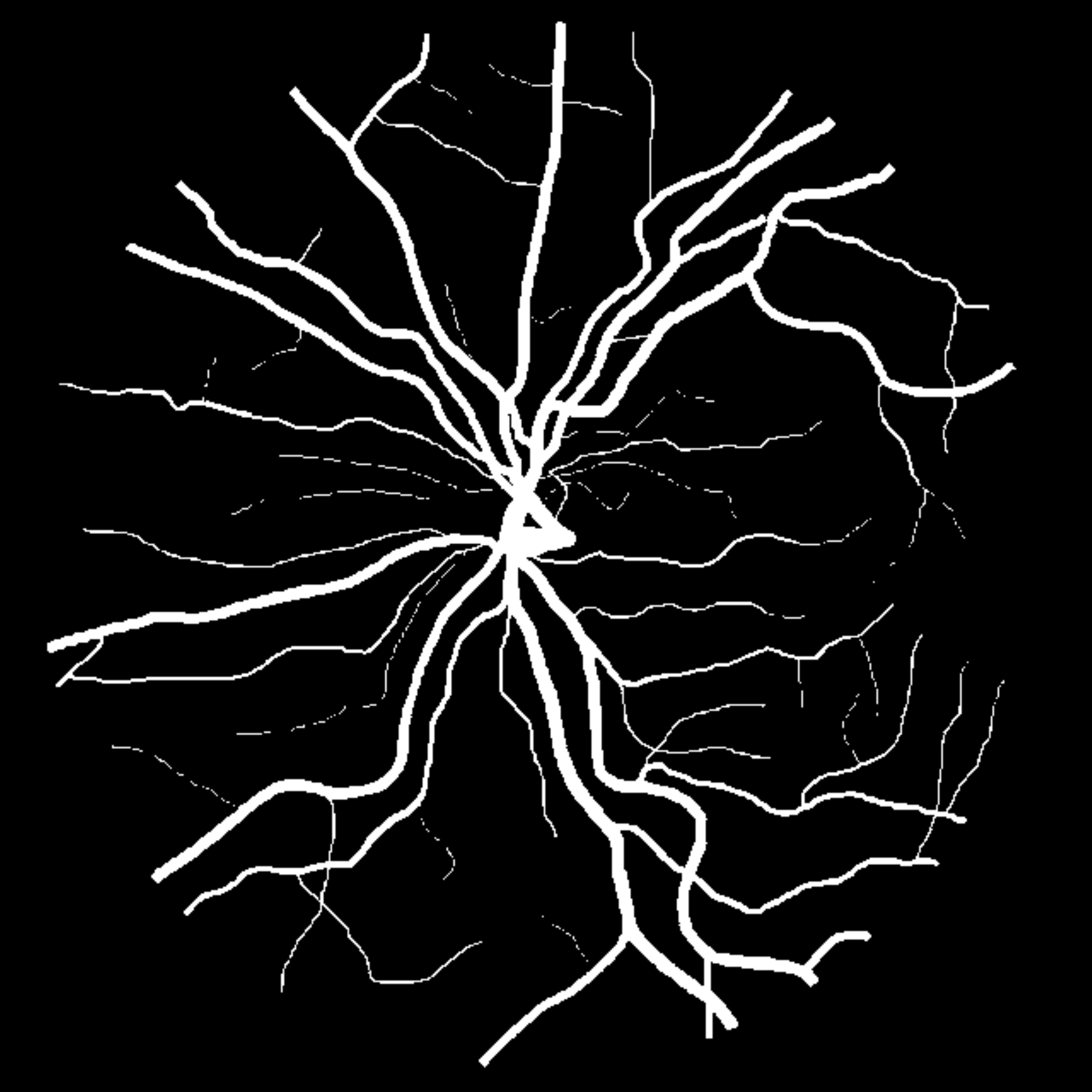}      &      \includegraphics[width=0.14\textwidth]{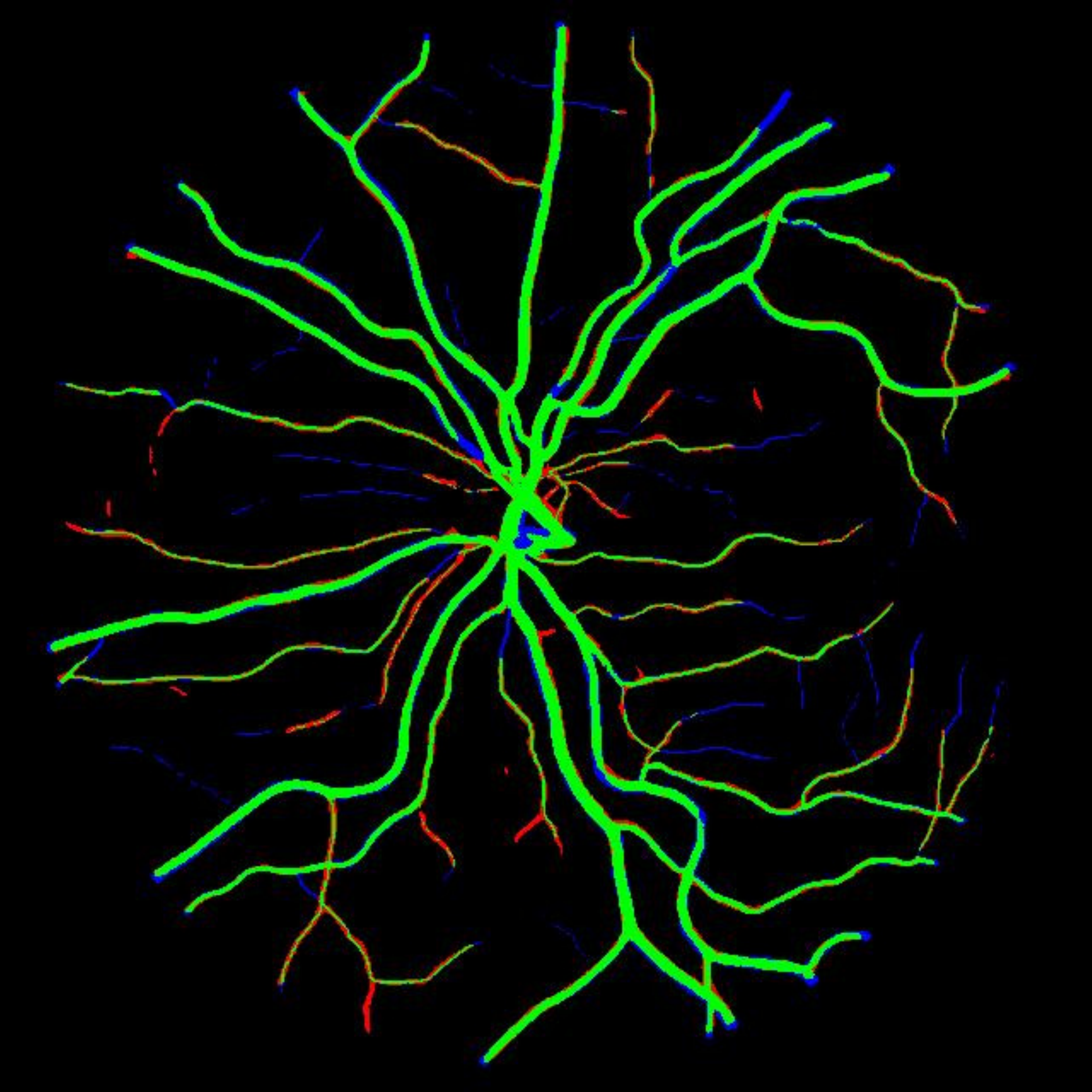}      & \includegraphics[width=0.14\textwidth]{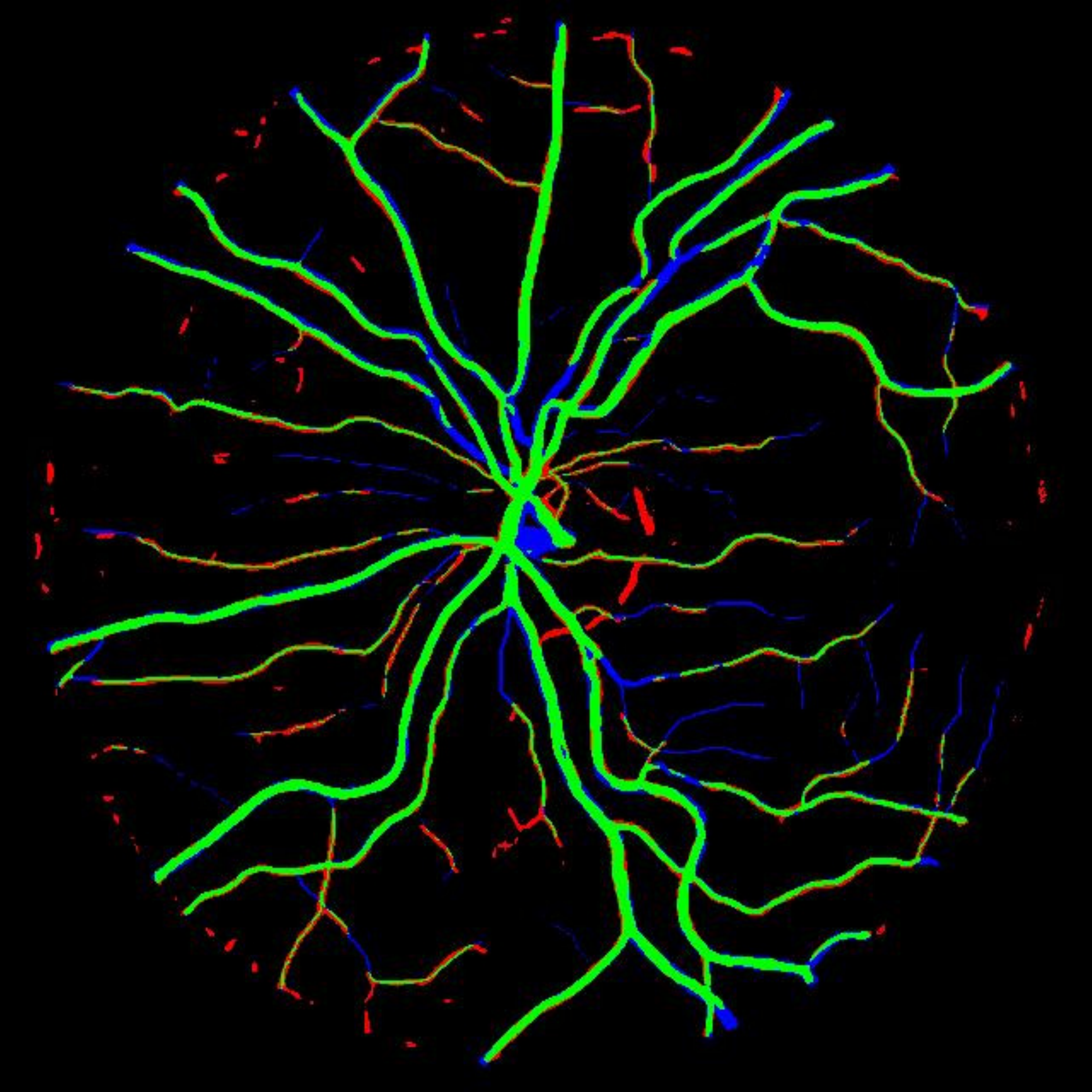}      & \includegraphics[width=0.14\textwidth]{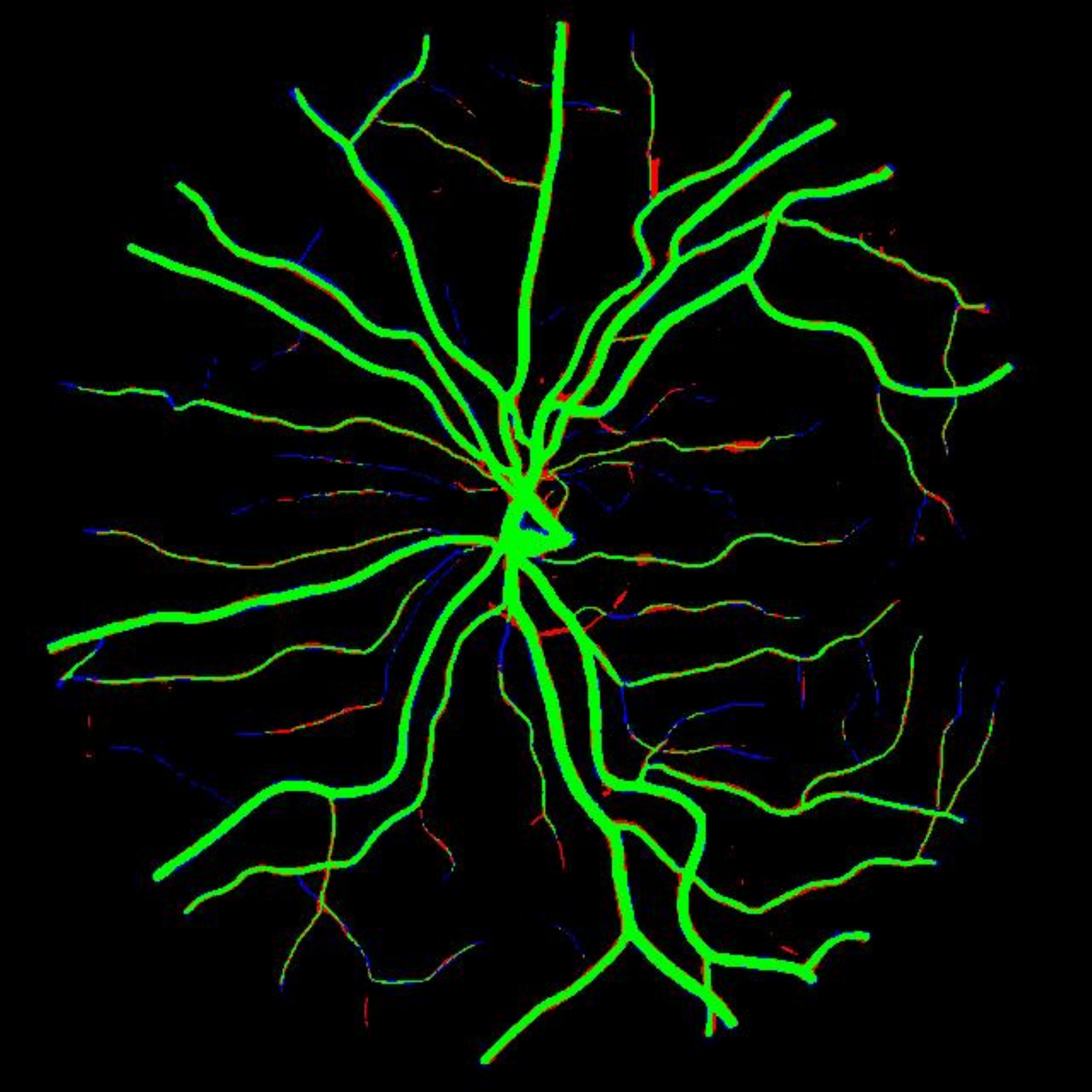} \\
			\includegraphics[width=0.14\textwidth]{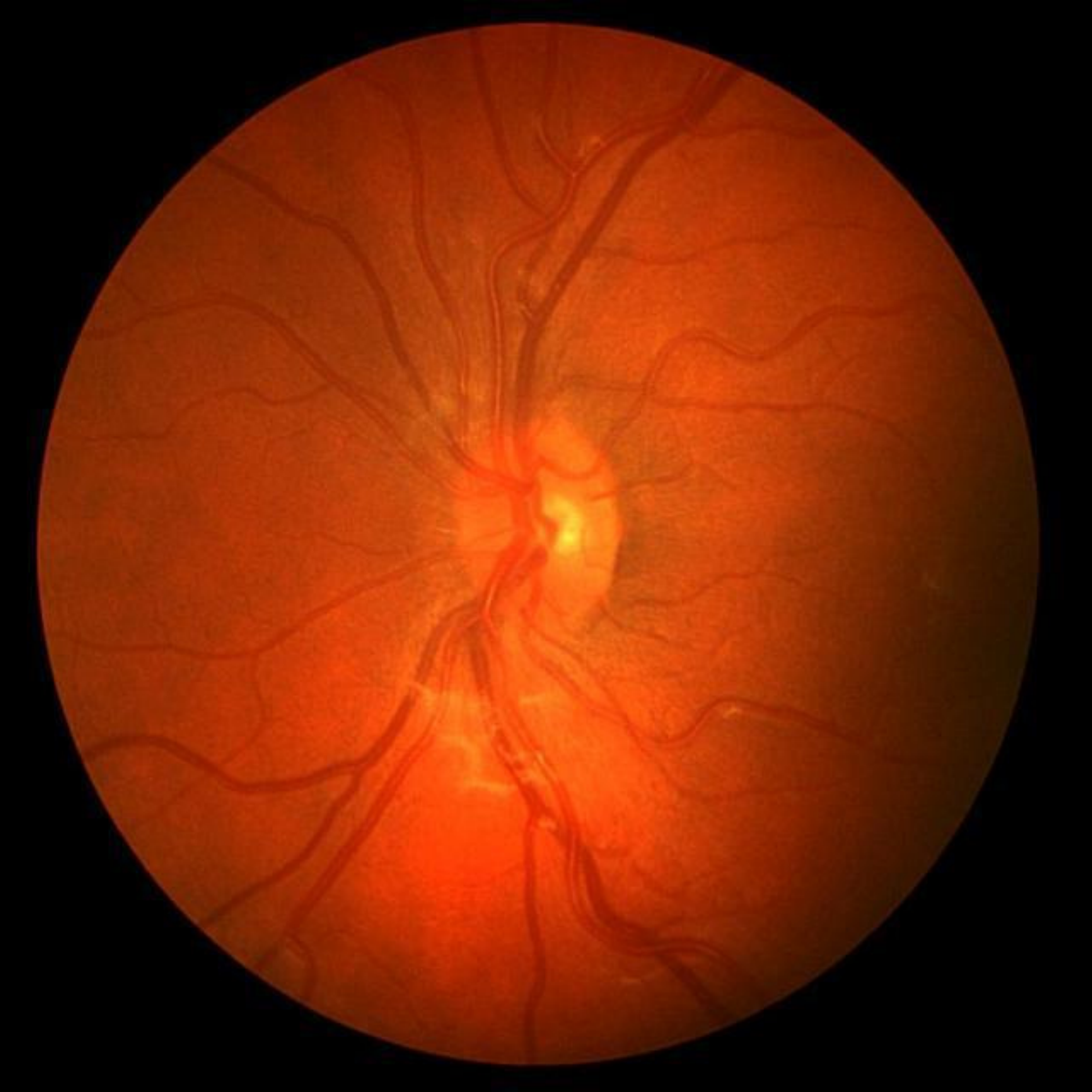}   & \includegraphics[width=0.14\textwidth]{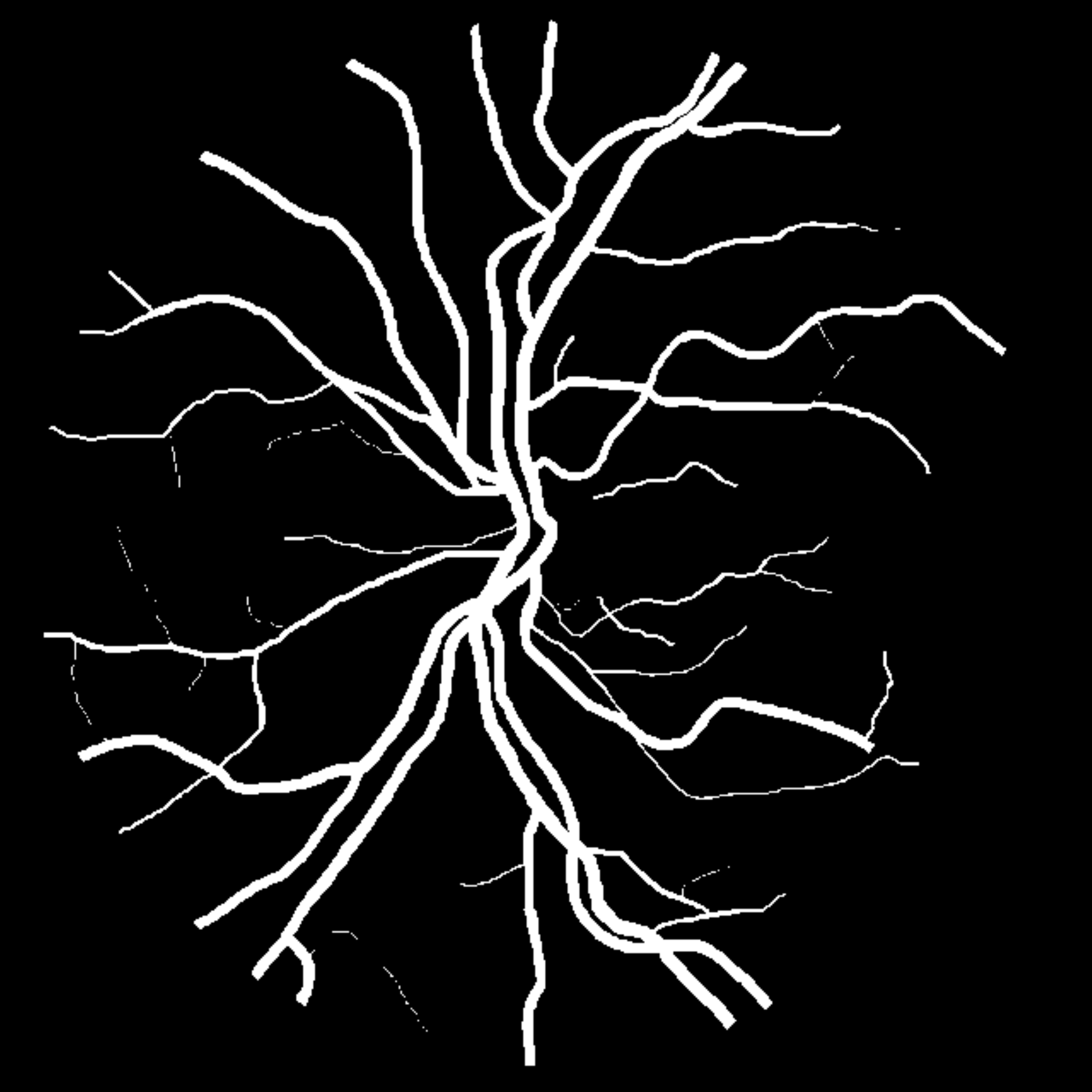}      &      \includegraphics[width=0.14\textwidth]{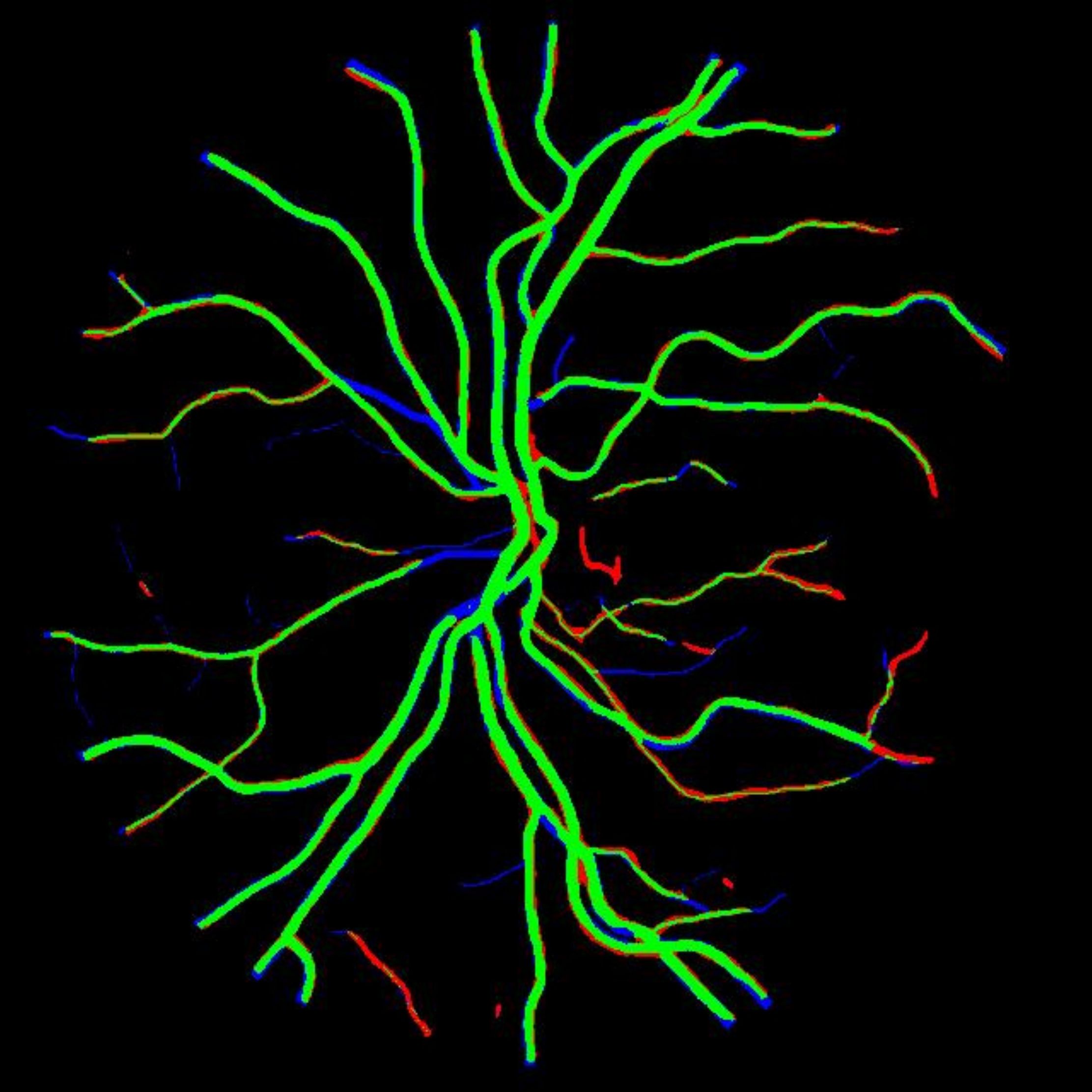}      & \includegraphics[width=0.14\textwidth]{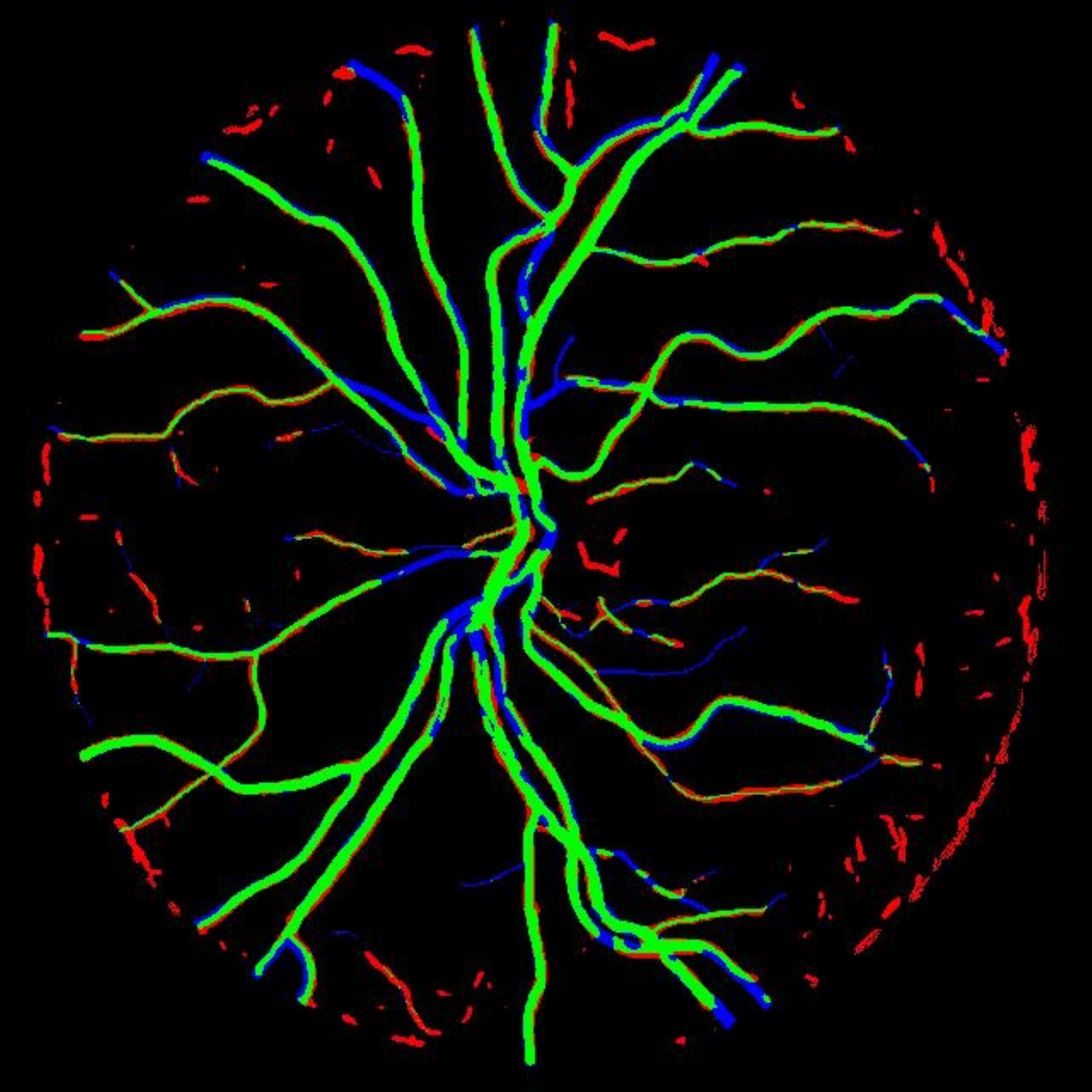}      & \includegraphics[width=0.14\textwidth]{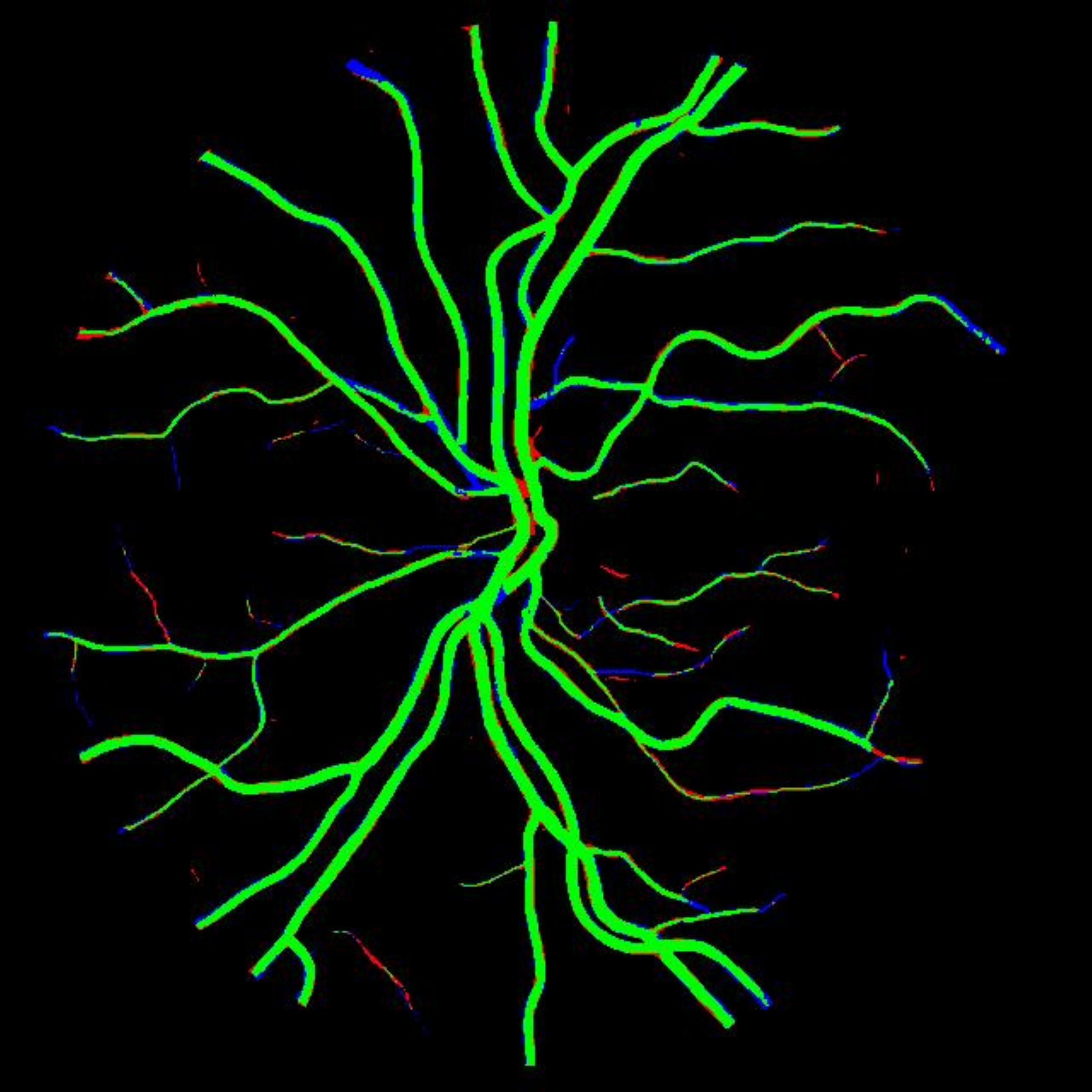} \\
		\end{tabular}%
	}
	\caption{Segmentation results delivered by our MRC-Net method on representative test images, i.e. image numbers 1, 2, 3, 5, and 7, from the CHASE dataset. From left to right, we show the input image, ground truth, and the results yielded by SegNet, VessSeg, and our network.}
	\label{visualCHASE}%
\end{figure*}%
\subsection{Evaluation Criteria}
The segmentation maps for vessels consist of binary images, where each pixel is classified as either part of a vessel or the background. These maps are created using manually labelled ground truth data by expert ophthalmologists, who categorize each pixel in the image as either vascular or nonvascular. When evaluating the accuracy of vessel segmentation methods, there are four possible outcomes for each image: true positive (TP) when vascular pixels are correctly identified as such, true negative (TN) when nonvascular pixels are correctly classified, false positive (FP) when nonvascular pixels are wrongly identified as vascular, and false negative (FN) when vascular pixels are mistakenly categorized as nonvascular. Therefore, to assess the effectiveness of different approaches in the literature, five commonly used parameters, including Sensitivity, Specificity, Accuracy, F1, and AUC, are utilized for comparison.






The AUC, Mathews correlation coefficient (Mathews) \cite{matthews1975}, Jaccard index (J), overlapping error (E) and balanced accuracy (Bacc) are also considered objective measures of the segmentation performance.

\label{ResultsandDiscussion}
\subsection{Ablation Study}
To investigate the contribution of the proposed multi-scale contextual feature extraction and the bi-directional recurrent feature fusion scheme to the overall performance of MRC-Net, we performed multiple experiments with architectural modifications. For the first experiment, we replaced the multi-resolution block with a simple convolutional layer and concatenated semantically different features directly. The resulting architecture was termed MRC-Net without Multi-resolution+BConvLSTM. The second architectural modification was termed as MRC-Net without Multi-resolution, which included the bi-directional recurrent block to minimize the semantic gap between features, whereas the multi-resolution block was replaced by a convolutional layer. The third modification included the multi-resolution block while the bi-directional recurrent block was removed. Table~\ref{tab:ablationMB} presents the results of the aforementioned experiments. It can be observed that the multi-resolution contextual feature block resulted in incremental improvement, whereas the robust feature fusion through a bi-directional recurrent block obtained more substantial improvements. The combined representational power of both modules affected performance improvements of 1.31\%, 0.94\%, and 2.32\% in terms of F1, Bacc, and J, respectively. 
 
The segmentation accuracy of the generator network in our proposed method was enhanced by incorporating adversarial training. Table~\ref{ablation1} displays a comparison of the segmentation performance achieved by the proposed method, with and without the application of adversarial training. The results demonstrate an improvement of 2.4\%, 1\%, and 3.37\% in terms of F1 measure for DRIVE, STARE, and CHASE datasets in support of the adversarial training. Improvements of 1.5\% and 5.65\% in terms of J on STARE and CHASE databases were also obtained. Additionally, consistent improvements in Bacc and E clearly demonstrate the effectiveness of the proposed end-to-end adversarial learning strategy. 

Table~\ref{losses} presents the contribution of the segmentation losses to the proposed MRC-Net method when applied to the DRIVE dataset. In the table, we have denoted MRC-Net+Dice as the option where our MRC-Net is implemented solely with the Dice loss. The MRC-Net+IoU denotes the case where the loss employs only the Intersection over Union (IoU). From the table, it can be observed that by introducing the segmentation-aware loss functions based on the IoU, the segmentation performance in terms of F1 and Jaccard index (J) is improved. It is also noticeable that the introduction of the Dice to the loss improves the balanced accuracy (Bacc). Although the IoU loss resulted in improved performance on the DRIVE dataset, the Dice loss exhibited better overall generalization ability on all the datasets in our experiments. 

{It can be observed from Tables}~\ref{tab:ablationMB}, \ref{ablation1} and \ref{losses} {that our method is designed through systematic evaluation of its stages including feature extraction, feature fusion, loss function, and adversarial framework. The methodical improvements are evident from the ablation with adversarial learning contributing most to the performance improvement. From the ablation results, we note that the proposed method focuses more on boosting the overall F1 and Jaccard index (J) scores while compromising on other measures. This can be attributed to the proposed adversarial framework settings and reinforces our design choice as discussed in Section}~\ref{overallArch}. {The contributions of the proposed feature extraction and fusion schemes to the overall performance are evident from Table}~\ref{tab:ablationMB}. 



\subsection{Comparison and Experiments}
\begin{table}
	\centering
	\caption{Performance of the proposed method in comparison to state-of-the-art approaches on the STARE database.}
	\resizebox{0.5\columnwidth}{!}{%
		\begin{tabular}{cccccc}
			\toprule
			\multicolumn{6}{c}{\textbf{Unsupervised methods}}\\
			\hline
			\textbf{Methods} & \textbf{Year} & \textbf{Se} & \textbf{Sp} & \textbf{Acc} & \textbf{AUC} \\
			\midrule
			\cite{khan2018robust} &2018 & 0.790 &0.965 &0.951 & -\\
			\cite{Khawaja2019a} &2019&0.7980&0.9732&0.9561 & -\\
			\cite{Khawaja2019a} &2019&0.7860&0.9725&0.9583 & -\\
			\cite{Khawaja2019} &2019&0.8011&0.9694&0.9545 & -\\
			\midrule
			\toprule
			\multicolumn{6}{c}{\textbf{Supervised methods}}\\
			\hline
			\textbf{Methods} & \textbf{Year} & \textbf{Se} & \textbf{Sp} & \textbf{Acc} & \textbf{AUC} \\
			\midrule
			\cite{Li2016} & 2016  & 0.7726 & 0.9844 & 0.9628 & 0.9879 \\
			\cite{Orlando2016} FC & 2017  & 0.7680 & 0.9738 & -   & - \\
			\cite{Orlando2016} UP & 2017  &0.7692 & 0.9675 & -   & - \\
			\cite{Yan2018} & 2018  & 0.7581 & 0.9846& 0.9612 & 0.9801 \\
			\cite{Olaf2015} U-Net & 2018 & 0.8270 & 0.9842 & 0.9690 & 0.9898 \\
			\cite{chen2018encoder} Deeplab v3++ & 2018 & 0.8320 & 0.9760 & 0.9650 & 0.9735 \\
			\cite{Soomro2019} Strided U-Net & 2019 & 0.8010 & 0.9690 & 0.9610 & 0.9450 \\
			\cite{Wang2019} DEU-Net &   2019    &   0.8074    & 0.9821     & 0.9661  & 0.9812\\
			\cite{Wu2019} Vessel-Net  &   2019    &   0.8132     & 0.9814     & 0.9661 & 0.9860\\
			\cite{Arsalan2019} VessNet &   2019    &   0.8526      & 0.9791   & 0.9697   & \textbf{0.9883}\\
			\cite{Wang2020} HAnet  &   2020    &   0.8239    & 0.9813  &  0.9670   & 0.9871\\
			\cite{IBTEHAZ202074} MultiRes UNet & 2020 & 0.7126 & \textbf{0.9908} & 0.9703 & 0.9444 \\
			\cite{M.Khan2020} RCED-Net & 2020  & \textbf{0.8397} & 0.9792 & 0.9659 & 0.9810\\
			MRC-Net   & 2020  & 0.8190 & 0.9874 & \textbf{0.9747} & 0.9706 \\
			\bottomrule
		\end{tabular}%
	}
	\label{STARE}%
\end{table}%
Our method and other state-of-the-art techniques were evaluated on three different datasets, namely STARE, DRIVE, and CHASE\_DB1. We first present a comprehensive comparison of various supervised and unsupervised methods. Subsequently, we compare our approach with U-Net \cite{Olaf2015} and SegNet \cite{Badrinarayanan2017} based methods. Our aim is to provide the reader with a detailed and self-contained comparison of our approach, while also demonstrating its performance against popular deep networks that are commonly used as benchmarks in the field.
\begin{table}
	\centering
	\caption{Quantitative results of our network in comparison to other approaches on the CHASE\_DB1 dataset.}
	\resizebox{0.5\columnwidth}{!}{%
		\begin{tabular}{cccccc}
			\toprule
			\multicolumn{6}{c}{\textbf{Unsupervised methods}}\\
			\hline
			\textbf{Methods} & \textbf{Year} & \textbf{Se} & \textbf{Sp} & \textbf{Acc} & \textbf{AUC} \\
			\midrule
			\cite{Roychowdhury2015} & 2015  & 0.7615 & 0.9575 & 0.9467 & 0.9623 \\
			\cite{Azzopardi2015} & 2015  & 0.7585 & 0.9587 & 0.9387 & 0.9487 \\
			\cite{Zhang2016} & 2016  & 0.7626 & 0.9661 & 0.9452 & 0.9606 \\
			\midrule
			\toprule
			\multicolumn{6}{c}{\textbf{Supervised methods}}\\
			\hline
			\textbf{Methods} & \textbf{Year} & \textbf{Se} & \textbf{Sp} & \textbf{Acc} & \textbf{AUC} \\
			\midrule
			\cite{Li2016} & 2016  & 0.7507 & 0.9793 & 0.9581 & 0.9716 \\
			\cite{Orlando2016} FC & 2017  & 0.7277 & 0.9712 & -   & - \\
			\cite{Jin2019} DUNet &   2019    & 0.7595    &   0.9878  &   0.9641   & 0.9832  \\
			\cite{Arsalan2019} VessNet &   2019    & 0.8206    &   0.9800     &   0.9726  & 0.9800  \\
			\cite{Yan2018} & 2018  & 0.7633 & 9809 & 0.9610 & 0.9781 \\
			\cite{Jiang2019c} D-Net & 2019  & 0.7839 &\textbf{0.9894} & 0.9721 & 0.9866 \\
			\cite{Shin2019} VGN & 2019  & \textbf{0.9463}  & 0.9364 & 0.9373 & - \\
			\cite{Wang2020} HAnet &   2020    & 0.8186    &   0.9844     &   0.9673    & \textbf{0.9881}  \\
			\cite{M.Khan2020} RCED-Net & 2020  & 0.8440 & 0.9810 & 0.9722 & 0.9830 \\
			MRC-Net & 2020  & 0.8485 & 0.9887 & \textbf{0.9779} & 0.9857 \\
			\bottomrule
		\end{tabular}%
	}
	\label{CHASEDB1}%
\end{table}%
To ensure a fair comparison with the alternatives, here we have conducted all experiments using the code from the authors where available. Note that there is no single framework to evaluate the methods shown here and, hence, we have followed the tuning and parameter settings specified by the authors and used the same standard training and testing splits for all datasets across all the alternatives in our experiments.     


We begin by presenting Tables \ref{DRIVE}, \ref{CHASEDB1}, and \ref{STARE}, which provide quantitative results for our network, MRC-Net, and a number of alternative methods. As can be observed from the tables, our network consistently outperforms existing methods in terms of accuracy. Additionally, for the CHASE dataset, the proposed network's sensitivity, AUC, and sensitivity are also highly competitive. It consistently ranks among the top models evaluated, and it's important to mention that there's no discernible pattern among the other techniques regarding the most optimal specificity, AUC, or sensitivity.



We now proceed to show a detailed comparison of our network with the results yielded by BCDUNet \cite{azad2019bi}, MultiResUNet \cite{IBTEHAZ202074}, SegNet \cite{Badrinarayanan2017} and U-Net++ \cite{Zongwei2018}. These are based upon U-Net \cite{Olaf2015} and SegNet \cite{Badrinarayanan2017} and have shown recently to deliver state-of-the-art performance.  We commence by showing qualitative results before focusing our attention on a more quantitative analysis. Figure~\ref{visualDRIVE} illustrates a visual comparison between our MRC-Net approach and other methods on the DRIVE database. It can be observed clearly from images 2 and 16 that MRC-Net delivers much fewer false positives on tiny vessels as compared with the state-of-the-art methods. Further, U-Net-based variants struggle on the boundary in image 4. SegNet seems to generate false tiny vessels in the majority of the images. The BCDUNet method tends to miss vessel information, which is robustly captured by the proposed MRC-Net method, while simultaneously suppressing false vessel information.

\begin{table}
	\centering
	\caption{The table presents a comprehensive performance evaluation of our network and several alternative methods on the DRIVE database.}
	\resizebox{0.5\columnwidth}{!}{%
		\begin{tabular}{cccccc}
			\toprule
			\multicolumn{6}{c}{\textbf{Unsupervised methods}}\\
			\hline
			\textbf{Methods} & \textbf{Year} & \textbf{Se} & \textbf{Sp} & \textbf{Acc} & \textbf{AUC} \\
			\midrule
			\cite{Azzopardi2015} & 2015  & 0.7655 & 0.9704 & 0.9442 & 0.9614 \\
			\cite{Roychowdhury2015} & 2015  & 0.7395 & 0.9782 & 0.9494 & 0.9672 \\
			\cite{Yin2015} & 2015  & 0.7246 & 0.979 & 0.9403 & N.A \\
			\cite{Zhang2016} & 2016  & 0.7743 & 0.9725 & 0.9476 & 0.9636 \\
			\midrule
			\midrule
			\multicolumn{6}{c}{\textbf{Supervised methods}}\\
			\hline
			\textbf{Methods} & \textbf{Year} & \textbf{Se} & \textbf{Sp} & \textbf{Acc} & \textbf{AUC} \\
			\midrule
			\cite{Cheng2014}  & 2014  & 0.7252 & 0.9798 & 0.9474 & 0.9648 \\
			\cite{Li2016} & 2016  & 0.7569 & 0.9816  & 0.9527 & 0.9738 \\
			\cite{Orlando2016} FC & 2016  & 0.7893 & 0.9792 & N.A   & 0.9507 \\
			\cite{Orlando2016} UP & 2017  & 0.7076 &0.9870& N.A   & 0.9474 \\
			\cite{Dasgupta2017} & 2017  & 0.7691   & 0.9801 & 0.9533 & 0.9744 \\
			\cite{Yan2018} & 2018  & 0.7653 & 0.9818 & 0.9542 &0.9752 \\
			\cite{Wang2019} DEU-Net  &   2019   &   0.7940    & 0.9816 & 0.9567 & 0.9772 \\
			\cite{Wu2019} Vessel-Net   &   2019    &   0.8038    & 0.9802  & 0.9578 & 0.9821 \\
			\cite{Arsalan2019} VessNet  &   2019    &   0.8022  & 0.9810    & 0.9655      & 0.9820 \\
			\cite{Wang2020} HAnet   &   2020    &   0.7991    & 0.9813   & 0.9581   & 0.9823 \\
			\cite{Yin2020}  &   2020    &   0.8038       & 0.9837    & 0.9578    & \textbf{0.9846} \\
			\cite{Adapa2020} & 2020  & 0.6994 & 0.9811  & 0.945 & N.A \\
			\cite{IBTEHAZ202074} MultiRes UNet & 2020 & 0.7928 & \textbf{0.9845} & 0.9677 & 0.9781 \\
			\cite{M.Khan2020} RCED-Net & 2020  & \textbf{0.8252} & 0.9787 & 0.9649 & 0.9780 \\
			MRC-Net  & 2020  & 0.8250 & 0.9837 & \textbf{0.9698} & 0.9825 \\
			\bottomrule
		\end{tabular}%
	}
	\label{DRIVE}%
\end{table}%

In Figure~\ref{visualSTARE}, a visual comparison of the STARE dataset is shown. It can be seen that the alternative methods generate a greater number of false positives, particularly around retinal boundaries, optic nerves, and small vessels, as evidenced by images 2 and 3. Conversely, MRC-Net exhibits greater robustness to these artifacts in these images.   
Similar results are observed when our method is applied to the CHASE dataset, as shown in Figure~\ref{visualCHASE}.



\begin{table*}
	\centering
	\caption{Performance comparison of our network with benchmark methods on the DRIVE database.}
	\begin{tabular}{lrrrrrrrrrr}
		\toprule
		\textbf{Methods} & \multicolumn{1}{c}{\textbf{Se}} & \multicolumn{1}{c}{\textbf{Sp}} & \multicolumn{1}{c}{\textbf{Acc}} & \multicolumn{1}{c}{\textbf{F1}} & \multicolumn{1}{c}{\textbf{Mathews}} & \multicolumn{1}{c}{\textbf{Dice}} & \multicolumn{1}{c}{\textbf{Bacc}} & \multicolumn{1}{c}{\textbf{J}} & \multicolumn{1}{c}{\textbf{E}} & \multicolumn{1}{c}{\textbf{AUC}} \\
		\midrule
		BCDUNet \cite{azad2019bi} & 0.8192 & 0.9833 & 0.9689 & 0.8217 & 0.8048 & 0.8217 & 0.9013 & 0.6977 & 0.3023 & 0.9775 \\
		MultiResUNet \cite{IBTEHAZ202074} & 0.79  & \textbf{0.9848} & 0.9678 & 0.8108 & 0.7936 & 0.8108 & 0.8874 & 0.6822 & 0.3178 & 0.9784 \\
		SegNet \cite{Badrinarayanan2017} & 0.8246 & 0.9804 & 0.9667 & 0.8127 & 0.7946 & 0.8127 & 0.9025 & 0.6847 & 0.3153 & 0.9752 \\
		U-Net++ \cite{Zongwei2018} & 0.8116 & 0.9823 & 0.9673 & 0.8126 & 0.7948 & 0.8126 & 0.8969 & 0.6851 & 0.3149 & 0.9815 \\
		MRC-Net & \textbf{0.8250} & 0.9837 & \textbf{0.9698} & \textbf{0.8270} & \textbf{0.8106} & \textbf{0.8270} & \textbf{0.9044} & \textbf{0.7055} & \textbf{0.2945} & \textbf{0.9825} \\
		\bottomrule
	\end{tabular}%
	\label{perfCompDRIVE}%
\end{table*}%

The quantitative comparison of the proposed MRC-Net method with the benchmark methods on the DRIVE, STARE, and CHASE databases is presented in Tables~\ref{perfCompDRIVE}, \ref{perfCompSTARE} and \ref{perfCompSTARE}, respectively. The quantitative results support the visual findings presented earlier. Our MRC-Net method outperforms all the alternatives in terms of all measures in the STARE data set. The CHASE and the DRIVE datasets, also consistently outperform all the other methods under consideration in all metrics except the specificity. 

\begin{table*}[]
	\centering
	\caption{Performance comparison of our network with benchmark methods on the STARE database.}
	\begin{tabular}{lrrrrrrrrrr}
		\toprule
		\textbf{Methods} & \multicolumn{1}{c}{\textbf{Se}} & \multicolumn{1}{c}{\textbf{Sp}} & \multicolumn{1}{c}{\textbf{Acc}} & \multicolumn{1}{c}{\textbf{F1}} & \multicolumn{1}{c}{\textbf{Mathews}} & \multicolumn{1}{c}{\textbf{Dice}} & \multicolumn{1}{c}{\textbf{Bacc}} & \multicolumn{1}{c}{\textbf{J}} & \multicolumn{1}{c}{\textbf{E}} & \multicolumn{1}{c}{\textbf{AUC}} \\
		\midrule
		BCDUNet \cite{azad2019bi} & 0.6778 & 0.989 & 0.9659 & 0.7401 & 0.7295 & 0.7401 & 0.8334 & 0.5981 & 0.4019 & 0.88 \\
		MultiResUNet \cite{IBTEHAZ202074} & 0.6989 & \textbf{0.9912} & 0.9694 & 0.7532 & 0.7524 & 0.7532 & 0.845 & 0.626 & 0.374 & 0.9452 \\
		U-Net++ \cite{Zongwei2018} & 0.8152 & 0.9858 & 0.973 & 0.8187 & 0.8041 & 0.8187 & 0.9005 & 0.6946 & 0.3054 & \textbf{0.9832} \\
		MRC-Net & \textbf{0.8190} & 0.9874 & \textbf{0.9747} & \textbf{0.8286} & \textbf{0.8150} & \textbf{0.8286} & \textbf{0.9032} & \textbf{0.7105} & \textbf{0.2895} & 0.9706 \\
		\bottomrule
	\end{tabular}%
	\label{perfCompSTARE}%
\end{table*}%


\subsection{Discussion}

It is worth noting that there are various architectural differences between our network and the alternatives presented in the previous section. One of the key observations is that our method achieves better results while having significantly fewer trainable parameters. In fact, our network is the most economical among the considered methods, with only 0.9 million trainable parameters, compared to 7 million of MultiRes UNet \cite{IBTEHAZ202074}, 9 million of Vess-Net \cite{M.Khan2020}, and 30 million of Strided UNet \cite{Soomro2019}. Moreover, the channel depth of our approach is smaller, which, along with the shortcut connections shown in Figure \ref{multResBlock}, enables our network to preserve fine-grained information. Moreover, our network only employs two pooling layers along with multi-resolution blocks. This not only prevents the loss of spatial information but also reduces the number of trainable parameters. The competitive performance of the proposed MRC-Net as compared with recent lightweight architectures can be observed from Table~\ref{lightweightComp}. Last, but not least, note our approach, in contrast with the alternatives, employs adversarial training and uses LSTM convolutional layers.
\begin{table*}[]
	\centering
	\caption{Performance comparison of the proposed MRC-NET with benchmark alternatives on the CHASE dataset.}
	\begin{tabular}{lrrrrrrrrrr}
		\toprule
		\textbf{Methods} & \multicolumn{1}{c}{\textbf{Se}} & \multicolumn{1}{c}{\textbf{Sp}} & \multicolumn{1}{c}{\textbf{Acc}} & \multicolumn{1}{c}{\textbf{F1}} & \multicolumn{1}{c}{\textbf{Mathews}} & \multicolumn{1}{c}{\textbf{Dice}} & \multicolumn{1}{c}{\textbf{Bacc}} & \multicolumn{1}{c}{\textbf{J}} & \multicolumn{1}{c}{\textbf{E}} & \multicolumn{1}{c}{\textbf{AUC}} \\
		\midrule
		SegNet \cite{Badrinarayanan2017} & 0.8111 & 0.9829 & 0.9698 & 0.8047 & 0.7884 & 0.8047 & 0.897 & 0.6734 & 0.3266 & 0.9805 \\
		vessSeg & 0.738 & 0.9763 & 0.958 & 0.7297 & 0.7072 & 0.7297 & 0.8572 & 0.5748 & 0.4252 & 0.913 \\
		MRC-Net & \textbf{0.8485} & \textbf{0.9887} & \textbf{0.9779} & \textbf{0.8548} & \textbf{0.8430} & \textbf{0.8548} & \textbf{0.9186} & \textbf{0.7466} & \textbf{0.2534} & \textbf{0.9857} \\
		\bottomrule
	\end{tabular}%
	\label{perfCompCHASE}%
\end{table*}%
Also note that, in the results shown in the previous sections, U-Net \cite{Olaf2015} and SegNet \cite{Badrinarayanan2017} consistently rank as the best methods in the publicly available datasets under consideration. This is somewhat expected, as it is expected that unsupervised methods would be among the worst performers. However, this must be taken into consideration, as supervised methods do have an inherent advantage. That being said, we decided to include them here for the sake of the completeness of the results.

In addition, recent studies have proposed modifications to the SegNet and UNet architectures to achieve state-of-the-art performance in retinal vessel segmentation, as reported in \cite{azad2019bi,Zongwei2018}. However, to attain state-of-the-art performance, these methods employ a larger number of trainable parameters, typically around ten times more than our network. Furthermore, in our experiments, we focused on standard datasets and commonly used performance metrics to facilitate better comparison with other published results in the literature. As far as we know, U-Net \cite{Olaf2015} and SegNet \cite{Badrinarayanan2017} remain widely used benchmarks, and as shown in our results, our method is still highly competitive against these benchmarks.
\begin{table}[htbp]
  \centering
  \caption{{Comparison with lightweight methods.}}
    \begin{tabular}{llccc}
    \hline
    \multicolumn{1}{c}{\textbf{{Method}}} & \multicolumn{1}{c}{\textbf{{Params}}} & \textbf{{Acc}} & \textbf{{F1}} & \textbf{{AUC}} \bigstrut\\
    \hline
    ERFNet \cite{Romera2018ERFNet} & 2.06M & 0.9598 & 0.7652 & 0.9633 \bigstrut[t]\\
    M2UNet \cite{Laibacher2019} & 0.55M & 0.9630 & 0.8091 & 0.9714 \\
    Little U-Net \cite{galdran2022state} & 0.034M &  -     & 0.8241 & 0.9798 \\
    Little W-Net \cite{galdran2022state} & 0.068M &   -    & \textbf{0.8279} & 0.9810 \\
    MRC-Net & 0.9M  & \textbf{0.9698} & 0.8270 & \textbf{0.9812} \bigstrut[b]\\
    \hline
    ERFNet \cite{Romera2018ERFNet}  & 2.06M & 0.9716 & 0.7872 & 0.9785 \bigstrut[t]\\
    M2UNet \cite{Romera2018ERFNet} & 0.55M & 0.9703 & 0.8006 & 0.9666 \\
    Little U-Net \cite{Romera2018ERFNet} & \textbf{0.034M} &    -   & 0.8020 & 0.9822 \\
    Little W-Net \cite{Romera2018ERFNet} & 0.068M &    -   & 0.8169 & \textbf{0.9874} \\
    MRC-Net & 0.9M  & \textbf{0.9779} & \textbf{0.8548} & 0.9857 \bigstrut[b]\\
    \hline
    \end{tabular}%
  \label{lightweightComp}%
\end{table}%
\section{Conclusions}
In this paper, we introduce a novel contextual network for the accurate identification of retinal vessels that considers the limitations of computing resources for deploying the network on devices with restricted resources such as smartphones. Our proposed network is an end-to-end trainable architecture that utilizes a multi-resolution feature extraction block to capture vessel information at different scales and preserve residual information with fine-grained detail. Furthermore, we present a feature fusion strategy that robustly combines encoder and decoder features without compromising important vessel information. Notably, our segmentation network has a relatively low parameter count of approximately 0.9 million, which is lower than several state-of-the-art approaches in the literature. We have presented results on three widely available datasets and compared them against several alternatives. In our experiments, our method was quite competitive on the DRIVE, STARE, and CHASE datasets. \hl{The proposed approach investigated bi-directional recurrent feature fusion and region-based compound losses for robust vessel segmentation. In future, we aim to investigate attention-incorporated feature fusion with joint optimization of boundary- and distribution-based losses for extending this approach to medical image segmentation.}

\bibliographystyle{IEEEtran}
\bibliography{References}

\end{document}